\documentclass[a4paper]{article}
\pdfoutput=1 

\usepackage{jheppub} 
\usepackage[T1]{fontenc} 
\usepackage[italian,english]{babel}
\usepackage{subfigure}
\usepackage{amssymb}
\usepackage{epsf}
\usepackage{rotating}
\usepackage{graphicx}
\usepackage{amsmath}
\usepackage{babel}
\usepackage{fancyhdr}
\usepackage{lineno}
\usepackage{color}
\usepackage{multirow}
\usepackage{mathrsfs}
\usepackage{url}
\usepackage{soul}
\usepackage{bm}

\usepackage{hyperref}
\hypersetup{
	colorlinks=true,
	citecolor=[rgb]{0.0,0.32,0.89},
	filecolor=magenta,      
	urlcolor=[rgb]{0.0,0.42,0.1},
	linkcolor=[rgb]{0.6,0.0,0.0},
	allcolors=[rgb]{0.0,0.42,0.24},
}

\usepackage{ifpdf}
\usepackage{subfigure}
\usepackage{tikz}
\usepackage[compat=1.1.0]{tikz-feynman}
\usetikzlibrary{arrows,shapes}
\usetikzlibrary{trees}
\usetikzlibrary{matrix,arrows} 				
\usetikzlibrary{positioning}				
\usetikzlibrary{calc,through}				
\usetikzlibrary{decorations.pathreplacing}  
\usepackage{pgffor}							

\usetikzlibrary{decorations.pathmorphing}	
\usetikzlibrary{decorations.markings}
\tikzset{
	vector/.style={decorate, decoration={snake}, draw},
	fermion/.style={draw=black, postaction={decorate}},
	scalar/.style={dashed,draw=black, postaction={decorate}}}
\tikzstyle{block} = [draw, rectangle,
minimum height=3em, minimum width=6em]

\definecolor{pantoneCB}{rgb}{0.0588235, 0.298039, 0.505882}
\hypersetup{bookmarks=true,unicode=true,pdftoolbar=true,pdfmenubar=true,
	pdffitwindow=false,
	pdfnewwindow=true,colorlinks=true,allcolors=pantoneCB}

\newcommand{\lsim}{\mathrel{\mathop{\kern 0pt \rlap
			{\raise.2ex\hbox{$<$}}}
		\lower.9ex\hbox{\kern-.190em $\sim$}}}
\newcommand{\gsim}{\mathrel{\mathop{\kern 0pt \rlap
			{\raise.2ex\hbox{$>$}}}
		\lower.9ex\hbox{\kern-.190em $\sim$}}}
\newcommand{\be}{\begin{equation}}
\newcommand{\ee}{\end{equation}}
\newcommand{\bea}{\begin{eqnarray}}
\newcommand{\eea}{\end{eqnarray}}

\newcommand*\xbar[1]{%
	\hbox{%
		\vbox{%
			\hrule height 0.65pt 
			\kern0.4ex
			\hbox{%
				\kern-0.05em
				\ensuremath{#1}%
				\kern0.0em
			}%
		}%
	}%
}

\definecolor{Gr}{rgb}{0.0, 0.66, 0.25}

\newcommand{\fbi}{fb$^{-1}$ }

\title{\boldmath Seesaw Models under the Lens of Angular Distributions at $\mu^+\mu^-,\, \mu^+ \mu^+, \, \mu^+ \gamma$ and $\mu^+ e^-$ Colliders}

\author[a,*]{Priyotosh Bandyopadhyay,}
\note[*]{Corresponding author}
\author[b]{Anirban Karan,}
\author[a, c]{and Chandrima Sen}

\affiliation[a]{Indian Institute of Technology Hyderabad, Kandi,  Sangareddy-502285, Telengana, India}
\affiliation[b]{INFN, Gruppo Collegato di Cosenz \& Dipartimento di Fisica, Università della Calabria, Arcavacata di Rende, I-87036, Cosenza, Italy}
\affiliation[c]{Theoretical Physics Division, Physical Research Laboratory, Ahmedabad-380009, Gujarat, India}

\emailAdd{bpriyo@phy.iith.ac.in}
\emailAdd{anirban.karan@unical.it}
\emailAdd{chandrima@prl.res.in}

\begin{document}

\abstract{The Seesaw extensions of the Standard Model {not only} provide a natural explanation for tiny neutrino masses, they also predict additional heavy states such as Majorana neutrinos, charged leptons, and scalar triplets. Distinguishing between these scenarios at future colliders will be essential if such particles are discovered. In this work, we investigate how leptonic colliders offer complementary avenues for this task, focusing on the role of reconstructed angular distributions, which encode information about the underlying matrix elements.
{One only needs to reconstruct the new particle (or SM charged lepton) in the final state and examine its angular distribution relative to the beam axis, without bothering about the other particles in the final state, and this is sufficient to reveal the underlying simple tree-level Seesaw scenario, as considered in this work.}
We perform a detailed PYTHIA8-based simulation for different Seesaw realizations: inverse Type-I, Type-II, and inverse Type-III at $\mu^+\mu^-$, $\mu^+\mu^+$, $\mu^+\gamma$, and $\mu^+e^-$ colliders. Characteristic angular patterns emerge that enable discrimination among the models, with muon colliders providing particularly promising reach. We also comment on the prospects of asymmetric $\mu e$ colliders in probing off-diagonal Yukawa structures.}

    \maketitle

\section{Introduction}

The discovery of neutrino oscillations has firmly established that neutrinos are massive, in contrast to the Standard Model (SM) prediction of massless neutrinos \cite{pdg}. A minimal extension with Dirac masses requires Yukawa couplings as small as $\mathcal{O}(10^{-12})$, which is highly fine-tuned. Seesaw mechanisms provide an elegant resolution to this problem by introducing additional states: heavy Majorana neutrinos in Type-I and Type-III, or an $SU(2)_L$ scalar triplet in Type-II \cite{Hambye:2012fh}.

While the primary motivation of these mechanisms is to explain the smallness of neutrino masses {at tree-level}, they also give rise to a variety of testable collider signatures. Extensive studies have explored such signals, including Majorana neutrinos in Type-I \cite{Minkowski:1977sc,Yanagida:1979as,Gell-Mann:1979vob,Glashow:1979nm,PhysRevLett.44.912}, scalar triplets in Type-II \cite{Konetschny:1977bn,PhysRevD.22.2227,Magg:1980ut,Cheng:1980qt,Lazarides:1980nt,Mohapatra:1980yp,Wetterich:1981bx,Gelmini:1980re,Chun:2013vma}, and fermionic triplets in Type-III \cite{Foot:1988aq,Ma:1998dn,Ma:2002pf,Hambye:2003rt,Abada:2007ux,Dorsner:2006fx,Abada:2008ea,He:2009tf,Bandyopadhyay:2009xa,Bandyopadhyay:2010wp}. {In this article, we adopt a complementary approach by studying the angular distributions of the associated Beyond Standard Model (BSM) particles (or SM charged leptons)}, namely Majorana neutrinos, heavy charged leptons, and triplet scalars, produced at various future colliders. {These distributions, reconstructed in the centre-of-mass frame}, carry information about the underlying matrix elements that depend on the spin and gauge charges. and therefore they {can serve as discriminants among the three simple tree-level Seesaw scenarios without knowing information about the rest of the particles in the final-state}. In some cases, even the off-shell behaviour of these particles leads to angular features that differ significantly from Standard Model expectations \cite{Cai:2017mow}.

{Such angular analyses complement the more conventional distinguishing signatures}, for example lepton number violation in inverse Seesaw models, the existence of doubly charged scalars in Type-II, or the presence of heavy charged fermions in Type-III. Angular observables are also powerful for extracting spin information and probing gauge representations of new states, as demonstrated in prior collider studies \cite{Datta:2005zs,Christensen:2013aua,Battaglia:2005ma,Bandyopadhyay:2020wfv,Bandyopadhyay:2020klr,Bandyopadhyay:2020jez}. For the collider analysis in this paper we focus on inverse Type-I and inverse Type-III models, where suitably small lepton-number violating parameters allow relatively large Yukawa couplings while keeping light neutrino masses suppressed. This feature enhances the collider visibility of the heavy states. Representative studies of inverse Seesaw setups and their collider implications can be found in \cite{Mondal:2016kof,Abada:2014vea,Das:2012ze}.

Although our analysis primarily focuses on muon colliders, the underlying features are common to other lepton colliders, including those with electrons. A high energy $\mu^+\mu^-$ collider has attracted particular attention in recent years \cite{Ankenbrandt:1999cta,Delahaye:2019omf,Bartosik:2019dzq,AlAli:2021let,Bartosik:2020xwr,Bandyopadhyay:2024gyg,deLima:2025ctj,Ghosh:2025dcv,Fan:2026yuh,Bandyopadhyay:2024plc,Ruhdorfer:2024dgz,Ma:2024ayr,Capdevilla:2024bwt,Bandyopadhyay:2021pld,Bandyopadhyay:2020otm}, because the heavier muon mass suppresses synchrotron radiation losses compared to electrons. This suppression enables more compact, power-efficient circular colliders reaching multi-TeV centre-of-mass energies and offering a favourable luminosity-to-power ratio \cite{MuonCollider:2022nsa}. In addition, for certain collider modes such as $\mu^+\gamma$ and $\mu^+\mu^+$, the use of positively charged muon beams is technically advantageous: $\mu^+$ beams are easier to cool and accumulate, and their reduced decay backgrounds enhance the feasibility of these collider configurations \cite{Hamada:2022mua,Hamada:2024ojj,Fukuda:2023yui,Fridell:2023gjx}.

Beyond $\mu^+\mu^-$ collisions, we also consider muon–photon and asymmetric $\mu^+e^-$ colliders. For example, the angular distribution of a doubly charged scalar produced in association with a muon at a $\mu^+\gamma$ collider can exhibit a pronounced dip structure, distinct from the Majorana-fermion driven signatures in Type-I and Type-III Seesaws. Off-shell exchange of a doubly charged scalar can also generate lepton number and flavour violating final states in $\mu^+\mu^+$ and $\mu^+e^-$ collisions, processes absent in the Standard Model and potentially sensitive to off-diagonal Yukawa couplings. The asymmetric $\mu^+e^-$ collider is therefore especially useful for probing flavour structure and off-diagonal Yukawa interactions \cite{Lu:2020dkx}.

The structure of this article is as follows. In \autoref{sec:model}, we provide a brief overview of the relevant Seesaw realizations and outline the parameter space consistent with present collider searches. \autoref{secang} briefs about the angular distributions in CM frame in general before we embark on different colliders. In \autoref{sec:mup_mum}, we describe the parton-level framework and simulation setup for $\mu^+\mu^-$ collisions, which serve as our baseline scenario. The subsequent sections, \autoref{sec:mup_mup}--\autoref{sec:mum_ep}, extend the analysis to alternative collider configurations, namely $\mu^+\mu^+$, $\mu^+\gamma$, and $\mu^+e^-$, highlighting their complementary sensitivity to different Seesaw signatures. Finally, we summarise our findings in the concluding section, emphasising the role of angular observables in disentangling Seesaw scenarios and commenting on the broader prospects of future muon-based facilities.


\section{Different Seesaw models}
\label{sec:model}

In this section, we provide a brief overview of the different Seesaw scenarios that are relevant for our study, together with the associated BSM particle spectrum accessible at future lepton colliders. Each Seesaw type is characterised by a distinct extension of the Standard Model field content: singlet Majorana neutrinos in Type-I, scalar triplets in Type-II, and fermionic triplets in Type-III. These new states not only play a central role in generating neutrino masses but also give rise to characteristic collider signatures. Our analysis focuses on how the angular distributions of such particles can serve as complementary tools to distinguish among the various Seesaw scenarios.

\subsection{Type-I seesaw}
\label{sec:I}

In the Type-I Seesaw framework \cite{Minkowski:1977sc,Yanagida:1979as,Gell-Mann:1979vob,Glashow:1979nm,PhysRevLett.44.912}, three generations of $SU(2)_L$ singlet, colorless, heavy right-handed neutrinos ($N_R : (1,1,0)$) with zero hypercharge are introduced. Their relevant interactions and mass terms in the Lagrangian are given by
\begin{equation}
\label{lagI}
-\mathscr{L}_I^{} \supset Y_{\nu}\, \xbar L\, \widetilde H N_{R} +\frac{1}{2} M_{R}\, \xbar {N^{\,c}_{R}} \, N_{R}\, +h.c.\,,
\end{equation} 
where $L$ and $H \equiv (H^+, H^0)^T$ are the SM lepton and Higgs doublets, $\widetilde H = i\sigma_2 H^*$, and $Y_\nu$ and $M_R$ are the Yukawa couplings and Majorana mass matrices with generation indices being suppressed. Here, the charge conjugate field is defined as $\psi^c_{L,R} \equiv (\psi_{L,R})^c$. 

After electroweak symmetry breaking (EWSB), the SM neutrinos mix with the Majorana fermions $N_R^c$, producing one set of light neutrinos and one set of heavy neutrinos. The mass matrix in the flavour basis and the masses for light ($\nu$) and heavy ($N^0$) neutrinos are given by:
\begin{equation}\label{massI}
\mathcal{M}=
\begin{pmatrix}
		0 & \frac{v_0 }{\sqrt 2}Y_\nu \\
		\frac{v_0 }{\sqrt 2}Y_\nu^T & M_R
\end{pmatrix},  \quad 	\quad
m_{\nu}\approx\, \frac{v_0^2}{2}\,Y_\nu  M_R^{-1}\, Y_\nu^T \quad 
\text{and} \quad M_{N^0}\approx M_R,
\end{equation}
under the assumption $|v_0 Y_\nu/\sqrt 2|\ll |M_R|$, with $H = (0,h+v_0)^T/\sqrt 2$ in unitary gauge and Higgs vacuum expectation value $v_0/\sqrt 2$. {Here, the mass-matrix is diagonalized to get the neutrino masses. One can adopt the Casas-Ibarra parameterization~\cite{Casas:2001sr} to construct the Yukawa matrices compatible with neutrino oscillation data.}
Once produced, these heavy neutrinos decay via $Z\nu$, $h\nu$, and $W^\pm l^\mp$ modes, with partial widths
\begin{equation}
\label{eq:decN0I}
\Gamma^{Z\nu}_{N^0}\approx\Gamma^{h\nu}_{N^0}\approx \frac{1}{2}\Gamma^{W l}_{N^0}\approx\frac{Y_\nu^2 M_{R}}{32\pi}~.
\end{equation}

From \autoref{massI}, it is apparent that achieving light neutrino masses of order $10^{-1},\text{eV}$ requires either very small Yukawa couplings, $Y_\nu \sim 10^{-6}$ for TeV-scale heavy neutrinos, or extremely heavy new states with $M_R \gtrsim 10^5$ TeV for moderately larger couplings $Y_\nu \gtrsim 10^{-3}$. This makes the collider signals of the conventional Type-I Seesaw very challenging to observe, even at future facilities. However, the variants such as the inverse Seesaw (ISS) provide an attractive alternative, where sizeable couplings and TeV-scale masses can be simultaneously accommodated. 

In the inverse Type-I (iType-I) Seesaw \cite{Mohapatra:1986aw,Nandi:1985uh,PhysRevD.34.1642}, three generations of two singlet fermions $(N_a, N_b)$ are introduced, with only $N_a$ coupling to SM leptons via Dirac mixing. In this construction, while the active neutrinos remain light, there exists a large off-diagonal coupling between $N_a$ and $N_b$ through the mass term $M_n$. The mass and Yukawa terms of the Lagrangian for this case is given by:
\begin{equation}
-\mathscr L_I^{in}= Y\,\xbar L\, \widetilde H\, N_a + M_n\, \xbar {N_a^{\,c}}\, N_b+\frac{1}{2}\,\mu_n \xbar {N_b^{\,c}}\, N_b\,+ h.c.,
\end{equation}
where the generation indices have been suppressed. Here, $Y$ is the Yukawa coupling of $N_a$ with SM leptons, large mass term $M_n$ indicates the mixing between $N_a$ and $N_b$, whereas, the tiny mass term $\mu_n$ (in general a complex symmetric matrix) signify the mass of  $N_b$. After EWSB and the mixing of the flavour states, there emerges one set of light neutrinos ($\nu$) and two sets of heavy neutrinos ($N^0,\widetilde{N}^0$), each set consisting of three generations. The entire mass matrix in the flavour basis along with masses for light and heavy neutrinos here can be expressed as
\begin{equation}
\begin{split}
\mathcal M{}&=\begin{pmatrix}
0&\frac{v_0 }{\sqrt 2}Y &0\\
\frac{v_0 }{\sqrt 2}Y^T&0& M_n\\
0&M_n^T&\mu_n
\end{pmatrix}, \\
m_{\nu}\approx\frac{v_0^2}{2}\, Y (M_n^{-1})^T {}& \mu_n\, M_n^{-1} Y^T, \quad
M_{N^0}^2\approx M_{\widetilde N^0}^2\approx M_n^{}\, M_n^T\,,
\end{split}
\end{equation} 
assuming $|v_0 Y/\sqrt 2| \ll |M_n|$. The smallness of the active neutrino masses is thus controlled not only by $Y$ and $M_n$ but also by the additional suppression parameter $\mu_n$. With an appropriate choice of $\mu_n$, the Yukawa couplings can be as large as $\mathcal O(10^{-1})$ for $M_n \sim \mathcal O(1\ \text{TeV})$, making the model testable at current and future colliders.

{Both Type-I and iType-I seesaw models can contribute to muon $(g-2)$ at one-loop mediated by $W$-boson and $\nu-N_a$ mixing as~\cite{Biggio:2008in}:
\begin{equation}
    \Delta a_\mu(I)=\frac{G_F m_\mu^2}{4\sqrt 2 \pi^2} \Big[\sum_{i}\frac{v_0^2}{2}(Y^\dagger M_n^{-1})_{\mu i} (M_n^{-1}Y)_{i\mu}\{D(x_i)-5/3\}-\sum_{i}\frac{1}{6}x_{\nu_i} U_{\mu i} U^\dagger_{i\mu}\Big]
\end{equation}
where, $D(x_i)=\frac{10-55x_i+90x_i^2-37 x_i^3-8x_i^4+6(7x_i-4)x_i^2\log x_i}{6(x_i-1)^4}$ with $x_i=M_{n,i}^2/M_W^2$ and $x_{\nu i}=m_{\nu,i}^2/M_W^2$, and $U$ is the PMNS matrix. For our choice of benchmark points, $\Delta a_\mu(I)\sim -10^{-12}$, which is well-below the current experimental limit of $(3.8 \pm 6.8)\times 10^{-11}$ \cite{Aliberti:2025beg, Muong-2:2025xyk}.
}

The heavy neutrinos eventually decay into light leptons accompanied by a Higgs or weak gauge bosons. The partial decay widths for $N^0$ and $\widetilde{N}^0$ are
\begin{equation}
\label{eq:decN0Iinv}
\Gamma^{Z\nu}_{N^0/\widetilde N^0}\approx\Gamma^{h\nu}_{N^0/\widetilde N^0}\approx \frac{1}{2}\Gamma^{W l}_{N^0/\widetilde N^0}\approx\frac{Y^2 M_n}{64\pi}~.
\end{equation}
Interestingly, the widths here are reduced by a factor of two compared to the usual Type-I Seesaw case in \autoref{eq:decN0I}. This reduction can be understood from the $h\nu$ mode: the scalar field $H$ couples to the right handed neutrinos through the $Y\,\xbar L\, \widetilde H\, N_a$ term, where only $N_a$ field contributes. Now, $N_a$ contains $\sim50\%$ of $N_0$ as well as  $\sim50\%$ of $\widetilde N_0$ unlike the usual Type-I seesaw, where $N^0$ is essentially a pure $N_R$. The same reasoning applies to the other decay channels as well.

Current collider bounds on heavy neutral leptons (HNL) are not very constraining. For these studies, the PMNS matrix is typically extended by introducing more elements $U_{lN}$, on which the bounds are presented. Depending on the mass of the HNLs,  these bounds vary significantly. For TeV-scaled HNLs this limit is: $U_{lN} \lsim 0.1$ \cite{CMS:2022hvh, ATLAS:2023tkz, ATLAS:2024rzi} which becomes even more relaxed with higher masses. In inverse or usual type-I seesaw $U_{lN}\propto (v_0 Y/M_n)$ easily satisfies this constraint with $\mathcal O (10^{-1})$ Yukawa couplings at $M_n\sim 1$ TeV.

\subsection{Type-II seesaw}

In the Type-II Seesaw framework \cite{Konetschny:1977bn,PhysRevD.22.2227,Magg:1980ut,Cheng:1980qt,Lazarides:1980nt,Mohapatra:1980yp,Wetterich:1981bx,Gelmini:1980re,Chun:2013vma}, a colorless $SU(2)_L$ triplet scalar $\Delta \sim (1,3,1)$ with hypercharge one is introduced. It is characterized by its mass $M_\Delta$, a trilinear coupling with the Higgs doublet $\mu_\Delta$, and Yukawa couplings $Y_\Delta$. The relevant part of the Lagrangian is
\small
\begin{equation}
\begin{split}	
\label{lagII}
-\mathscr{L}_{II}^{} {}& \supset M_\Delta^2\,Tr\,(\Delta^\dagger\Delta)+ \Big[ Y_{\Delta}\, \xbar L^{\,c}(i\sigma_2\,\Delta^\dagger)\,L 
- \mu_\Delta^{}\,\widetilde H^\dagger\Delta^\dagger H +h.c.\Big]\\
{}&\text{with} \quad \Delta=\begin{pmatrix}
\Delta^+/\sqrt 2&\Delta^{++}\\
\Delta^0&-\Delta^+/\sqrt 2
\end{pmatrix}.
\end{split}
\end{equation}

\normalsize
Though, there could be other gauge invariant terms in the Lagrangian involving quartic couplings of triplet and the SM doublet as well as the self-quartic coupling of the triplet, for our analysis these terms remain irrelevant and hence we do not write them down explicitly. After electroweak symmetry breaking, the neutral component $\Delta^0$ acquires a small vev, $v_\Delta/\sqrt 2$, generating Majorana masses for the light neutrinos as
\begin{equation}
\label{massII}
m_{\nu}=\sqrt 2\, v_\Delta^{}\,Y_\Delta^{} \quad \text {where} \quad v_\Delta^{}\approx\frac{\mu_\Delta^{} v_0^2}{\sqrt 2\, M_\Delta^{2}}\,.
\end{equation}
The doubly charged scalar $\Delta^{++}$ can decay into either same-sign di-lepton (SSD) or di-boson channels \cite{ATLAS:2021jol,Padhan:2019jlc,Jia:2024wqi}. The partial width to di-lepton channel is proportional to $Y_\Delta^2$, while that to di-boson varies with $v_\Delta^2$.

As seen from \autoref{massII}, with a scalar triplet at the TeV scale and $\mu_\Delta \sim 20$ eV, one can achieve $m_{\nu}\sim 0.1$ eV for Yukawa couplings as large as $\mathcal O(10^{-1})$. This opens up the possibility of probing the model at colliders.

The presence of the scalar triplet modifies the tree-level masses of the weak gauge bosons and hence contributes  to the $\rho$-parameter too.
The $\rho$ parameter in this model can be expressed as
\begin{equation}
\rho \,=\frac{m_W^2}{m_Z^2 \cos^2\theta_w}=\frac{1+(2v_\Delta^2/v_0^2)}{1+(4v_\Delta^2/v_0^2)}~.
\end{equation}
The current global fit on electroweak precision observables results in $\rho = 1.00031 \pm 0.00019$ \cite{pdg} which requires $v_\Delta \lsim 2.5$ GeV at $3\sigma$ level. In this model, the self energies of the weak gauge bosons also acquire additional contributions modifying the oblique parameters. However, these contributions can be kept under control by considering large mass of the triplet scalar ($M_\Delta$) with small mass gap among its components.

Type-II seesaw model can contribute to lepton flavour violating decays $\tau\to \bar\ell_j\ell_k\ell_m$ or $\mu\to \bar e e e$ at tree level via doubly charged scalar. Again, it can also enhance radiative lepton flavour violating process like $\mu\to e \gamma$ through singly and doubly charged scalars at one loop level. The decay width for these channels can be written as\cite{Kakizaki:2003jk,Antusch:2018svb,Primulando:2019evb,Lindner:2016bgg,Han:2021nod}:

\small
\begin{align}
&\Gamma(l_i\to \bar l_j l_k l_m)=\frac{1}{2(1+\delta_{km})}\bigg(\frac{m_{l_i}^5}{192\pi^3\, M_{\Delta^{++}}^4}\bigg)\Big|(Y_\Delta)_{km}(Y_\Delta)_{ij}\Big|^2,\nonumber\\
&\Gamma(l_i\to l_j\gamma)=\frac{ \alpha\,m_{l_i}^5}{(192\pi^2)^2}\;\Big|Y_\Delta^\dagger Y_\Delta\Big|^2_{ij}\;\bigg(\frac{1}{M_{\Delta^+}^2}+\frac{8}{M_{\Delta^{++}}^2}\bigg)^2,
\end{align}
\normalsize
where $M_{\Delta^{++}}$ and $M_{\Delta^+}$ are the masses of $\Delta^{++}$ and $\Delta^+$, $\delta_{km}$ is the Kronecker-delta and $\alpha$ is the electromagnetic coupling constant. However, for our analysis, we choose diagonal coupling $Y_\Delta$ and hence, the bounds from lepton flavour violating decays can safely be ignored. Type-II seesaw {also} contributes to muonium anti-muonium oscillation \cite{Pontecorvo:1957cp,Han:2021nod,Chang:1989uk} at tree level via $\Delta^{++}$. The probability of muonium to anti-muonium transition is given by:

\begin{align}
\begin{split}
\mathcal P(\mathscr M\to\bar {\mathscr M}){}&=64^3\Big(\frac{3\pi^2\alpha^3}{G_f m_\mu^2}\Big)^2 \Big(\frac{m_e}{m_\mu}\Big)^6 \Big(\frac{G_{\mathscr M\bar {\mathscr M}}}{G_f}\Big)^2 \\
\text{with} \quad G_{\mathscr M\bar {\mathscr M}}{}&=\frac{(Y_\Delta)_{ee}(Y_\Delta^*)_{\mu\mu}}{16\sqrt 2 M_{\Delta^{++}}^2},
\end{split}
\end{align}
where $G_f$ is the Fermi constant. The upper bound on this probability, given by the PSI experiment, is  $\mathcal P(\mathscr M\to\bar {\mathscr M})\leq 8.3 \times 10^{-11}$ \cite{Willmann:1998gd} at 90\% confidence level which is expected to be improved beyond the level of $10^{-13}$ in MACE experiment \cite{MACE}. For our choice of benchmark points this probability becomes less than $\sim10^{-12}$, which is well-below the current experimental limit by PSI. Models with heavy neutrinos (like Type-I, Type-III and inverse seesaw) also contribute to this transition probability, but the effects come at one loop-level \cite{Clark:2003tv,Cvetic:2005gx}.

Through the doubly and the singly charged scalars, Type-II seesaw affects $(g-2)$ of muon as:
\begin{align}
&\Delta a_{\mu} (\Delta^{++})=-\,\frac{1}{4\pi^2}\frac{m_\mu^2}{M^2_{\Delta^{++}}}\sum_l\Big(\frac{4}{3}-\frac{m_l}{m_\mu}\Big)\Big|(Y_\Delta)_{l\mu}\Big|^2~,\nonumber\\
&\Delta a_{\mu} (\Delta^{+})=-\,\frac{1}{48\pi^2}\frac{m_\mu^2}{M^2_{\Delta^{+}}}\sum_l\Big|(Y_\Delta)_{l\mu}\Big|^2~.
\end{align} 
It is interesting to notice that if we assume diagonal coupling, then the total contribution of triplet to muon $(g-2)$ is negative. For our choice of benchmark points, $\Delta a_\mu\sim-10^{-12}$, which is quite smaller than the observed experimental value, $(3.8 \pm 6.8)\times 10^{-11}$ \cite{Aliberti:2025beg, Muong-2:2025xyk}. We do not enforce neutrino oscillation constraints, since our main goal is to study angular distributions of Seesaw scenarios at lepton colliders rather than to scan the full parameter space consistent with all low-energy observables.
 
The experimental lower bound on the masses of  neutral and singly charged scalars are not very stringent. One can easily evade these experimental bounds by considering neutral and singly charged scalars heavier than 500 GeV \cite{Karan:2023kyj}. The doubly charged scalar gets severe constraint from multi-lepton final state searches at LHC. Although the precise lower bound on the mass of such scalars depends on the underlying model, one can consider it to be around 1 TeV \cite{ATLAS:2022pbd}, as found by ATLAS. Our benchmark choices, as will be presented later, are therefore comfortably consistent with current experimental constraints.

\subsection{Type-III seesaw}
\label{subsec:model}

Likewise, the Type-III seesaw model \cite{Foot:1988aq,Ma:1998dn,Ma:2002pf,Hambye:2003rt,Abada:2007ux,Dorsner:2006fx,Abada:2008ea,He:2009tf,Bandyopadhyay:2009xa,Bandyopadhyay:2010wp} consists of all the SM fields in addition to the three generations of colourless $SU(2)_L$ triplet (adjoint) fermions $(\Sigma_L\,({1,3},0))$ with hypercharge zero, mass $M_\Sigma^{}$ and Yukawa coupling $\sqrt 2\,Y_\Sigma^{}$. The apposite pieces of Lagrangian are the following:
\begin{equation}
\begin{split}	
\label{lagIII}
-\mathscr{L}_{III}^{} \supset \frac{1}{2} {}& Tr[\xbar\Sigma_{L}M_{\Sigma}\,\Sigma_{L}^{\,c}]+\sqrt 2\,Y_{\Sigma}\,\xbar L\,\Sigma_{L}^{c}\,\widetilde{H} +h.c.\,,\\
\text{with} {}& \quad \Sigma_L=\begin{pmatrix}
\Sigma^0/\sqrt 2&\Sigma^+\\
\Sigma^-&-\Sigma^0/\sqrt 2
\end{pmatrix}.
\end{split}
\end{equation} 
After EWSB, the SM neutrinos mix with the neutral components of the fermionic triplet, while the charged  ones mingle with the charged leptons. The mass matrices in flavour basis and the masses for all the leptons in energy basis are given by:
\begin{equation*}
\mathcal M=\begin{pmatrix}
0&\frac{v_0 }{\sqrt 2}Y_\Sigma \\
\frac{v_0 }{\sqrt 2}Y_\Sigma^T& M_\Sigma
\end{pmatrix}, \quad 
\mathcal M_{l'}=\begin{pmatrix}
m_{l'}&v_0Y_\Sigma \\
0& M_\Sigma
\end{pmatrix}, \quad 
\end{equation*}
\begin{equation}
\label{massIII}
m_{\nu}\approx\frac{v_0^2}{2} \, Y_\Sigma\, M_\Sigma^{-1}\, Y_\Sigma^T\,, \quad m_{l}\approx m_{l'},\quad M_{N^0}\approx M_{N^\pm}\approx M_\Sigma\,.
\end{equation}
Here, $l'$ denotes the SM charged leptons in flavour basis while $l$ and $N^\pm$ symbolise the same with the light and heavy masses in the mass-basis of Type-III scenario. Now, these heavy leptons will eventually decay to the lighter particles. While the heavy neutral particle $N^0$ will decay through $Z\nu$, $h\nu$ and $W^\pm l^\mp$ with the decay widths given by: 
\begin{equation}
\label{eq:decN0}
\Gamma^{Z\nu}_{N^0}\approx\Gamma^{h\nu}_{N^0}\approx \frac{1}{2}\Gamma^{W l}_{N^0}\approx\frac{Y_\Sigma^2 M_\Sigma}{32\pi}~,
\end{equation}
the heavy charged leptons $N^\pm$ decays to $Zl^\pm$, $hl^\pm$ and $W^\pm\nu$ with the following decay widths:
\begin{equation}
\label{eq:decNpm}
\Gamma^{Zl}_{N^\pm}\approx\Gamma^{hl}_{N^\pm}\approx \frac{1}{2}\Gamma^{W\nu}_{N^\pm}\approx \frac{Y_\Sigma^2 M_\Sigma}{32\pi}~,
\end{equation}
where we have assumed that $M_\Sigma\gg m_h$ with $m_h$ being the mass of the SM Higgs boson around $125.5$ GeV. At this point, it is interesting to mention that though the masses of $N^0$ and $N^\pm$ are same at tree level, there can emerge a mass splitting of $\Delta M\approx166$ MeV \cite{Cirelli:2005uq} while considering the loop corrections and it can open up some decay channels of $N^\pm$ to $N^0$ with the decay width as follows:
\begin{align}
\begin{split}
\label{eq:NpmtoN0}
\Gamma_{N^\pm}^{N^0\pi^{\pm}}=\frac{2G_F^2 V_{ud}^2\Delta M^3 f_\pi^2}{\pi}\sqrt{1-\frac{m_\pi^2}{\Delta M^2}} \quad \text{and} \quad  \Gamma_{N^\pm}^{N^0e\,\nu_e}=\frac{25}{3}\Gamma_{N^\pm}^{N^0\mu\,\nu_\mu}=\frac{2G_F^2 \Delta M^5}{15\pi^3}~,
\end{split}
\end{align}
where $f_\pi$ is the pion form factor, $m_\pi$ is the mass of pion, $G_F$ is the Fermi constant and $V_{ud}$ is the CKM matrix element. One should also notice here that none of the decay widths mentioned in \autoref{eq:NpmtoN0} depends on mass or Yukawa coupling of the $SU(2)$ triplet. However, for a TeV scale fermionic triplet with Yukawa greater than $10^{-7}$, one can safely neglect these modes due to very small branching fractions with $\lesssim 1 \%$ \cite{Sen:2021fha}.

Like the Type-I seesaw, in this case also, the demand of light neutrino mass being less than $0.1$ eV pushes the Yukawa coupling of TeV-scaled fermionic triplet to $\mathcal O (10^{-6})$, which is quite challenging to observe in the colliders. Therefore, we investigate for inverse Type-III scenario (iType-III), where three generations of two left handed fermionic triplets $(\Sigma_a,\Sigma_b)$ with only one of them interacting to SM leptons are introduced \cite{Ma:2009kh,Eboli:2011ia,Agostinho:2017biv,Bandyopadhyay:2020djh}. Here too, the flavour states $\Sigma_a$ and $\Sigma_b$ mix substantially through the large mass term $M_n$. The corresponding relevant portions of the Lagrangian are given by:
\begin{equation}
-\mathscr L_{III}^{in}= \sqrt 2 Y\xbar L \Sigma_a^{c} \widetilde H + Tr[\;\xbar {\Sigma_a^c} M_n \Sigma_b+\frac{1}{2}\;\xbar{\Sigma_b^c}\,\mu_n\, \Sigma_b]+h.c.,
\end{equation}
where the generation indices have been suppressed. Here, $Y$ denotes the Yukawa coupling of the triplet $\Sigma_a$ with SM leptons and scalar doublet, $M_n$ represents the large mass term that mixes $\Sigma_a$ and $\Sigma_b$, and the small parameter $\mu_n$ (a complex symmetric matrix) characterises the Majorana mass of $\Sigma_b$. After EWSB and mixing, distinct mass eigenstates arise for both neutral and charged leptons. As in the inverse Type-I scenario, the spectrum contains one set of light neutrinos $(\nu)$ and two sets of heavy neutrinos $(N^0,\widetilde{N}^0)$, each comprising three generations. Similarly, in the charged sector, one set of light charged leptons $(l)$ coexists with two sets of heavy charged leptons $(N^\pm,\widetilde{N}^\pm)$. The corresponding mass matrices in the flavour basis, together with the physical masses, are given by,
\begin{equation}
\begin{split}
\mathcal M=\begin{pmatrix}
0&\frac{v_0 }{\sqrt 2}Y&0 \\
\frac{v_0 }{\sqrt 2}Y^T&0& M_n\\
0&M_n^T&\mu_n
\end{pmatrix}, {}&\quad 
\mathcal M_{l'}=\begin{pmatrix}
m_{l'}&v_0Y&0 \\
0&0& M_n\\
0&M_n^T&\mu_n
\end{pmatrix}, \\
m_{\nu}\approx\frac{v_0^2}{2}\, Y (M_n^{-1})^T \mu_n\, M_n^{-1}{}& Y^T,\quad m_l\approx m_{l'} \quad \text{ and } \\
M_{N^0/\widetilde N^0}^2\approx M_{N^\pm/\widetilde N^\pm}^2{}&\approx M_n^{}\, M_n^T\,,
\end{split}
\end{equation}
where $l'$ denotes the flavour states of SM charged leptons. Similar to the Type-I case, one can easily satisfy the neutrino bounds with suitable choice of $\mu_n$ while considering $Y\sim\mathcal{O} (0.1)$ and $M_n\sim \mathcal O (1$ TeV).

{The Type-III and iType-III scenarios both generate contribution to muon $(g-2)$ at one-loop level mediated by $W$, $Z$, $h$ and the BSM femions as~\cite{Biggio:2008in, Escribano:2021css}:
\begin{align}
    \Delta a_\mu(III)= \frac{G_F m_\mu^2}{4\sqrt 2 \pi^2}\bigg[-\frac{1}{2}\sum_i x_{\nu,i} U_{li} U^\dagger_{il}+ & \sum_i \frac{v_0}{2}(Y^\dagger M_n^{-1})_{\mu i} (M_n^{-1} Y)_{i\mu} \times \nonumber\\
    & \Big\{\frac{7}{6}-\frac{16}{3}+\frac{8}{3}\cos^2 \theta_W+A(x_i)+B(y_i)+C(Z_i)\Big\}\bigg]
\end{align}
\begin{align}
    \text{where, } &A(x_i)=\frac{-38+185x_i-246x_i^2+107x_i^3-8x_i^4+18(4-3x_i)x_i^2\log x_i}{6(x_i-1)^4}\nonumber\\
    & B(y_i)=\frac{40-46y_i-3y_i^2+2y_i^3+7y_i^4+18(4-3y_i)y_i \log y_i}{6(y_i-1)^4}\nonumber\\
    & C(z_i)=\frac{-16z_i+45z_i^2-36z_i^3+7z_i^4+6(3z_i-2)z_i\log z_i}{6(z_i-1)^4}
\end{align}
with $x_{\nu i}=m_{\nu,i}^2/M_W^2$, $x_{i}=M_{n,i}^2/M_W^2$, $y_{i}=M_{n,i}^2/M_Z^2$ and $z_{i}=M_{n,i}^2/M_h^2$. For our choice of benchmark points, $\Delta a_\mu(III)\sim -10^{-12}$, like the other two seesaw scenario,  which is well-below the current experimental limits~\cite{Aliberti:2025beg, Muong-2:2025xyk}.
}

The heavy leptons ultimately decay into light (SM) leptons accompanied by either the Higgs boson or weak gauge bosons. The partial decay widths of $N^0$ and $\widetilde{N}^0$ for the various channels are given by:
\begin{equation}
\label{eq:decN0inv}
\Gamma^{Z\nu}_{N^0/\widetilde N^0}\approx\Gamma^{h\nu}_{N^0/\widetilde N^0}\approx \frac{1}{2}\Gamma^{W l}_{N^0/\widetilde N^0}\approx\frac{Y^2 M_n}{64\pi}~.
\end{equation}
As already mentioned in \autoref{sec:I}, the partial decay widths of the heavy neutrinos in the inverse Type-III scenario are reduced by a factor of two compared to the usual Type-III case.
An additional interesting feature arises in the charged sector: while the neutral states $N^0/\widetilde N^0$ couple to $Z$, $h$, and $W^{\pm}$ (\autoref{eq:decN0inv}), the charged states show a split behaviour. One of them, $N^\pm$, couples only to $Z\ell$ and $h\ell$, leading to decays with 50\% branching ratio into each channel, whereas the other, $\widetilde N^\pm$, couples exclusively to $W^{\pm}\nu$ and hence decays entirely through this mode (see \autoref{appen1}). The corresponding partial decay widths for the charged components are:
\begin{equation}
\Gamma^{Zl}_{N^\pm}\approx\Gamma^{hl}_{N^\pm}\approx \frac{1}{2}\Gamma^{W\nu}_{\widetilde N^\pm}\approx \frac{Y^2 M_n}{32\pi}~.
\end{equation}

The ATLAS and CMS collaborations have already searched for heavy charged and neutral leptons as well as type-III seesaw heavy leptons at the LHC. The non-observation of such states has led to experimental upper bounds on the pair-production cross-section as a function of the heavy lepton mass. This in turn sets a lower bound of 910 GeV \cite{ATLAS:2022yhd,CMS:2019lwf,ATLAS:2020wop} on the mass of type-III seesaw heavy leptons.
However, the theoretical predictions in these analyses assume only a single generation of heavy leptons, whereas in our case there are six copies of heavy charged and neutral states. We therefore scale the theoretical estimates of Refs.~\cite{ATLAS:2022yhd,CMS:2019lwf,ATLAS:2020wop} by a factor of six and obtain a lower bound of about 1100 GeV on the mass of the fermionic triplet in the inverse Type-III case (three generations) at the $2\sigma$ level.

In summary, the Seesaw scenarios predict heavy neutrinos, charged leptons, and scalar triplets with masses in the TeV range, consistent with current experimental limits. {There are several gauge extended seesaw scenarios like in the context of $U_{B-L}, U(1)_{\psi}, U(1)_{\chi}$ where the basic Type-I \cite{Nomura:2021adf,Baek:2013fsa,Bandyopadhyay:2011qm,Bandyopadhyay:2014sma,Bandyopadhyay:2017bgh,Bandyopadhyay:2022mej}, Type-II \cite{Das:2024yvt,Mahapatra:2020dgk,Mishra:2025llv} and Type-III scenarios are extended . The other phenomenological extensions with dark matter and Majorana neutrino mass generation also explored these mechanism \cite{Baek:2014awa,Avila:2025qsc,Karan:2025pud,Parashar:2022wrd,Bandyopadhyay:2018qcv,Bandyopadhyay:2020qpn,Bandyopadhyay:2019xfb,Bandyopadhyay:2012px,Das:2019pua,Jangid:2020dqh,Das:2019fee,Bandyopadhyay:2011aa,Bandyopadhyay:2010wp,Bandyopadhyay:2020djh,Bandyopadhyay:2009xa}.} {However, we keep ourselves restricted to the minimal seesaw scenarios at the tree-level. It is also important to mention that for simplicity we assumed the diagonal Yukawa couplings, but for a generic scenario one can consider non-diagonal Yukawa couplings too. Since the flavor of the neutrinos cannot be tagged at the colliders, summing of the corresponding cross-sections over the neutrino flavors must be considered. In that case, $\sum_i|Y_{\mu i}|^2$ would be the relevant quantity affecting the cross-sections at a particular muon collider ~\cite{Bandyopadhyay:2022mej}. But the effect of PMNS matrix would not be very drastic since the $3\times3$ block of the PMNS matrix is nearly unitary.}

However, in the following, we investigate how angular distributions at future lepton colliders can provide a complementary means of distinguishing among these possibilities.


\section{Angular Distribution}\label{secang}

It is well known that angular distributions in the centre-of-mass (CM) frame encode valuable information about the underlying matrix elements, including the spins of the initial and final state particles, as well as the Lorentz structure of the propagators and interaction vertices. Previous studies have shown that the spin of various BSM particles can be extracted from such distributions \cite{Datta:2005zs, Battaglia:2005ma, Christensen:2013aua,delAguila:2008cj}. At hadron colliders such as the LHC, however, this approach is limited because the CM frame is not directly accessible due to the unknown boost of the initial partons along the beam axis. Reconstruction is only possible if the final state is fully visible.

An interesting feature arises when a massless gauge boson participates in the initial or final state: the amplitude may vanish at specific angles, a phenomenon known as Radiation Amplitude Zeros (RAZ) \cite{Brodsky:1982sh,D0:2008abl}. These zeros, determined by the charges and momenta of the external particles, cause the angular distribution to vanish at certain values of $\theta$. Such behaviour has been proposed as a way to distinguish particles with fractional electromagnetic charges, e.g. leptoquarks, through their characteristic angular patterns \cite{Bandyopadhyay:2020jez, Bandyopadhyay:2020klr, Bandyopadhyay:2020wfv}.

In this work, however, our focus is on the angular distributions of BSM states arising from Seesaw scenarios. Leptonic colliders provide an ideal environment for such studies, since they naturally operate in the CM frame for symmetric beams.

\begin{figure}[h!]
	\begin{center}
		\centering
		\includegraphics[width=0.4\linewidth]{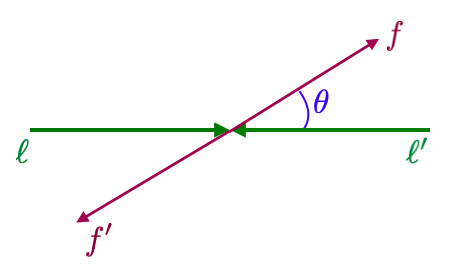}
		\caption{Schematic diagram of a generic lepton collision. The polar angle $\theta$ of one of the final-state particles with respect to the beam axis (the $Z$-axis) is defined in the CM frame.}\label{AngDist}
	\end{center}
\end{figure}

A typical topology is illustrated in \autoref{AngDist}, where initial-state leptons $\ell \ell'$ scatter into a final state $f f'$. Here $\theta$ denotes the angle between $f$ and the beam axis. The differential distribution in $\cos\theta$ exhibits distinctive features that depend on the Seesaw mechanism under consideration. Constructing this angle requires knowledge of the on-shell BSM particle momenta, making the reconstruction of heavy on-shell masses crucial for this study. Different final-state topologies are employed for this reconstruction in each Seesaw scenario.

While the angular distributions are universal in form, we focus on those collider processes where the cross-sections are enhanced, providing better sensitivity. In the following, we present detailed results for different muon-based colliders such as $\mu^+\mu^-$, $\mu^+\mu^+$, $\mu^+\gamma$, and $\mu^+e^-$, and discuss how their angular distributions can be used to distinguish between the different Seesaw models.


\section{At $\mu^+\mu^-$ collider}
\label{sec:mup_mum}

Lepton colliders, such as $\ell^+\ell^-$ machines, have played a pivotal role in testing the Standard Model. For instance, the Large Electron-Positron Collider (LEP) \cite{LEPWG_Higgs:2003ing} was highly successful in validating the SM through precision electroweak measurements \cite{ALEPH:2005ab,ALEPH:2013dgf}, while future $e^+e^-$ facilities such as the ILC \cite{Asner:2013psa,Bambade:2019fyw}, CLIC \cite{Aicheler:2018arh,Kemppinen:2021urj}, and FCC-ee \cite{Blondel:2021ema,Blondel:2019yqr,Zimmermann:2015mea,Boscolo:2021hsq} are designed to push the frontiers of high energy physics even further. Nevertheless, in circular $e^+e^-$ colliders, the small electron mass causes significant synchrotron radiation losses, constraining the maximum achievable centre-of-mass energy. Linear colliders circumvent this limitation but face substantial technical challenges at multi-TeV energies.

Muon colliders \cite{Ankenbrandt:1999cta,Delahaye:2019omf,Bartosik:2019dzq,AlAli:2021let,Bartosik:2020xwr} offer a promising alternative. The muon's much larger mass suppresses synchrotron radiation, enabling compact circular designs at multi-TeV energies, with proposals reaching up to $\sqrt{s}\sim30$ TeV \cite{Delahaye:2019omf,Han:2020pif,Costantini:2020stv}. Since muons are fundamental leptons, their collisions provide a clean environment compared to hadron colliders, and unlike protons, the entire beam energy is available for hard scattering. With projected integrated luminosities of order $10$ ab$^{-1}$ \cite{Abe:2019thb}, such facilities are highly suited for exploring heavy states beyond the SM.

Despite challenges associated with muon decays and the resulting beam-induced background, recent design studies suggest these issues can be mitigated with advanced shielding and detector technology. Therefore, a high-energy $\mu^+\mu^-$ collider represents an ideal environment to probe the angular distributions of BSM states, including those predicted in Seesaw scenarios. In the following, we discuss the parton-level angular distributions at a $\mu^+\mu^-$ collider.

\renewcommand{\arraystretch}{1.5}
\begin{table}[h!]
    \centering
    \begin{tabular}{|c||cc|}
    \hline\hline
    \multicolumn{3}{|c|}{\textbf{At $\bm{\mu^+\mu^-}$ collider}}\\
    \hline\hline
    \multirow{-3.5}{*}{\rotatebox{90}{\textbf{iType-I/III}}} &
		\begin{tikzpicture}
			\begin{feynman}
				\vertex (a1);
				\vertex [left=1cm of a1] (a0){$\mu^+$};
				\vertex [right=1cm of a1] (a2){$\nu$};
				\vertex [below=1.85 cm of a1] (b1);
				\vertex [left=1cm of b1] (b0){$\mu^-$};
				\vertex [right=1cm of b1] (b2){$N^0/\widetilde{N}^0$};
				\diagram {(a0)--[anti fermion](a1)--[](a2),
					(b0)--[fermion](b1)--[](b2),
					(a1)--[boson, edge label'=$W$](b1)};
			\end{feynman}
		\end{tikzpicture}
		
        &
        
		\begin{tikzpicture}
			\begin{feynman}
				\vertex (a1);
				\vertex [left=1.0cm of a1] (a0){$\mu^+$};
				\vertex [right=1.0cm of a1] (a2){$\nu$};
				\vertex [below=1.85 cm of a1] (b1);
				\vertex [left=1.0cm of b1] (b0){$\mu^-$};
				\vertex [right=1.0cm of b1] (b2){$N^0/\widetilde{N}^0$};
				\diagram {(a0)--[anti fermion](a1)--[](b2),
					(b0)--[fermion](b1)--[](a2),
					(a1)--[boson, edge label'=$W$](b1)};
			\end{feynman}
		\end{tikzpicture}
\\
\hline
\multirow{-4}{*}{\rotatebox{90}{\textbf{Type-II}}} & \begin{tikzpicture}
			\begin{feynman}
				\vertex (a1);
				\vertex [above left=1cm of a1] (a0){$\mu^+$};
				\vertex [right=1.5cm of a1] (a2);
				\vertex [above right=1 cm of a2] (a3){$\Delta^{--}$};
				\vertex [below left=1cm of a1] (b0){$\mu^-$};
				\vertex [below right=1cm of a2] (b3){$\Delta^{++}$};
				\diagram {(a0)--[anti fermion](a1)--[boson,edge label'=$\gamma/Z$](a2)--[charged scalar](a3),
					(b0)--[fermion](a1),(a2)--[anti charged scalar](b3)};
			\end{feynman}
		\end{tikzpicture}

         &
        
		\begin{tikzpicture}
			\begin{feynman}
				\vertex (a1);
				\vertex [left=1cm of a1] (a0){$\mu^+$};
				\vertex [right=1cm of a1] (a2){$\Delta^{++}$};
				\vertex [below=1.85 cm of a1] (b1);
				\vertex [left=1cm of b1] (b0){$\mu^-$};
				\vertex [right=1cm of b1] (b2){$\Delta^{--}$};
				\diagram {(a0)--[anti fermion](a1)--[anti charged scalar](a2),
					(b0)--[fermion](b1)--[charged scalar](b2),
					(a1)--[fermion, edge label=$\ell$](b1)};
			\end{feynman}
		\end{tikzpicture}
\\
\hline
\multirow{-4}{*}{\rotatebox{90}{\textbf{iType-III}}} &  \begin{tikzpicture}
			\begin{feynman}
				\vertex (a1);
				\vertex [above left=1cm of a1] (a0){$\mu^+$};
				\vertex [right=1.5cm of a1] (a2);
				\vertex [above right=1 cm of a2] (a3){$\mu^+/N^+$};
				\vertex [below left=1cm of a1] (b0){$\mu^-$};
				\vertex [below right=1cm of a2] (b3){$N^-/\mu^-$};
				\diagram {(a0)--[anti fermion](a1)--[boson,edge label'=$Z$](a2)--[fermion](a3),
					(b0)--[fermion](a1),(a2)--[anti fermion](b3)};
			\end{feynman}
		\end{tikzpicture}
        
		&
        
		\begin{tikzpicture}
			\begin{feynman}
				\vertex (a1);
				\vertex [left=1cm of a1] (a0){$\mu^+$};
				\vertex [right=1cm of a1] (a2){$\mu^+/N^+$};
				\vertex [below=1.85 cm of a1] (b1);
				\vertex [left=1cm of b1] (b0){$\mu^-$};
				\vertex [right=1cm of b1] (b2){$N^-/\mu^-$};
				\diagram {(a0)--[anti fermion](a1)--[anti fermion](a2),
					(b0)--[fermion](b1)--[fermion](b2),
					(a1)--[boson, edge label'=$Z$](b1)};
			\end{feynman}
		\end{tikzpicture}
        \\
     \hline\hline   
    \end{tabular}
		\caption{{Feynman diagrams for dominant  channels distinguishing the seesaw scenarios at $\mu^+\mu^-$ collider.}}
    \label{tab:feyn_mupm}
\end{table}

\vspace*{3mm}
\noindent
$\bullet$ \underline{\textbf{iType-I seesaw :}} 

In order to investigate the signature of iType-I seesaw at $\mu^+\mu^-$ collider, we look for the process $\mu^+\mu^-\to \nu N^0/\widetilde{N}^0$. This mode mainly occurs through $W$-boson mediated t- and u-channel diagrams as shown in the first row of \autoref{tab:feyn_mupm}. Though there is a possibility for occurrence of this process via $h$ and $Z$ mediated s-channel, but due to large centre-of-mass energy both of the propagators will be highly off-shell and hence contributions from those two diagrams can safely be neglected. In principle one can consider the pair production of the BSM neutrinos also but the cross-section is heavily  suppressed  as compared to associated one. It is worth mentioning that the same mode will be present for iType-III seesaw scenario too.

\vspace*{3mm}
\noindent
$\bullet$ \underline{\textbf{Type-II Seesaw :}}

Compared to iType-I, Type-II Seesaw models involve scalars, which are spin-zero particles, and are therefore expected to exhibit different angular distributions. To investigate this, we focus on the pair production channel of doubly charged scalars, which is advantageous due to its larger cross-section, thereby increasing the likelihood of detectability. The Feynman diagrams for this process are shown in the second row of \autoref{tab:feyn_mupm}. This process can occur via photon and $Z$ mediated s-channel diagrams, or through a lepton-mediated t-channel diagram.

\vspace*{3mm}
\noindent
$\bullet$ \underline{\textbf{iType-III Seesaw :}}

As we have mentioned while discussing the iType-I scenario, the observation of the mode $\mu^+\mu^-\to \nu N^0/\widetilde{N}^0$ indicates the existence of iType-I or iType-III seesaw; however, it cannot differentiate these two models. Here we consider the charged  triplet fermion  pair  production, which is absent in iType-I and explore the angular distribution at $\mu^+\mu^-$ collider. For this purpose we investigate the channel $\mu^+\mu^-\to \ell^\pm N^\mp$ owing to higher cross-section as compared to pair production. Detection of this mode accompanied by $\nu N^0/\widetilde{N}^0$ channel will point out the presence of iType-III scenario, whereas, observation of only the second channel will indicate the existence of iType-I case. The relevant Feynman diagrams for the mentioned channel, which are s- and t-channel processes mediated by $Z$ boson, are displayed in the third row of \autoref{tab:feyn_mupm}. However, it is important to mention that the dominant contribution to this mode comes from the t-channel only. By charge conjugation symmetry we see that the cross-section is same for $\ell^+\ell^-\to \ell^- N^+$ process also. 

\begin{figure*}[hbt]
	\begin{center}
		\hspace*{-0.7cm}
		\mbox{\subfigure[{iType-I/III}]{\includegraphics[width=0.35\linewidth,angle=-0]{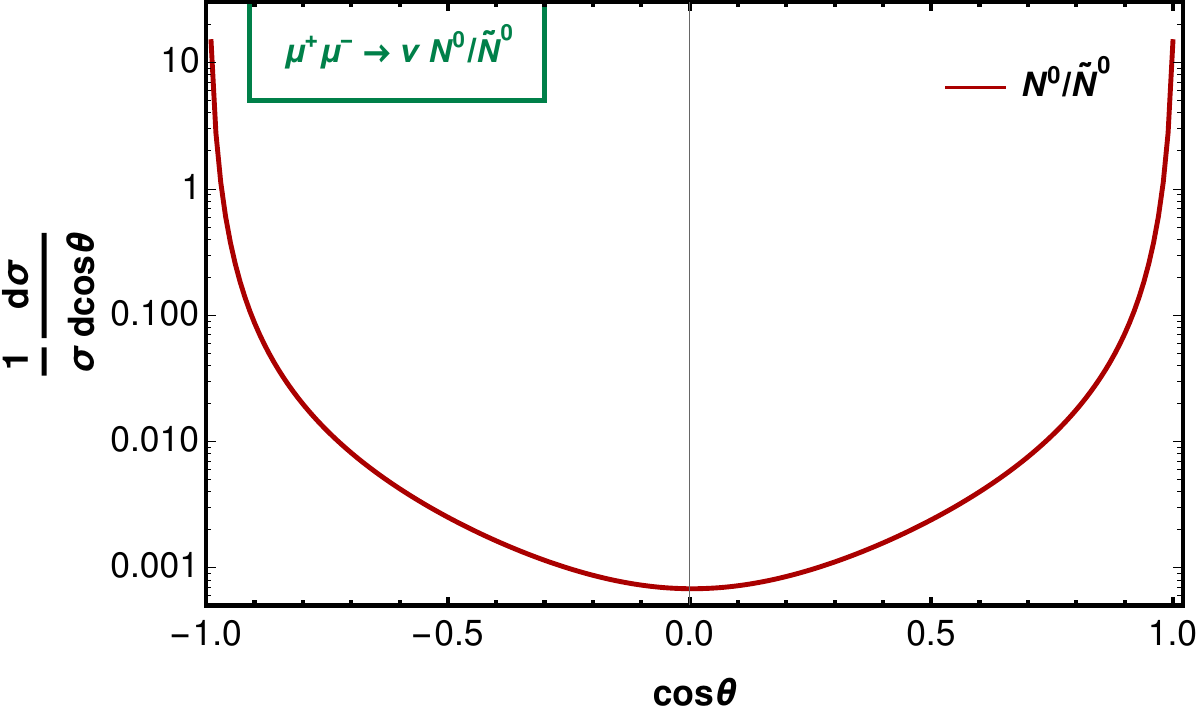}\label{}}
			\subfigure[{Type-II}]{\includegraphics[width=0.35\linewidth,angle=-0]{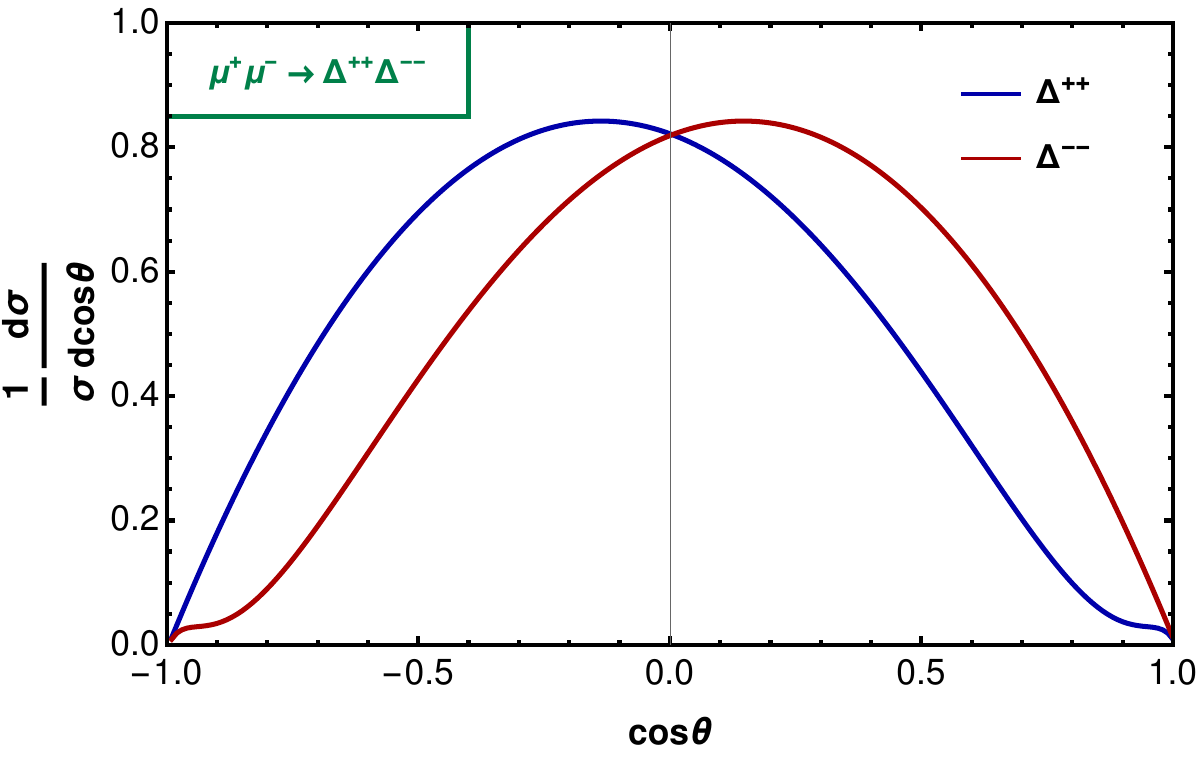}\label{}}
			\subfigure[{iType-III}]{\includegraphics[width=0.35\linewidth,angle=-0]{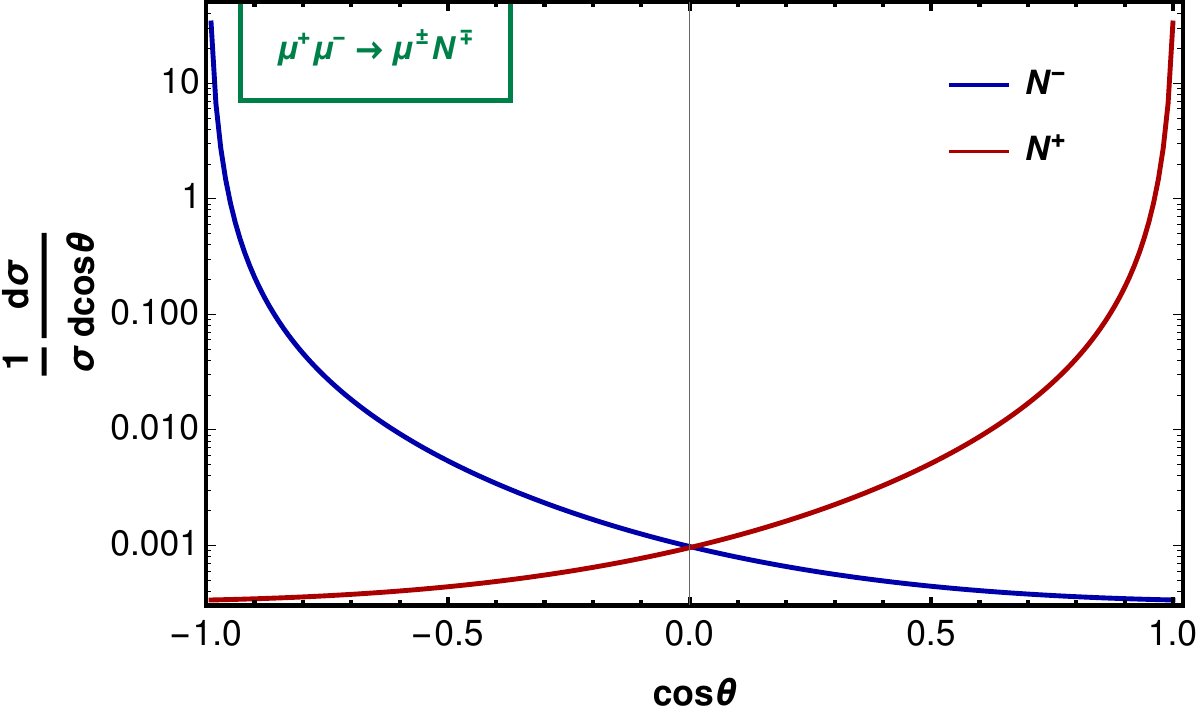}\label{}}}		
		\caption{Angular distributions for the BSM particles in iType-I (a), Type-II (b) and iType-III (c) Seesaw with $Y=Y_\Delta =0.2$, $M_n=M_\Delta=1.2$\,TeV and $\sqrt{s}=3.0$\,TeV in the $\mu^+ \mu^-$ collider.}\label{mu+mu-ang_anal}
	\end{center}
\end{figure*}

In \autoref{mu+mu-ang_anal} we depict the angular distributions at  the parton level of the BSM particles relevent to the different Seesaw mechanisms i.e. $\nu N^0/\widetilde{N}^0$  (for iType-I), $\Delta^{++/--}$ (for  Type-II) and $N^\pm$ (for  iType-III) in the centre-of-mass frame in $\mu^+ \mu^-$ collider with the centre-of-mass energy of  $\sqrt{s}=3.0$\,TeV. It is evident that the bowl shape behaviour in case of iType-I depicts a minima around $\cos{\theta}=0$. This feature  remains the same even for the $\nu N^0/\widetilde{N}^0$ pair production, which is very distinctive compared to the Type-II and iType-III attributed to the Lorentz structure of the matrix elements, determined by the spins of the propagators and the final state particles. Compared to \autoref{mu+mu-ang_anal}(a), in \autoref{mu+mu-ang_anal}(b) we see the angular distribution for $\Delta^{++/--}$ in the case of Type-II. We see a dome shape structure with the asymmetries for $\Delta^{++}$ and $\Delta^{--}$ around $\cos{\theta}\simeq 0, \,  \pm 1$. Finally, we see that for iType-III case the minima reach towards $\cos{\theta}\simeq \pm 1$ for $N^+$ and $N^-$, respectively. The mathematical expressions for angular distributions and total cross-sections concerning these processes at leading order are presented in \autoref{sec:formula}.

In the next few subsection we will try to replicate the parton level distributions of \autoref{mu+mu-ang_anal} via reconstructing the on-shell BSM particles via suitable final state  topologies. A PYTHIA8 \cite{Sjostrand:2014zea} level simulation is carried out for such reconstruction as they are summarized below.

\subsection{Collider simulation} \label{sec:collmupmum}

For the collider simulation via PYTHIA8, we chose the following benchmark points, where the masses of the BSM particles are kept the same for all three Seesaw mechanisms considering the respective collider search bounds obtained from CMS \cite{CMS:2017pet,CMS:2019lwf}, ATLAS \cite{ATLAS:2017xqs,ATLAS:2020wop}, and suitable final state topologies are chosen to probe such particles. For $M_n$ and $M_{\Delta}=1.2, \, 2.0$ TeV, the centre-of-mass energies are chosen as 3.0, 6.0 TeV, respectively as shown in \autoref{crs_mupmum}.

\begin{table}[h!]
	\renewcommand{\arraystretch}{1.5}
	\centering
	\begin{tabular}{|c|c|c|c|c|c|}
		\cline{4-6}
		\multicolumn{3}{c|}{}&\multicolumn{3}{c|}{Cross-section (in fb)} \\
		\hline 
		Benchmark& $M_n$ or, $M_{\Delta}$ & $E_{CM}$& Type-I  & Type-II & Type-III \\ 
		Points &in TeV &in TeV & $\mu^+\mu^- \to \nu N^0/\widetilde{N}^0$ & $\mu^+\mu^- \to \Delta^{++} \Delta^{--}$ & $\mu^+\mu^- \to \mu^{\pm}N^{\mp}$ \\
		\hline \hline
		BP1	& 1.2 &  3.0 & 77.4 & 1.7 & 25.0  \\ \hline
		BP2	& 2.0 & 6.0 & 29.7 & 0.9 & 9.6 \\ \hline
	\end{tabular}
	\caption{Masses corresponding to different benchmark points, energy of collision in CM frame and the hard scattering cross-sections (in fb) for iType-I, Type-II and iType-III Seesaw models in $\mu^+ \mu^-$ collider. ($Y_\Delta$ and $Y =0.2$, $\mu_\Delta$ and $\mu_n=10$ eV)}  \label{crs_mupmum}
\end{table}

\begin{figure*}[hbt]
	\begin{center}
		\hspace*{-0.7cm}
		\mbox{\subfigure[{iType-I/III}]{\includegraphics[width=0.35\linewidth,angle=-0]{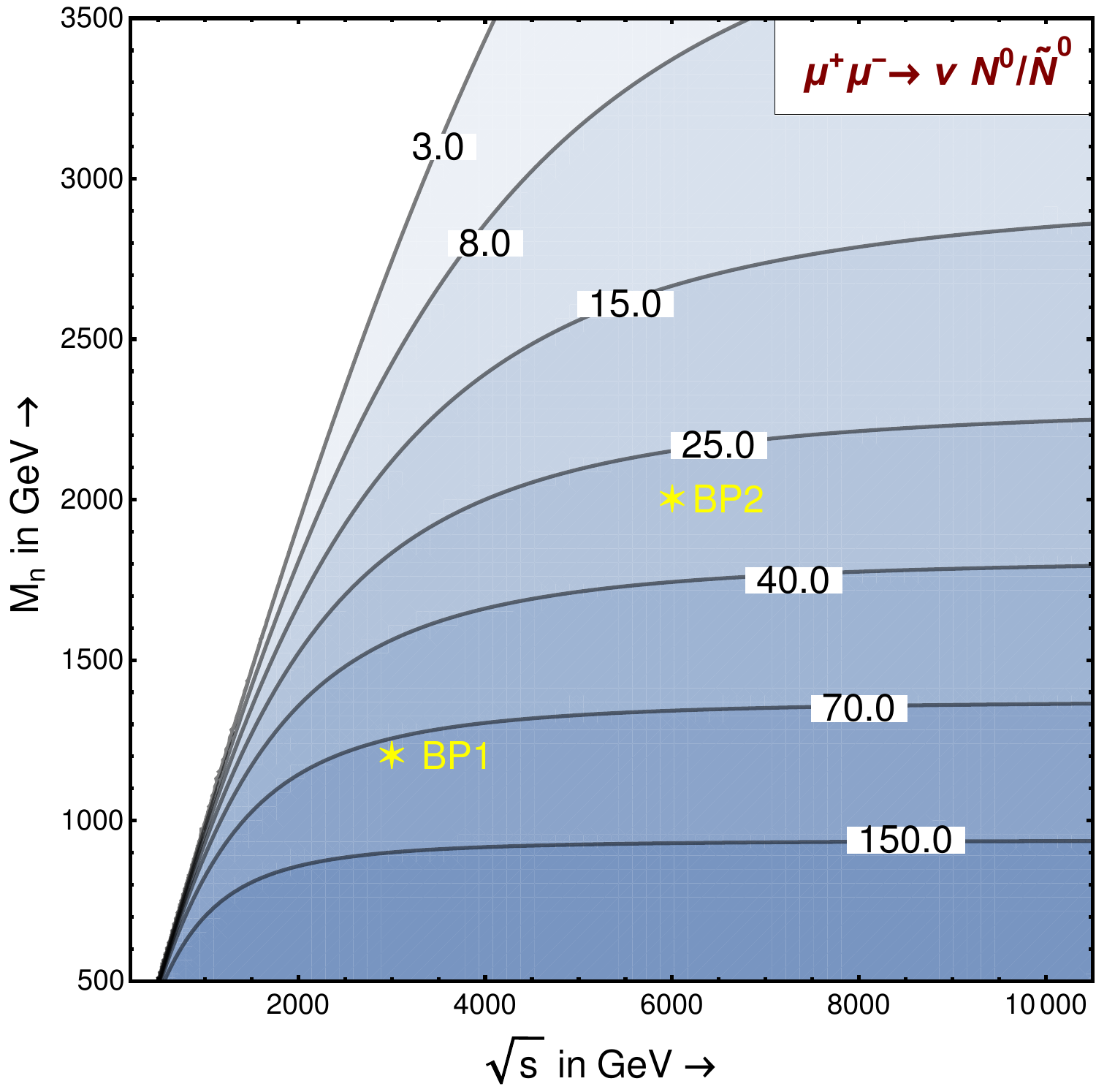}\label{}}
			\subfigure[{Type-II}]{\includegraphics[width=0.35\linewidth,angle=-0]{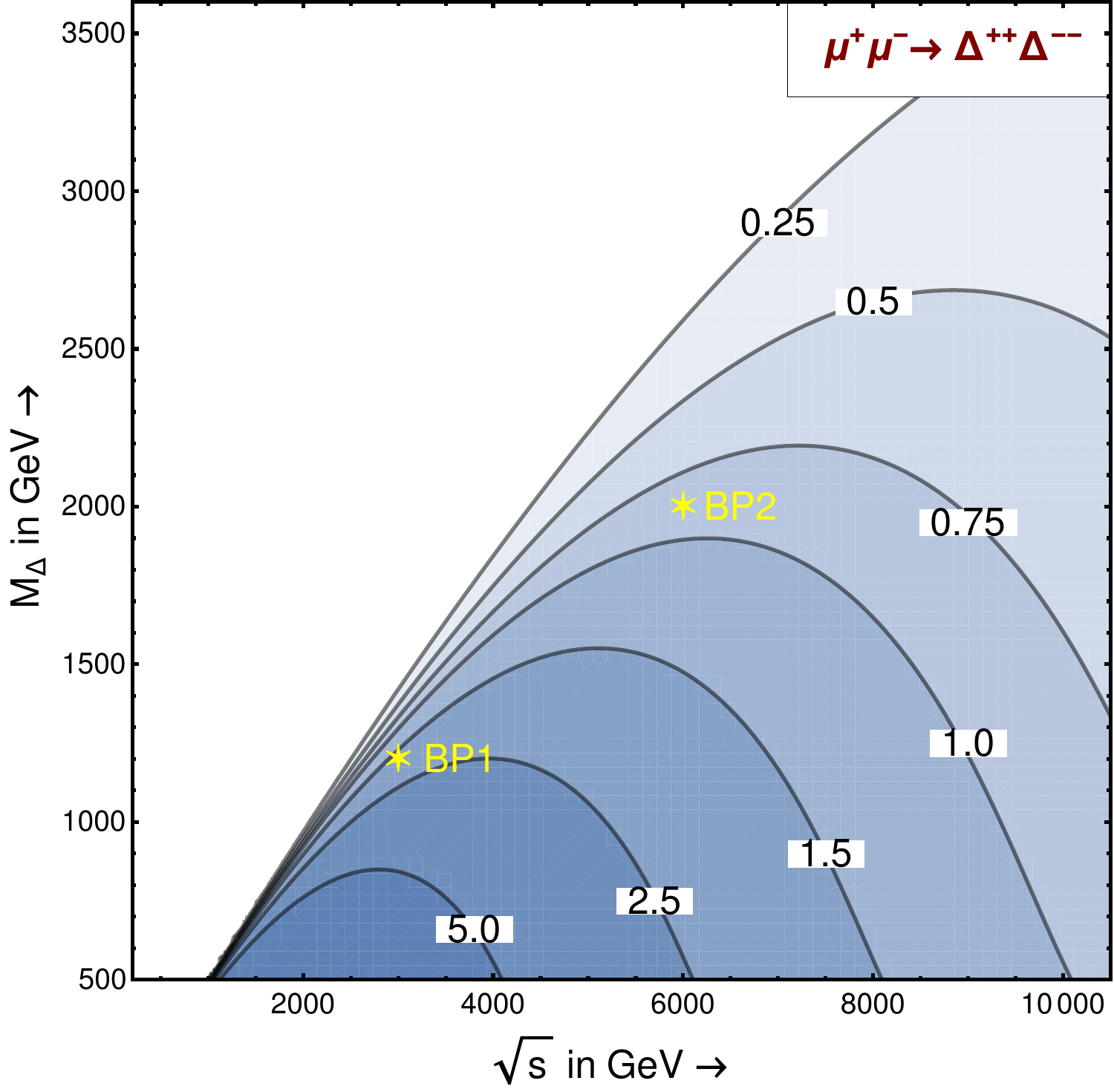}\label{}}
			\subfigure[{iType-III}]{\includegraphics[width=0.35\linewidth,angle=-0]{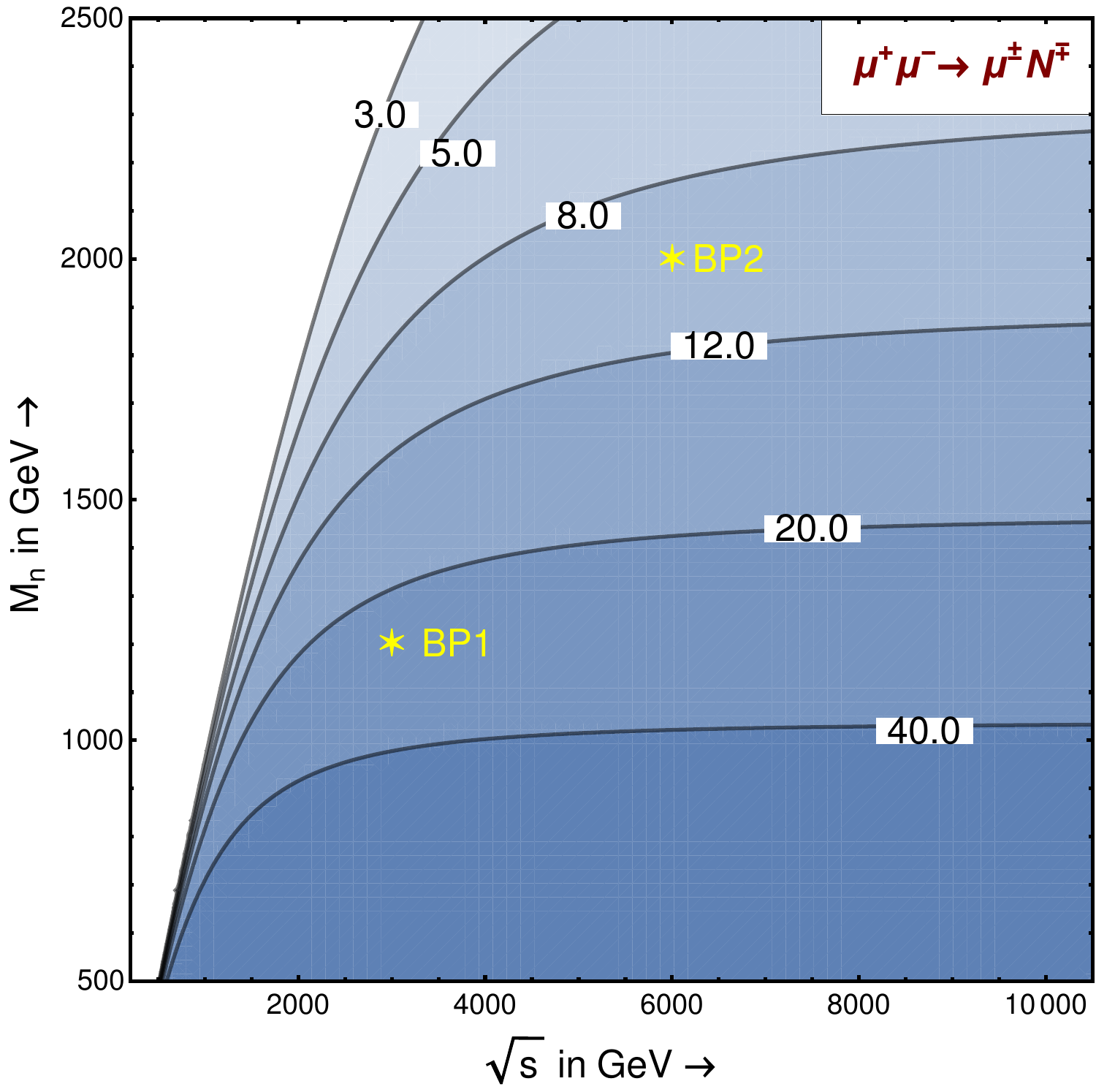}\label{}}}		
		\caption{Contours of the total cross-sections (fb) for $\mu^+ \mu^-$ collider in $M_n$ or $M_{\Delta}$ versus the centre-of-mass energy ($\sqrt{\hat{s}}$) plane for $Y_\Delta$ and $Y =0.2$, $\mu_\Delta$ and $\mu_n=10$ eV.}\label{crosec_mu+mu-}
	\end{center}
\end{figure*}

The comparative cross-sections with respect to the centre-of-mass energy are plotted in \autoref{crosec_mu+mu-} for the three seesaw processes. Different contours depict different values of the cross-sections in fb, where the Yukawa couplings are chosen $Y=0.2$, $\mu_n=10$ eV (the same values are taken for $Y_\Delta$ and $\mu_\Delta$ in the Type-II case). The larger choices of Yukawa in case of iType-I and iType-III are motivated from the inverse Seesaw scenarios as discussed before. The darker to fainter blue regions show higher to lower values of cross-sections, respectively. The benchmark points are shown by yellow stars. 

\subsubsection{Set up for collider simulation}\label{sec:col_sim}	
To perform the collider analysis, at first, the models are implement  at SARAH-4.14.2 \cite{Staub:2013tta} and model files for CalcHEP \cite{Belyaev:2012qa} are prepared. The branching fractions and production cross-sections for different BSM particles are estimated through CalcHEP. Then we generate events for different relevant modes via CalcHEP and use the generated ``.lhe'' files as an input to PYTHIA8 \cite{Sjostrand:2014zea} where the events are simulated with FastJET-3.0.3 \cite{Cacciari:2011ma}. The following criteria are being maintained during the simulation:

\begin{enumerate}
	\item Though we are interested in the angular distribution of various channels for the whole region, to avoid beam line events we restrict the calorimeter coverage for $|\eta|< 2.5$. 
	
	\item Regarding jet formation, we use:
	\begin{itemize}
		\item[$\bullet$] The {\tt ANTI-KT} algorithm with the jet radius $R=0.5$.
		\item[$\bullet$]  The minimum transverse momentum for jets $p_{T,\text{min}}^{\text{jet}}= 20\, \text{GeV}$.
	\end{itemize}
	
	\item The stable leptons are detected with following cuts:
	\begin{itemize}
		\item[$\bullet$] The minimum transverse momentum of the leptons $p_{T,min}=10\,\text{GeV}$ with $|\eta_{\text{max}}|=2.5$.
		\item[$\bullet$] The leptons are isolated from the jet with $\Delta R_{\ell j}\geq 0.4 $, where $\Delta R_{\ell j}=\sqrt{\Delta\eta_{\ell j}^2+\Delta\phi_{\ell j}^2}$~.
		\item[$\bullet$] For a selection of clean lepton, we put an additional cut i.e., the total transverse momentum of the hadrons within the cone $\Delta R =0.3$ will be $\leq 0.15\; p_{T}^{\ell}$. Here $p_{T}^{\ell}$ is the transverse momentum for the leptons within that specified cone.
	\end{itemize}
	
	\item We have already denoted $l$ as all the three generations of SM charged leptons in mass basis. However, most of our upcoming discussions are based on electron and muon only. Therefore, we symbolize these two leptons as $\ell$. 
\end{enumerate}

The cuts used here are generically  true for other colliders used in this article unless specified or added otherwise.

\subsection{Final states in reconstructing the resonance particles}

In this sub-section we discuss the relevant final states for three different Seesaw mechanisms that enable us to construct the distinguishing angular distributions that we are  looking for.  Equipped  with  collider cuts and thresholds we  only show the invariant mass distribution constructing the BSM particle respective to different  Seesaw mechanisms.

\subsubsection{iType-I Seesaw at $\mu^+ \mu^-$ Collider} 

\begin{figure*}[h]
	\begin{center}
		\mbox{\subfigure[BP1 (iType-I/III)]{\includegraphics[width=0.42\linewidth,angle=-0]{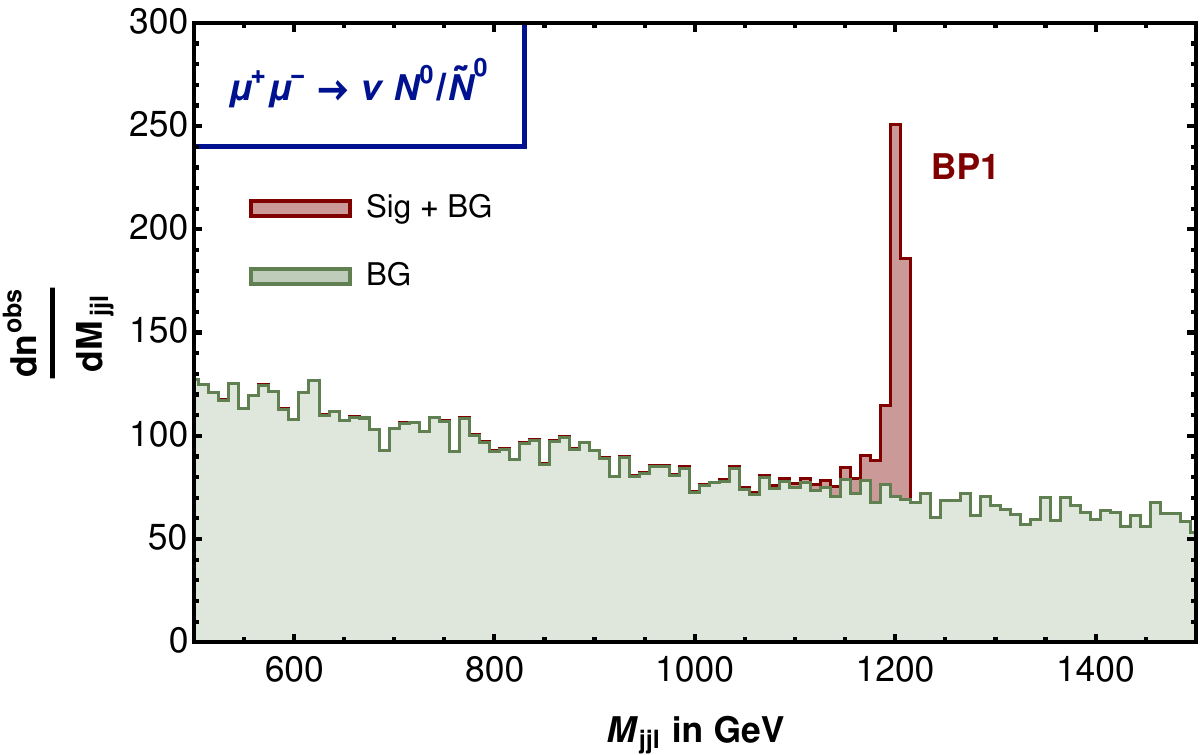}\label{}}
			\hspace*{1.0cm}
			\subfigure[BP2 (iType-I/III)]{\includegraphics[width=0.42\linewidth,angle=-0]{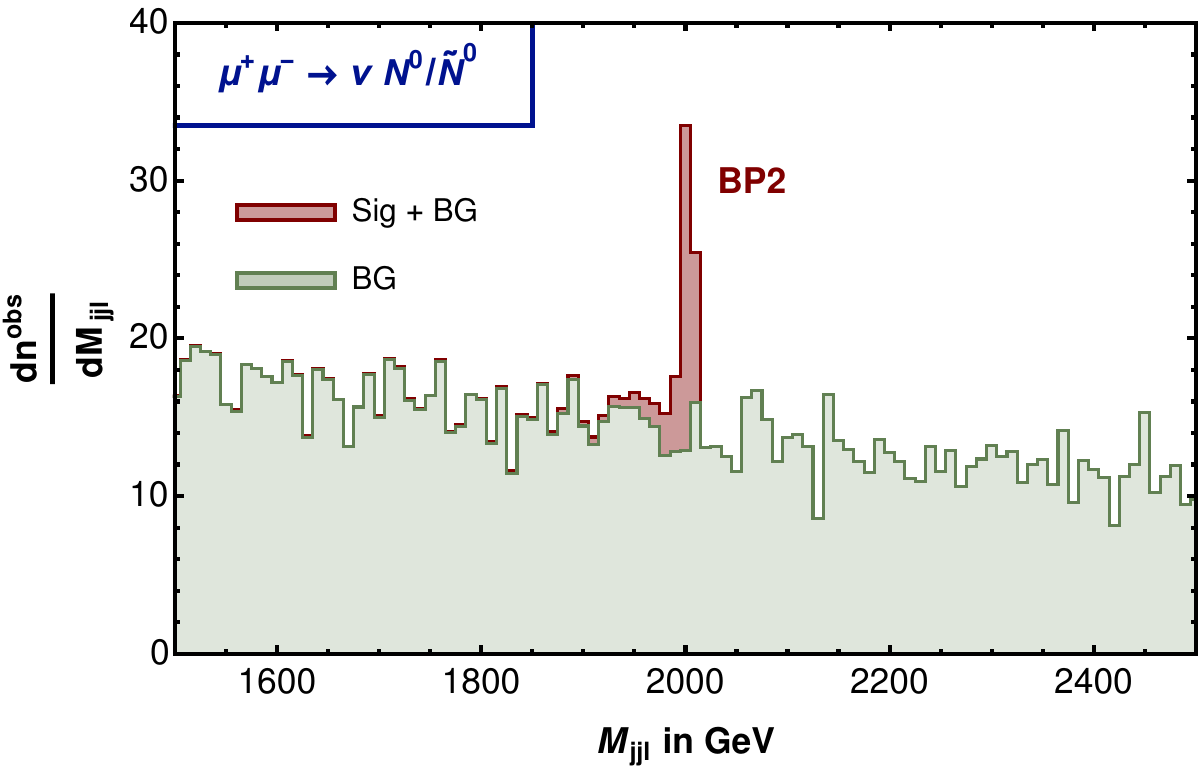}\label{}}}		
		\caption{Di-jet-mono-lepton invariant mass distribution ($M_{jjl}$) for (a) BP1 and (b) BP2 at the centre-of-mass energies of 3.0, 6.0\, TeV, respectively with the integrated luminosity of 1000\,fb$^{-1}$.  The total (signal $+$ SM background) signature is depicted in brown and the SM background is in olive green.}\label{mu+mu-Ty1invM}
	\end{center}
\end{figure*}

We study the associated production of $N^0/\widetilde{N}^0$ at a $\mu^+\mu^-$ collider, which enhances the signal cross-section. These heavy states decay through $\ell^\pm W^\mp$, leading to $jj\ell$ final states. Since the neutrino remains invisible, our focus is on reconstructing the leg of $N^0/\widetilde{N}^0$. In the iType-I Seesaw, this can be done using the invariant mass $M_{jj\ell}$ reconstruction, where we impose $|M_{jj}-M_W|\leq 10$ GeV to ensure an on-shell $W^\pm$ in the final state. In \autoref{mu+mu-Ty1invM}, we show the $M_{jj\ell}$ distributions for two benchmark points, plotted against the SM backgrounds, at $\sqrt{s}=3$ and $6$ TeV, respectively, with an integrated luminosity of 1000 fb$^{-1}$. The SM background is shown in olive green, while the total contribution (signal + background) is displayed in brown. Distinct invariant mass peaks corresponding to the benchmark points are clearly visible.

\begin{table*}[hbt]
	\renewcommand{\arraystretch}{1.5}
	\centering
	\hspace*{-0.4cm}
	\scalebox{0.87}{\begin{tabular}{|c|c|c||c|c|c|c|c|c|c|c|}
			\cline{2-11}
			\multicolumn{1}{c|}{}&\multirow{2}{*}{Final states}& \multirow{2}{*}{Signal} &\multicolumn{8}{c|}{Backgrounds} \\ 
			\cline{4-11}
			\multicolumn{1}{c|}{}& & & $t\bar{t}$ & $VV$ & $VVV$ & $t\bar{t}V$ & $\ell^{\pm}W^{\mp}\overset{\textbf{\fontsize{1.5pt}{1.5pt}\selectfont(--)}}{\nu}$ & $\tau^+ \tau^-$ & $\tau^{\pm}W^{\mp}\overset{\textbf{\fontsize{1.5pt}{1.5pt}\selectfont(--)}}{\nu}$ & $\tau^+ \tau^- Z$ \\ 
			\hline
			\multirow{5}{*}{BP1} & $1\ell +2j + p_T^{\rm miss}\geq 350\,\rm GeV$ & 227.73 & 310.84 & 1485.65 & 1116.56 & 14.65 & 1180.0 & 151.05 & 200.57 & 59.06 \\ 
			\cline{3-11}
			& $+|M_{jj\ell} -M_{n}|\leq 10\,$GeV & 92.95 & 0.00 & 6.20 & 1.12 & 0.00 & 24.82 & 0.00 & 5.16 & 0.02  \\ 
			\cline{2-11}
			& Total & 92.95 &\multicolumn{8}{c|}{ 37.32 } \\		
			\cline{2-11}	
			& $\rm{S}_{\text{sig}} (\mathcal{L}_{\rm int}=1000\, \rm fb^{-1})$  &  \multicolumn{9}{c|}{8.14} \\  
			\cline{2-11}
			& $\int \mathcal{L}_{5\sigma}\,[\rm{fb}^{-1}]$ &  \multicolumn{9}{c|}{377.30} \\ 
			\hline \hline
			\multirow{5}{*}{BP2} & $1\ell +2j + p_T^{\rm miss}\geq 350\,\rm GeV$ & 91.58 & 23.01 & 840.42 & 920.02 & 4.83  & 573.34 & 67.23 & 80.27 & 31.63 \\ 
			\cline{3-11}
			& $+|M_{jj\ell} -M_{n}|\leq 10\,$GeV & 12.62 & 0.00 & 1.59 & 0.21 & 0.00 & 27.06 & 0.00 & 0.39 & 0.00 \\ 
			\cline{2-11}
			& Total & 12.62  &\multicolumn{8}{c|}{ 29.25 } \\
			\cline{2-11}	
			& $\rm{S}_{\text{sig}} (\mathcal{L}_{\rm int}=1000\, \rm fb^{-1})$  &  \multicolumn{9}{c|}{1.95} \\  
			\cline{2-11}
			& $\int \mathcal{L}_{5\sigma}\,[\rm{fb}^{-1}]$ &   \multicolumn{9}{c|}{6574.62} \\ 
			\hline 
	\end{tabular}}
	\caption{Number of events for signal and background corresponding to $1\ell +2j+p_T^{\rm miss} \geq 350\,$GeV final state for the benchmark points mentioned in \autoref{crs_mupmum}  \textcolor{black}{with the integrated luminosity of 1000\,fb$^{-1}$.}}  \label{sigbgTy1}
\end{table*}

After the successful reconstruction of  $N^0/\widetilde{N}^0$  via  $M_{jj\ell}$ invariant mass  distribution we focus on the final state defined in \autoref{sigbgTy1} where we look at $1\ell +2j + p_T^{\rm miss}\geq 350\,\rm {GeV} +|M_{jj\ell} -M_{n}|\leq 10\,\rm{GeV}$.  The dominant  SM background  numbers  viz. $t\bar{t},  \,VV, \,  VVV,\, t\bar{t}V,   \,  W^\pm W^\mp, \tau^+ \tau^-, \, \tau^{\pm}W^{\mp}\overset{\textbf{\fontsize{1.5pt}{1.5pt}\selectfont(--)}}{\nu}$  and $\tau^+ \tau^- Z$ are also shown at an integrated  luminosity  of 1000 fb$^{-1}$.  The SM processes involving  gauge boson contribute dominantly. $\tau$ decays to electron and muon can contribute to this final state. However,  an invariant mass cut of $|M_{jj\ell} -M_{n}|\leq 10\,$GeV removes  such background. The event number looks healthy  for BP1 and a $5\sigma$ signal significance  needs around 377 fb$^{-1}$ of integrated luminosity. For BP2 it takes around 6500 fb$^{-1}$ of  integrated luminosity for $5\sigma$ significance.

\subsubsection{Type-II Seesaw at $\mu^+ \mu^-$ Collider} 

\begin{figure*}[h]
	\begin{center}
		\mbox{\subfigure[BP1 (Type-II)]{\includegraphics[width=0.42\linewidth,angle=-0]{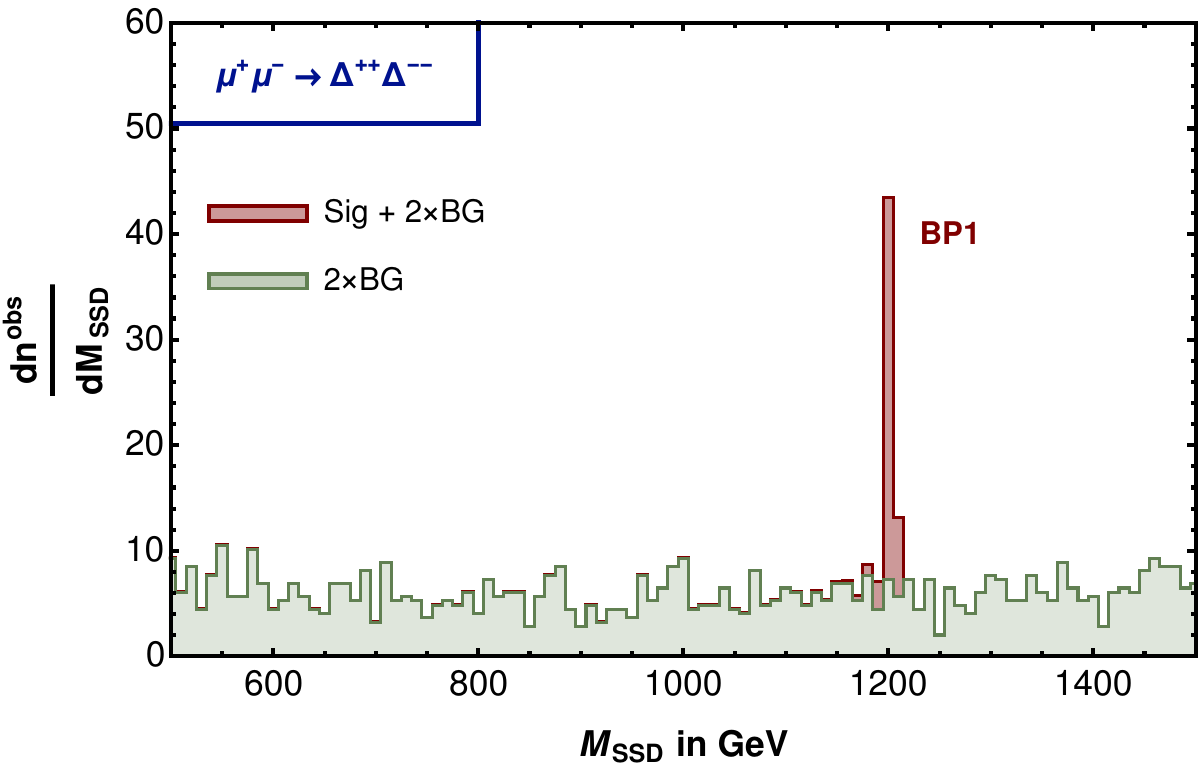}}
			\hspace*{1.0cm}
			\subfigure[BP2 (Type-II)]{\includegraphics[width=0.42\linewidth,angle=-0]{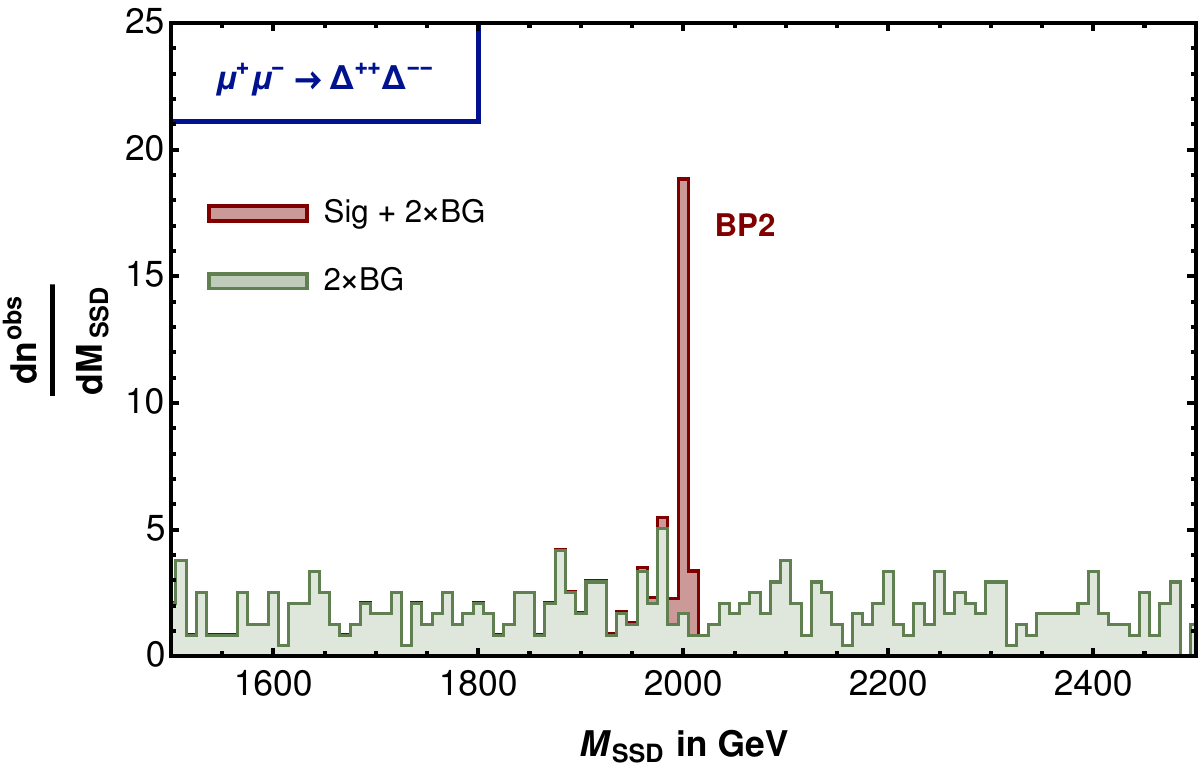}}}		
		\caption{Same sign di-lepton invariant mass distribution ($M_{\rm SSD}$) for (a) BP1 and (b) BP2 at the centre-of-mass energies of 3.0, 6.0\, TeV, respectively with the integrated luminosity of 1000\,fb$^{-1}$.  The total (signal $+$ SM background scaled by 2) signature is depicted in brown and the SM background (scaled by 2) is in olive green. }\label{mu+mu-Ty2invM}
	\end{center}
\end{figure*}

Similarly, in the Type-II Seesaw model, we consider the process $\mu^+\mu^- \to \Delta^{++} \Delta^{--}$, where the $\Delta^{\pm\pm}$ particles decay into same sign di-leptons (SSD). It is important to note that the doubly charged scalars can decay into either same sign di-lepton or di-boson channels \cite{FileviezPerez:2008jbu}. The decay width to the di-lepton channel is proportional to $Y_{\Delta}^2$, while the width to the di-boson mode varies with $v_\Delta^2$. In our analysis, we assume a Yukawa coupling of 0.2, and the selection of a small triplet scalar VEV ensures that the doubly charged scalar decays exclusively into the di-lepton mode with a 100\% branching ratio. The construction of the invariant mass distribution of $\ell^{\pm}\ell^{\pm}$ can reveal the mass of $\Delta^{\pm\pm}$, as shown in \autoref{mu+mu-Ty2invM}. The SM background (scaled with two) is presented in olive green while the signal plus background (scaled by two) is shown by brown colour. Distinctive peaks in brown colour near the BSM scalar mass are clearly visible across the two benchmark points, while the background shows more or less flat distributions.

For the final state, in \autoref{sigbgTy2}, we consider $2\ell^+ + 2\ell^- + |M_{\ell \ell} - M_{\Delta}| \leq 10\, \text{GeV}$ at a $\mu^+\mu^-$ collider with centre-of-mass energies of 3 TeV and 6 TeV, corresponding to BP1 and BP2, respectively. The event numbers are presented for an integrated luminosity of 1000 fb$^{-1}$, where the contributions from potential SM backgrounds are also shown, with the most dominant being $\ell^\pm \ell^\mp Z$. Background events involving vector bosons also contribute significantly; however, reconstructing the di-lepton invariant mass peak and identifying it at the BSM scalar mass can effectively reduce these contributions. We observe that for BP1 and BP2, an early data set with approximately 61 fb$^{-1}$ and 127 fb$^{-1}$ of integrated luminosity, respectively, will be sufficient to achieve a $5\sigma$ significance.

\begin{table*}[h!]
	\renewcommand{\arraystretch}{1.3}
	\centering
	\scalebox{0.9}{\begin{tabular}{|c|c|c||c|c|c|c|c|c|}
			\cline{2-9}
			\multicolumn{1}{c|}{}&\multirow{2}{*}{Final states}& \multirow{2}{*}{Signal} &\multicolumn{6}{c|}{Backgrounds} \\ 
			\cline{4-9}
			\multicolumn{1}{c|}{}&& & $t\bar{t}$ & $VV$ & $VVV$ & $t\bar{t}V$ & $\ell^+\ell^-Z$ & $\tau^+ \tau^- Z$ \\ 
			\hline
			\multirow{5}{*}{BP1} & $2\ell^+ +2\ell^-$ & 539.11 & 0.20 & 120.92 & 113.25 & 2.24 & 12553.93  & 13.11  \\ 
			\cline{3-9}
			& $+|M_{\ell \ell} -M_{\Delta}|\leq 10\,$GeV & 521.13 & 0.00 & 0.52 & 0.83 & 0.03 & 136.08 & 0.22 \\ 
			\cline{2-9}
			& Total & 521.13  &\multicolumn{6}{c|}{ 137.68 } \\
			\cline{2-9}	
			& $\rm{S}_{\text{sig}} (\mathcal{L}_{\rm int}=1000\, \rm fb^{-1})$  &  \multicolumn{7}{c|}{20.30} \\  
			\cline{2-9}
			& $\int \mathcal{L}_{5\sigma}\,[\rm{fb}^{-1}]$ &   \multicolumn{7}{c|}{60.67 } \\ 
			\hline \hline
			\multirow{5}{*}{BP2} & $2\ell^+ +2\ell^-$ & 231.85 & 0.01 & 30.86 & 53.09 & 0.40 & 6909.14 & 3.88 \\ 
			\cline{3-9}
			& $+|M_{\ell \ell} -M_\Delta|\leq 10\,$GeV & 223.88 & 0.00 & 0.32 & 0.11 & 0.00 & 30.62 & 0.04  \\ 
			\cline{2-9}
			& Total & 223.88  &\multicolumn{6}{c|}{ 31.09 } \\
			\cline{2-9}	
			& $\rm{S}_{\text{sig}} (\mathcal{L}_{\rm int}=1000\, \rm fb^{-1})$  &  \multicolumn{7}{c|}{14.02} \\  
			\cline{2-9}
			& $\int \mathcal{L}_{5\sigma}\,[\rm{fb}^{-1}]$ &   \multicolumn{7}{c|}{127.18} \\ 
			\hline 
	\end{tabular}}
	\caption{Number of events for signal and background corresponding to $2\ell^+ +2\ell^-$ final state for the benchmark points mentioned in \autoref{crs_mupmum} \textcolor{black}{with the integrated luminosity of 1000\,fb$^{-1}$.}}  \label{sigbgTy2}
\end{table*}

\subsubsection{iType-III Seesaw at $\mu^+ \mu^-$ Collider} 

\begin{figure*}[h]
	\begin{center}
		\mbox{\subfigure[BP1 (iType-III)]{\includegraphics[width=0.42\linewidth,angle=-0]{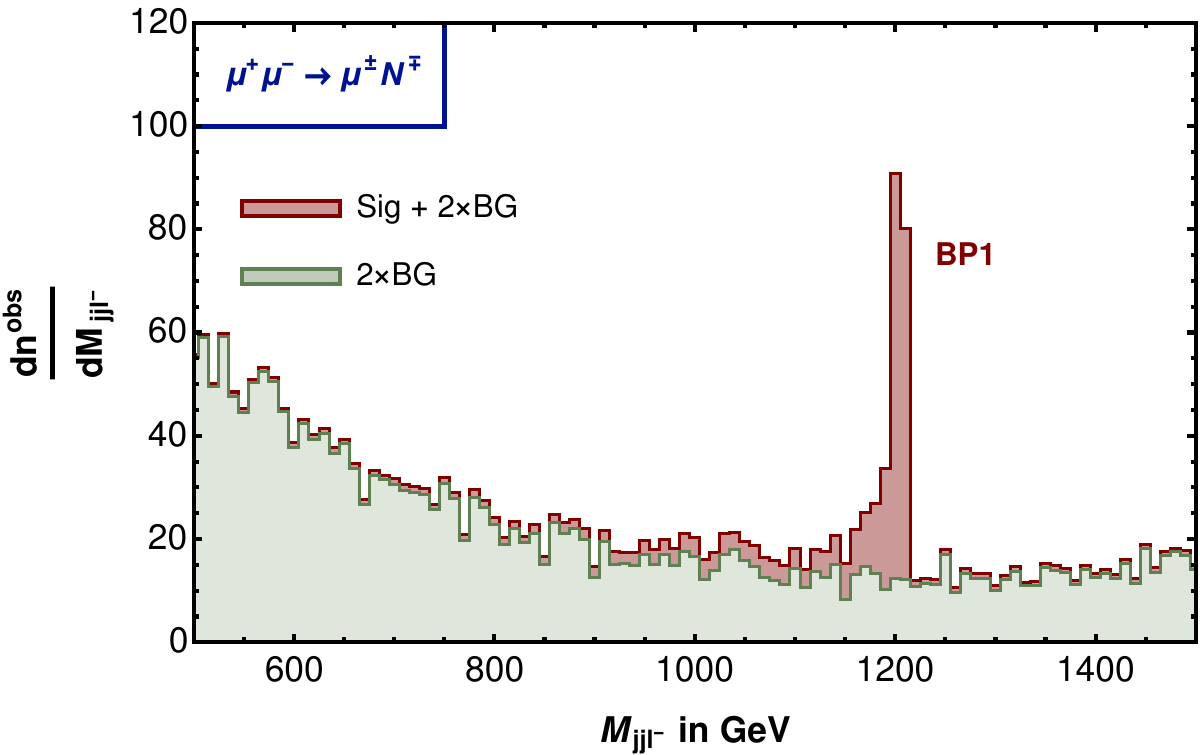}\label{}}
			\hspace*{1.0cm}
			\subfigure[BP2 (iType-III)]{\includegraphics[width=0.42\linewidth,angle=-0]{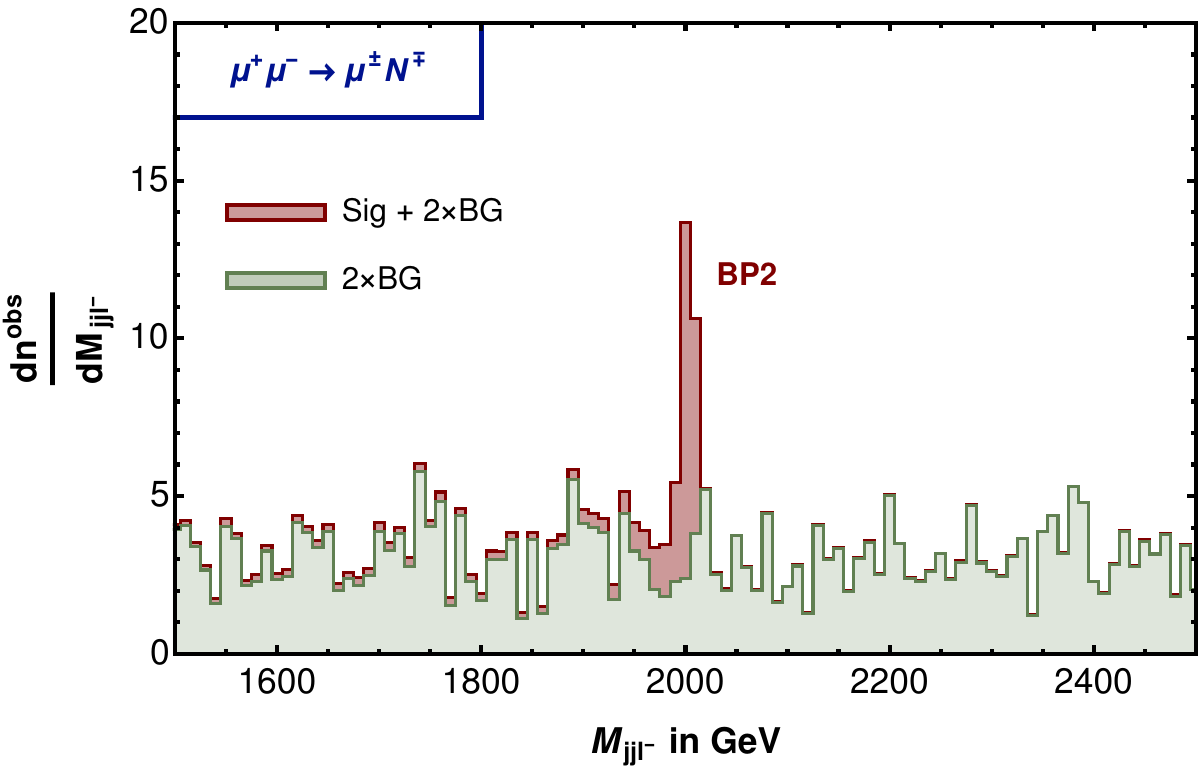}\label{}}}		
		\caption{Di-jet-mono-lepton invariant mass distribution ($M_{jjl^-}$) for (a) BP1 and (b) BP2 at the centre-of-mass energies of 3.0, 6.0\, TeV, respectively with the integrated luminosity of 1000\,fb$^{-1}$.  The total (signal $+$ SM background scaled by 2) signature is depicted in brown and the SM background (scaled by 2) is in olive green. }\label{mu+mu-Ty3invM}
	\end{center}
\end{figure*}

Finally, we also consider the inverse Type-III Seesaw scenario, where we investigate the process $\mu^\pm N^\mp$ to enhance the cross-section compared to $N^\pm$ pair production for a TeV-scale Type-III fermion. The reconstruction of $N^\pm$ is accomplished by first reconstructing the $Z$ boson through the requirement $|M_{jj} - M_Z| \leq 10$ GeV. The resulting jet pairs, involved in $Z$ boson mass reconstruction, are then used to plot the invariant mass distribution of $jj\ell$, as shown in \autoref{mu+mu-Ty3invM}. The same colour coding is used here for signal and total events (signal + background). In the brown regions, clear peaks are observed around the mass of the charged fermion for all benchmark points.

\autoref{sigbgTy3} presents the event numbers for the final state $\rm{OSD} + 2j + |M_{jj\ell} - M_{N}| \leq 10$ GeV for the benchmark points and dominant SM backgrounds at centre-of-mass energies of 3 TeV and 6 TeV, respectively, with an integrated luminosity of 1000 fb$^{-1}$. The dominant background contributor is $\ell^+\ell^- Z$. However, successful reconstruction of the charged triplet fermion can significantly reduce this background. For a $5\sigma$ signal significance, approximately 10 fb$^{-1}$ and 118 fb$^{-1}$ of luminosity are required for BP1 and BP2, respectively.


\begin{table*}[h!]
	\renewcommand{\arraystretch}{1.3}
	\centering
	\scalebox{0.9}{\begin{tabular}{|c|c|c||c|c|c|c|c|c|}
			\cline{2-9}
			\multicolumn{1}{c|}{}&\multirow{2}{*}{Final states}& \multirow{2}{*}{Signal} &\multicolumn{6}{c|}{Backgrounds} \\ 
			\cline{4-9}
			\multicolumn{1}{c|}{}&& & $t\bar{t}$ & $VV$ & $VVV$ & $t\bar{t}V$ & $\ell^+\ell^-Z$ & $\tau^+ \tau^- Z$ \\ 
			\hline
			\multirow{5}{*}{BP1} &  $\rm{OSD} + 2j$ & 6976.33 & 58.37 & 1233.99 & 1090.81 & 16.50 & 86270.6 & 99.07 \\ 
			\cline{3-9}
			& $+|M_{jj \ell} -M_{n}|\leq 10\,$GeV & 2576.34 & 0.04 & 7.75 & 4.88 & 0.01 & 154.71 & 1.08  \\ 
			\cline{2-9}
			& Total & 2576.34  &\multicolumn{6}{c|}{ 168.47 } \\
			\cline{2-9}	
			& $\rm{S}_{\text{sig}} (\mathcal{L}_{\rm int}=1000\, \rm fb^{-1})$  &  \multicolumn{7}{c|}{49.17} \\  
			\cline{2-9}
			& $\int \mathcal{L}_{5\sigma}\,[\rm{fb}^{-1}]$ &   \multicolumn{7}{c|}{10.33} \\ 
			\hline \hline
			\multirow{5}{*}{BP2} &  $\rm{OSD} + 2j$ & 816.39 & 2.51 & 242.92 & 529.99 & 7.55 & 47987.10 & 23.21 \\ 
			\cline{3-9}
			& $+|M_{jj \ell} -M_n|\leq 10\,$GeV & 243.90 & 0.00 & 0.64 & 0.61 & 0.00 & 36.91 & 0.13 \\ 
			\cline{2-9}
			& Total & 243.90  &\multicolumn{6}{c|}{ 38.29 } \\
			\cline{2-9}	
			& $\rm{S}_{\text{sig}} (\mathcal{L}_{\rm int}=1000\, \rm fb^{-1})$  &  \multicolumn{7}{c|}{14.52} \\  
			\cline{2-9}
			& $\int \mathcal{L}_{5\sigma}\,[\rm{fb}^{-1}]$ &   \multicolumn{7}{c|}{118.59} \\ 
			\hline 
	\end{tabular}}
	\caption{Number of events for signal and background corresponding to $\rm{OSD} + 2j$ final state for the benchmark points mentioned in   		\autoref{crs_mupmum} \textcolor{black}{with the integrated luminosity of 1000\,fb$^{-1}$.}} \label{sigbgTy3}
\end{table*}	


\subsection{Reconstructed angular distribution at $\mu^+ \mu^-$ collider}

\begin{figure*}[hbt]
	\begin{center}
		\mbox{\subfigure[iType-I/III]{\includegraphics[width=0.325\linewidth,angle=-0]{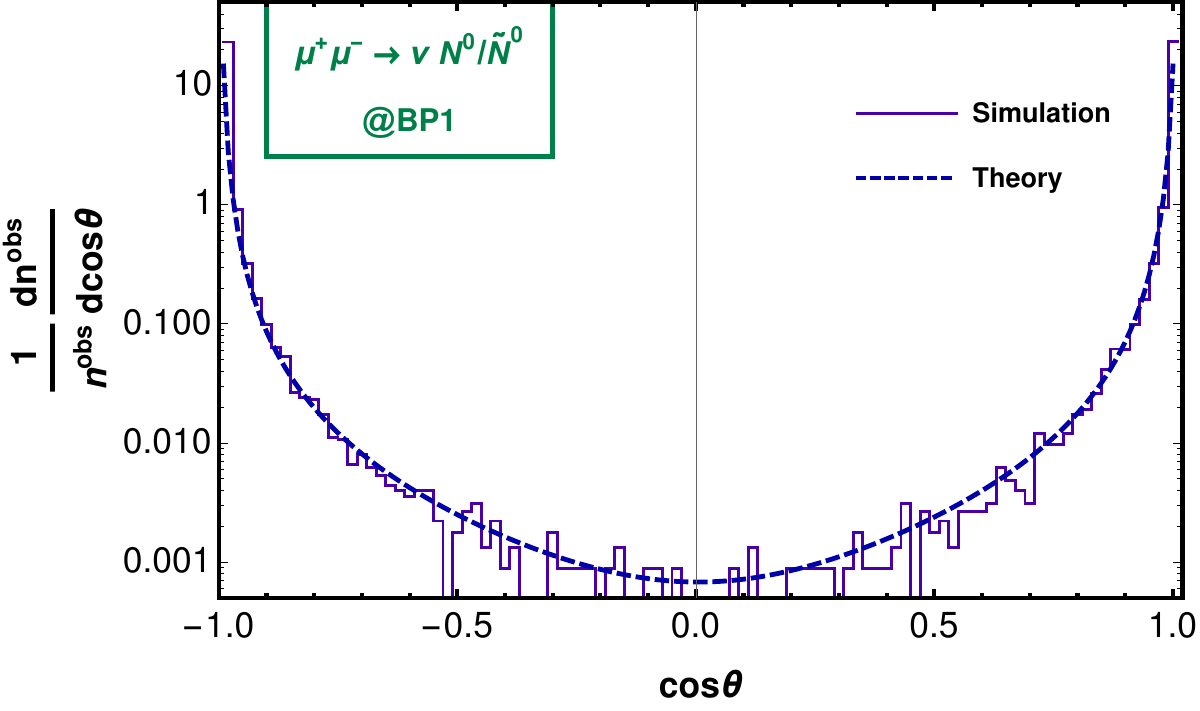}}
			\subfigure[Type-II]{\includegraphics[width=0.32\linewidth,angle=-0]{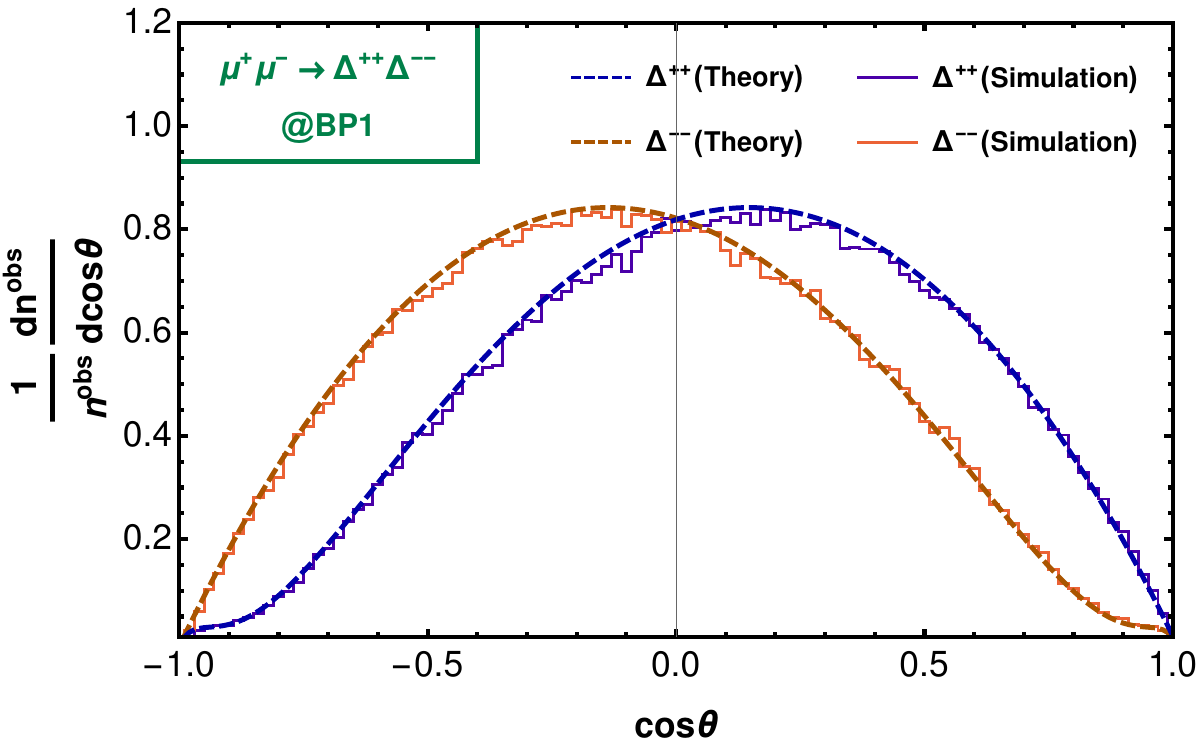}}
			\subfigure[iType-III]{\includegraphics[width=0.325\linewidth,angle=-0]{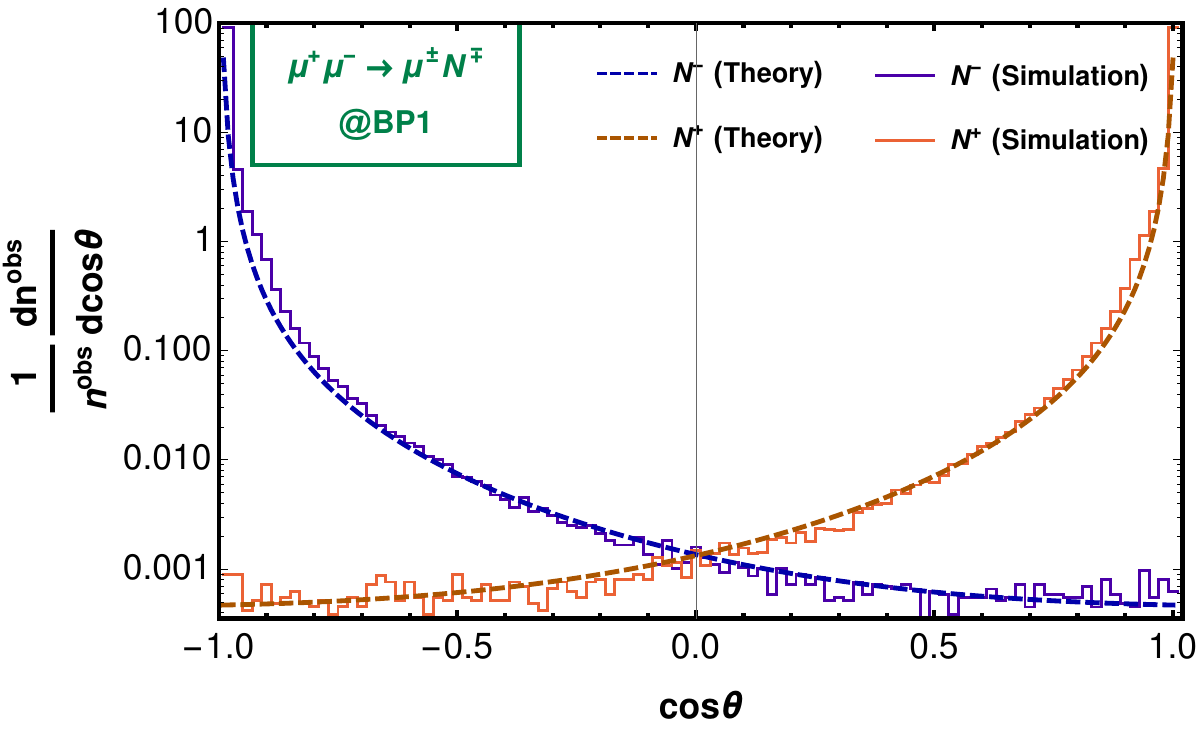}}}		
		\caption{Angular distribution of  the theory and simulation  at $\mu^+ \mu^-$ collider for the seesaw scenarios for BP1.   }\label{angdismupmumF}
	\end{center}
\end{figure*}

Finally, we address the reconstruction of angular distributions in the centre-of-mass frame, which can be distinctive for different scenarios. \autoref{angdismupmumF} displays the angular distributions of BSM particles ($N^0/\widetilde{N}^0,\, \Delta^{\pm\pm},\, N^\pm$) in the centre-of-mass frame for the three different Seesaw scenarios. \autoref{angdismupmumF}(a) shows the reconstructed angular distribution of $N^0/\widetilde{N}^0$, presented as a histogram, which closely resembles \autoref{typ1} and aligns with the parton-level prediction indicated by the dashed curve. Since the collision occurs in the centre-of-mass frame, constructing the angular distribution in this frame is straightforward. \autoref{angdismupmumF}(b) illustrates a similar angular distribution for the production of $\Delta^{\pm\pm}$ pairs in the Type-II Seesaw scenario, as also described in \autoref{typII}. The simulated histograms for the reconstructed $\Delta^{++}$ (blue) and $\Delta^{--}$ (orange) closely match the parton-level distributions, highlighting the distinct differences from the $N^0/\widetilde{N}^0$ distribution in the iType-I case. \autoref{angdismupmumF}(c) presents the angular distributions for iType-III Seesaw, corresponding to $N^+$ (blue) and $N^-$ (orange) with reconstructed $N^\pm$. These histograms perfectly match the parton-level distributions, as described in \autoref{lNds}. It is evident that the angular distributions differ significantly from those of the iType-I and Type-II Seesaw scenarios for the corresponding BSM particles. Furthermore, we get a typical asymmetric angular distribution in the iType-III
scenario, for the two different singly-charged fermions ($N^+$ and $N^-$).

\begin{table}[h!]
	\renewcommand{\arraystretch}{1.5}
	\centering
    \hspace*{-1.0cm}
	\begin{tabular}{|c|c|c|c|c|c|c|c|c|c|c|}
		\cline{2-11}
		\multicolumn{1}{c|}{} & \multicolumn{2}{c|}{iType-I} & \multicolumn{4}{c|}{Type-II} & \multicolumn{4}{c|}{iType-III} \\
		\cline{2-11}
		\multicolumn{1}{c|}{} & \multicolumn{2}{c|}{$\cos \theta$} & \multicolumn{4}{c|}{$\cos \theta$} & \multicolumn{4}{c|}{$\cos \theta$} \\
		\cline{2-11}
		\multicolumn{1}{c|}{} & ($-0.7 \to 0.0$) & ($0.0 \to 0.7$) & \multicolumn{2}{c|}{($-0.7 \to 0.0$)} & \multicolumn{2}{c|}{($0.0 \to 0.7$)} & \multicolumn{2}{c|}{($-0.7 \to 0.0$)} & \multicolumn{2}{c|}{($0.0 \to 0.7$)} \\
		\cline{2-11}
		\multicolumn{1}{c|}{} & $N^0/\widetilde{N}^0$ & $N^0/\widetilde{N}^0$ & $\Delta^{++}$ & $\Delta^{--}$ & $\Delta^{++}$ & $\Delta^{--}$ & $N^+$ & $N^-$ & $N^+$ & $N^-$ \\
		\hline 
		Signal & 51.90 & 48.8 & 209.8 & 277.3 & 280.4 & 210.9 & 92.8 & 791.1 & 802.0 & 89.4  \\
		\hline 
		Background & 33.3 & 39.5 & 89.6 & 92.1  & 77.1 & 76.4 & 140.4 & 137.7 & 127.1 & 125.9  \\
		\hline 
	\end{tabular}
	\caption{Number of events for $\cos{\theta}:(-0.7,0),\,(0,0.7)$ for the BP1 of three different seesaw scenarios, corresponding to the final states discussed earlier at an integrated luminosity of 1000~\fbi.}  \label{tab:assymtry1}
\end{table}

{As the difference among  the different seesaw scenarios are evident from \autoref{angdismupmumF}, the symmetry or asymmetry around $\cos{\theta}=0$, can give rise a quantitative estimate of these segregation. For a quantitative estimates we show the number of events for the final states shown in \autoref{sigbgTy1}, \autoref{sigbgTy2}, \autoref{sigbgTy3} for two regions of $\cos{\theta}$ i.e. $(0:0.7)$ and $(0:-0.7)$ in \autoref{tab:assymtry1} with backgrounds, respectively at an integrated luminosity of 1000 fb$^{-1}$. As the distribution for iType-I in \autoref{angdismupmumF}(a) we see a symmetric distributions and for the events within $(-0.7:0.7)$ of $\cos{\theta}$ giving rise to signal significance of $9\sigma$. However, as the distribution is symmetric around $\cos{\theta}=0$ giving rise to no asymmetry. On the contrary for Type-II \autoref{angdismupmumF}(b) we see the asymmetry for $\Delta^{++}$ and $\Delta^{--}$. This asymmetry gives rise to events around 70 and 67 events, and signal significances of $9.3\sigma$ and $7.3 \sigma$ for $\Delta^{++}$ and $\Delta^{--}$, respectively. The asymmetry for BP1 of iType-III gives rise to signal event number difference of 709 and 701 for $N^+$ and $N^-$, respectively with signal significances of $27 \sigma$ and $26.3$ for the chosen final states as given in \autoref{sigbgTy3}. Thus, as a novel feature, comparing the asymmetric signal events in all three scenarios one can easily distinguish these BSM scenarios from other possibilities.  }

\section{At $\mu^+\mu^+$ collider}\label{sec:mup_mup}

In addition to the conventional $\mu^+\mu^-$ setup, proposals for a $\mu^+\mu^+$ collider have recently gained attention \cite{Hamada:2022mua,Chen:2024tqh}. One of the key advantages lies in the relative ease of producing and cooling $\mu^+$ beams compared to $\mu^-$ beams, as demonstrated at facilities like J-PARC through ultra-cold muon technology, which has successfully cooled and accelerated a $\mu^+$ beam via muonium formation and laser ionization techniques \cite{Mibe:2011zz}. A high-quality $\mu^+$ beam not only improves luminosity but also reduces beam-induced backgrounds, thereby making a $\mu^+\mu^+$ configuration technically appealing. From a physics perspective, the $\mu^+\mu^+$ collider provides a unique environment for probing BSM signatures that are absent in the Standard Model. Since the initial state carries a net electric charge of $+2$, it naturally favours processes involving doubly charged particles or lepton-number–violating interactions. This makes it an excellent setup for testing Seesaw scenarios that predict heavy Majorana neutrinos or doubly charged scalars. Motivated by these considerations, we now explore the angular distributions of BSM particles in different Seesaw realizations at a $\mu^+\mu^+$ collider.

Inverse Type-I Seesaw is particularly challenging to probe in a $\mu^+\mu^+$ collider. However, the process $\mu^+\mu^+ \to W^+W^+$, where BSM particles can play a significant role, becomes relevant. This process is entirely absent in the SM via Drell-Yan production, due to the non-existence of Majorana fermions or doubly charged particles. Therefore, any deviation from SM predictions, along with differences in angular distributions, could signal new physics. This is illustrated in \autoref{tab:WW}, where $\mu^+\mu^+ \to W^+W^+$ is mediated by $N^0/\widetilde{N}^0$ in iType-I and iType-III Seesaw scenarios, and by $\Delta^{++}$ in the Type-II Seesaw scenario via the s-channel.

\renewcommand{\arraystretch}{1.5}
\begin{table}[h!]
    \centering
    \begin{tabular}{|c||cc|}
    \hline\hline
     \multicolumn{3}{|c|}{\textbf{$\bm{\mu^+\mu^+\to W^+W^+}$}}\\
     \hline\hline
    \multirow{-1}{*}{\rotatebox{90}{\textbf{iType-I/III}}} &
		\begin{tikzpicture}
			\begin{feynman}
				\vertex (a1);
				\vertex [left=1cm of a1] (a0){$\mu^+$};
				\vertex [right=1cm of a1] (a2){$W^+_\rho$};
				\vertex [below=1.5 cm of a1] (b1);
				\vertex [left=1cm of b1] (b0){$\mu^+$};
				\vertex [right=1cm of b1] (b2){$W^+_\sigma$};
				\diagram {(a0)--[anti fermion](a1)--[boson](a2),
					(b0)--[anti fermion](b1)--[boson](b2),
					(a1)--[majorana, edge label'=$N^0/\widetilde N^0\,$](b1)};
			\end{feynman}
		\end{tikzpicture}
		
        &
		
        \begin{tikzpicture}
			\begin{feynman}
				\vertex (a1);
				\vertex [left=1.0cm of a1] (a0){$\mu^+$};
				\vertex [right=1.0cm of a1] (a2){$W^+_\rho$};
				\vertex [below=1.5 cm of a1] (b1);
				\vertex [left=1.0cm of b1] (b0){$\mu^+$};
				\vertex [right=1.0cm of b1] (b2){$W^+_\sigma$};
				\diagram {(a0)--[anti fermion](a1)--[boson](a2),
					(b0)--[anti fermion](b1)--[boson](b2),
					(a1)--[anti majorana, edge label'=$N^0/\widetilde N^0\,$](b1)};
			\end{feynman}
		\end{tikzpicture}
        
		\\
        
		& \begin{tikzpicture}
			\begin{feynman}
				\vertex (a1);
				\vertex [left=1.0cm of a1] (a0){$\mu^+$};
				\vertex [right=1.0cm of a1] (a2){$W^+_\rho$};
				\vertex [below=1.5 cm of a1] (b1);
				\vertex [left=1.0cm of b1] (b0){$\mu^+$};
				\vertex [right=1.0cm of b1] (b2){$W^+_\sigma$};
				\diagram {(a0)--[anti fermion](a1)--[boson](b2),
					(b0)--[anti fermion](b1)--[boson](a2),
					(a1)--[majorana, edge label'=$N^0/\widetilde N^0\,$](b1)};
			\end{feynman}
		\end{tikzpicture}
		
        &
        
		\begin{tikzpicture}
			\begin{feynman}
				\vertex (a1);
				\vertex [left=1.0cm of a1] (a0){$\mu^+$};
				\vertex [right=1.0cm of a1] (a2){$W^+_\rho$};
				\vertex [below=1.5 cm of a1] (b1);
				\vertex [left=1.0cm of b1] (b0){$\mu^+$};
				\vertex [right=1.0cm of b1] (b2){$W^+_\sigma$};
				\diagram {(a0)--[anti fermion](a1)--[boson](b2),
					(b0)--[anti fermion](b1)--[boson](a2),
					(a1)--[anti majorana, edge label'=$N^0/\widetilde N^0\,$](b1)};
			\end{feynman}
		\end{tikzpicture}
        \\
        \hline
        
		\multirow{-4}{*}{\rotatebox{90}{\textbf{Type-II}}} 
        &
        \multicolumn{2}{c|}{\begin{tikzpicture}
			\begin{feynman}
				\vertex (a1);
				\vertex [above left=1cm of a1] (a0){$\mu^+$};
				\vertex [right=1.5cm of a1] (a2);
				\vertex [above right=1 cm of a2] (a3){$W_\rho^+$};
				\vertex [below left=1cm of a1] (b0){$\mu^+$};
				\vertex [below right=1cm of a2] (b3){$W_\sigma^+$};
				\diagram {(a0)--[anti fermion](a1)--[anti charged scalar,edge label'=$\Delta^{++}$](a2)--[ boson](a3),
					(b0)--[anti fermion](a1),(a2)--[ boson](b3)};
			\end{feynman}
		\end{tikzpicture}}
 \\
     \hline\hline   
    \end{tabular}
		\caption{Feynman diagrams for $\mu^+\mu^+\to W^+W^+$ for the seesaw scenarios at $\mu^+\mu^+$ collider.}
    \label{tab:WW}
\end{table}

In the case of the iType-I Seesaw, the $\mu^+\mu^+ \to W^+W^+$ cross-section is proportional to $(Y_\nu^2/M_N)^2$. As previously mentioned, either $Y_\nu$ must be very small or $M_N$ must be very large to maintain the neutrino mass at $\mathcal{O}(10^{-1} \text{ eV})$, which results in a vanishingly small cross-section. In the inverse Seesaw, destructive interference occurs between the t-channel diagrams mediated by $N^0$ and $\widetilde{N}^0$, further reducing the cross-section. A similar situation arises in the iType-III scenario for the same channel. For the Type-II scenario, the additional contribution comes from the doubly charged scalar particle in the s-channel. However, the $\mu^+\mu^+\Delta^{--}$ vertex is proportional to $Y_\Delta$, while the $W^+W^+\Delta^{--}$ vertex scales with $v_\Delta$, leading to a cross-section proportional to $v_\Delta^2Y_\Delta^2 \sim \frac{1}{2}m_{\nu}^2$ (as detailed in \autoref{massII}). Consequently, in the Type-II case, this cross-section is also vanishingly small.

However, it is interesting to note that a $\mu^+ \mu^+$ collider can still be instrumental in distinguishing between Type-II and iType-III Seesaw mechanisms. This can be achieved through final states involving leptons of different flavors or those involving at least one heavy BSM lepton (see \autoref{tab:feyn_mupp}), as detailed in the following subsections.

\renewcommand{\arraystretch}{1.5}
\begin{table}[h!]
    \centering
    \begin{tabular}{|c||cc|}
    \hline\hline
     \multicolumn{3}{|c|}{\textbf{At $\bm{\mu^+\mu^+}$ collider}}\\
     \hline\hline
    \multirow{-4}{*}{\rotatebox{90}{\textbf{Type-II}}} &
		\multicolumn{2}{c|}{\begin{tikzpicture}
			\begin{feynman}
				\vertex (a1);
				\vertex [above left=1cm of a1] (a0){$\mu^+$};
				\vertex [right=1.5cm of a1] (a2);
				\vertex [above right=1 cm of a2] (a3){$e^+$};
				\vertex [below left=1cm of a1] (b0){$\mu^+$};
				\vertex [below right=1cm of a2] (b3){$e^+$};
				\diagram {(a0)--[anti fermion](a1)--[anti charged scalar,edge label'=$\Delta^{++}$](a2)--[anti fermion](a3),
					(b0)--[anti fermion](a1),(a2)--[anti fermion](b3)};
			\end{feynman}
		\end{tikzpicture}}
 \\
 \hline
  \multirow{-3.3}{*}{\rotatebox{90}{\textbf{iType-III}}}      &
        \begin{tikzpicture}
			\begin{feynman}
				\vertex (a1);
				\vertex [left=1cm of a1] (a0){$\mu^+$};
				\vertex [right=1cm of a1] (a2){$\mu^+$};
				\vertex [below=1.5 cm of a1] (b1);
				\vertex [left=1cm of b1] (b0){$\mu^+$};
				\vertex [right=1cm of b1] (b2){$N^+$};
				\diagram {(a0)--[anti fermion](a1)--[anti fermion](a2),
					(b0)--[anti fermion](b1)--[anti fermion](b2),
					(a1)--[boson, edge label'=$Z$](b1)};
			\end{feynman}
		\end{tikzpicture}
		
        &
        
		\begin{tikzpicture}
			\begin{feynman}
				\vertex (a1);
				\vertex [left=1.0cm of a1] (a0){$\mu^+$};
				\vertex [right=1.0cm of a1] (a2){$\mu^+$};
				\vertex [below=1.5 cm of a1] (b1);
				\vertex [left=1.0cm of b1] (b0){$\mu^+$};
				\vertex [right=1.0cm of b1] (b2){$N^+$};
				\diagram {(a0)--[anti fermion](a1)--[anti fermion](b2),
					(b0)--[anti fermion](b1)--[anti fermion](a2),
					(a1)--[boson, edge label'=$Z$](b1)};
			\end{feynman}
		\end{tikzpicture}
        \\
     \hline\hline   
    \end{tabular}
		\caption{Feynman diagrams for dominant  channels distinguishing the seesaw scenarios at $\mu^+\mu^+$ collider.}
    \label{tab:feyn_mupp}
\end{table}

\vspace*{3mm}
\noindent
$\bullet$ \underline{\textbf{Type-II Seesaw :}}

In the Type-II Seesaw scenario at a $\mu^+ \mu^+$ collider, the process $\mu^+\mu^+ \to e^+e^+$ occurs via an s-channel diagram mediated by $\Delta^{++}$, as depicted in {the first row of \autoref{tab:feyn_mupp}}. Unlike the $W^+W^+$ channel, the advantage of this mode is that both vertices are proportional solely to $Y_\Delta$. However, since this process is mediated only through the s-channel, the cross-section decreases significantly if we move away from resonance production.

\vspace*{3mm}
\noindent
$\bullet$ \underline{\textbf{iType-III Seesaw :}}

Now, we isolate the signature of the inverse Type-III Seesaw at a $\mu^+\mu^+$ collider. A distinctive feature of the iType-III Seesaw model is the presence of heavy charged leptons ($N^\pm$) alongside heavy neutral neutrinos ($N^0$). By probing these heavy charged leptons, one can differentiate the iType-III scenario. To this end, the scattering process $\mu^+\mu^+ \to \mu^+ N^+$, which occurs through $Z$-boson mediated t- and u-channel diagrams as shown in second row of \autoref{tab:feyn_mupp}, should be examined.

While the process could, in principle, occur via $h$-mediation, the $\mu^+\mu^-h$ coupling is very small for light leptons, making it negligible. It is important to note that the inverse Type-III Seesaw model includes six copies of heavy charged leptons. Unlike the neutral components $N^0/\widetilde{N}^0$, which couple to $Z$, $h$, and $W^{\pm}$, one of the charged components, $N^{\pm}$, couples exclusively to $Z\ell^\pm$ and $h\ell^\pm$, while the other, $\widetilde{N}^{\pm}$, interacts only with $W^{\pm}\nu$. Consequently, $N^\pm$ decays via the $hl^\pm$ and $Zl^\pm$ modes, each with a 50\% branching ratio, while $\widetilde{N}^\pm$ decays entirely into $W^\pm \nu$. Since the process studied here, is predominantly $Z$-mediated, $\widetilde{N}^\pm$ will not be produced in $\mu^+\mu^+$ collisions. Additionally, only diagonal couplings are considered, so for a given initial state, only one type of heavy charged lepton will be produced. 

\begin{figure*}[hbt]
	\begin{center}
		\hspace*{-0.7cm}
		\mbox{\subfigure[Type-II]{\includegraphics[width=0.40\linewidth,angle=-0]{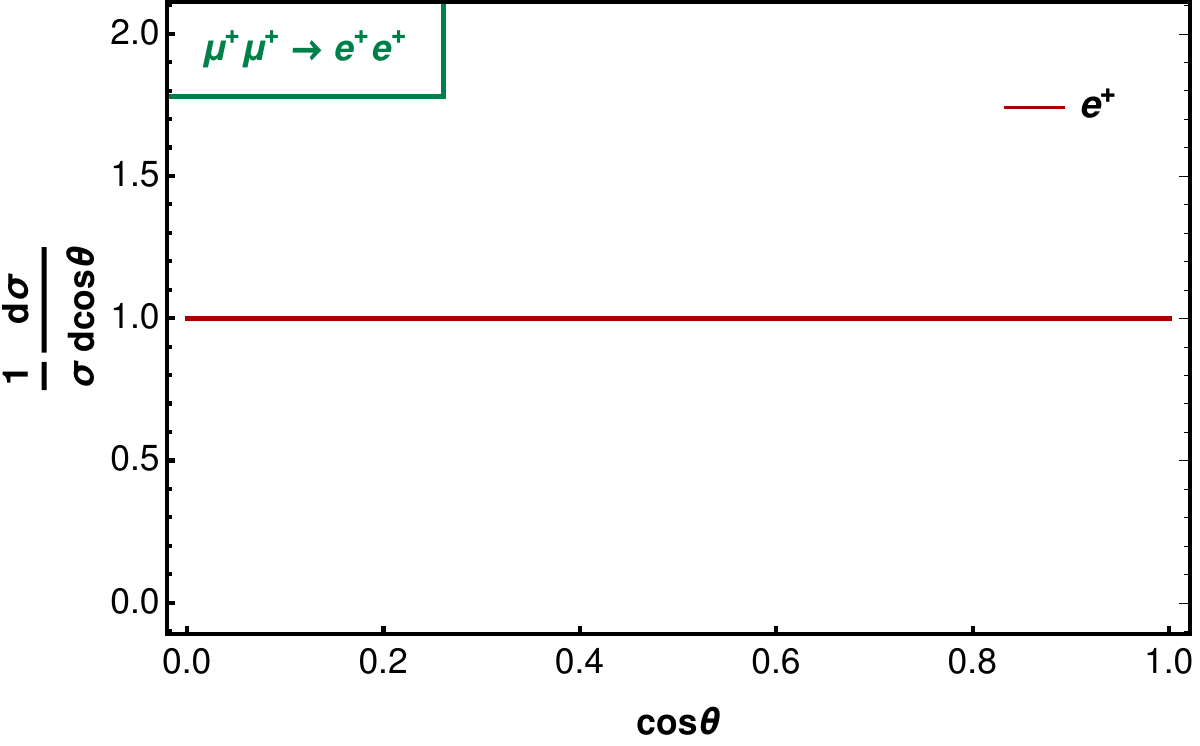}}
			\hspace*{1.0cm}
			\subfigure[iType-III]{\includegraphics[width=0.40\linewidth,angle=-0]{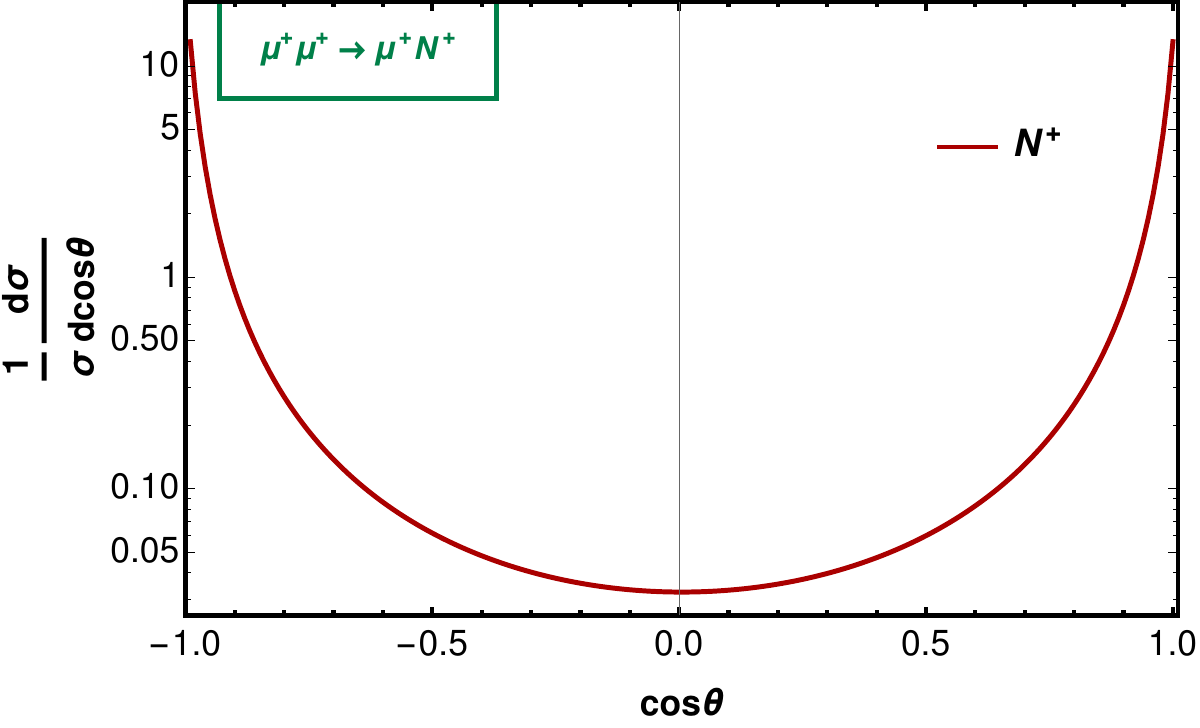}}}		
		\caption{Angular distributions of $e^+$ in Type-II (a) and $N^+$ in iType-III (b) Seesaw models with $Y$ and $Y_\Delta =0.2$, $M_\Delta$ and $M_n =1.25$\,TeV and $\sqrt{s}=1.5$\,TeV in the $\mu^+ \mu^+$ collider.} \label{mu-mu-ang_anal}
	\end{center}
\end{figure*}

The parton-level angular distributions for the final state particles, namely positrons in the Type-II Seesaw scenario and $N^+$ in the iType-III scenario, are illustrated in \autoref{mu-mu-ang_anal} (a) and (b) respectively, within the CM frame at a $\mu^+ \mu^+$ collider with a centre-of-mass energy of $\sqrt{s} = 1.5$ TeV. Notably, the angular distribution for the final state $e^+$ relative to the beam axis in the Type-II Seesaw scenario exhibits a flat profile. This flat distribution is a distinctive signature of the doubly charged scalar $\Delta^{++}$ in the s-channel propagator, which would, in turn, indicate the presence of the Type-II Seesaw mechanism. In contrast, the angular distribution of $N^+$ in the iType-III Seesaw scenario resembles a tub shape, with an excessive number of events in the peripheral region (i.e., $|\cos \theta| \sim 1$) and fewer events in the central region (i.e., $|\cos \theta| \sim 0$). In the following subsection, we reconstruct the BSM particle from its fully visible final state and verify whether the angular distribution of the reconstructed BSM particle matches the parton-level distribution. The cross-sections and the angular distributions for these modes in the centre-of-mass frame of $\mu^+\mu^+$ collider at the leading order are mentioned in \autoref{sec:formula}.

\subsection{Collider simulation} \label{sec:collmummum}

We now turn to the collider simulation of the Type-II and iType-III Seesaw mechanisms at the $\mu^+ \mu^+$ collider. In the Type-II scenario, a doubly charged scalar acts as the mediator, while in the iType-III case, an $SU(2)$ triplet charged fermion is produced. For both models, we assume the BSM particle masses to be $M_n$ and $M_\Delta = 1.25$ TeV and 2.2 TeV for the two benchmark points, with corresponding centre-of-mass energies of 1.5 TeV and 2.5 TeV, respectively. \autoref{crs_mummum} presents the production cross-sections for the processes $\mu^+\mu^+ \to e^+ e^+$ and $\mu^+\mu^+ \to \mu^{+}N^{+}$ for the benchmark points in the centre-of-mass frame.

\begin{table}[h!]
	\renewcommand{\arraystretch}{1.6}
	\centering
	\begin{tabular}{|c|c|c|c|c|}
		\cline{4-5}
		\multicolumn{3}{c|}{}&\multicolumn{2}{c|}{Cross-section (in fb)} \\
		\hline 
		Benchmark& $M_{\Delta}$ or $M_n$ & $E_{CM}$  & Type-II & iType-III \\ 
		Points &in TeV &in TeV &  $\mu^+\mu^+ \to e^+ e^+$ & $\mu^+\mu^+ \to \mu^{+}N^{+}$ \\
		\hline \hline
		BP1	& 1.25 & 1.5 & 7.4 & 8.8  \\ \hline
		BP2	& 2.2 & 2.5 & 4.9 & 2.1 \\ \hline
	\end{tabular}
	\caption{Masses corresponding to different benchmark points, energy of collision in CM frame and the hard scattering cross-sections (in fb) for Type-II and iType-III Seesaw models in $\mu^+ \mu^+$ collider. ($Y_\Delta$ and $Y_N =0.2$, $\mu_\Delta$ and $\mu= 10$ eV)}  \label{crs_mummum}
\end{table}

\begin{figure*}[hbt]
	\begin{center}
		\hspace*{-0.7cm}
		\mbox{\subfigure[Type-II]{\includegraphics[width=0.35\linewidth,angle=-0]{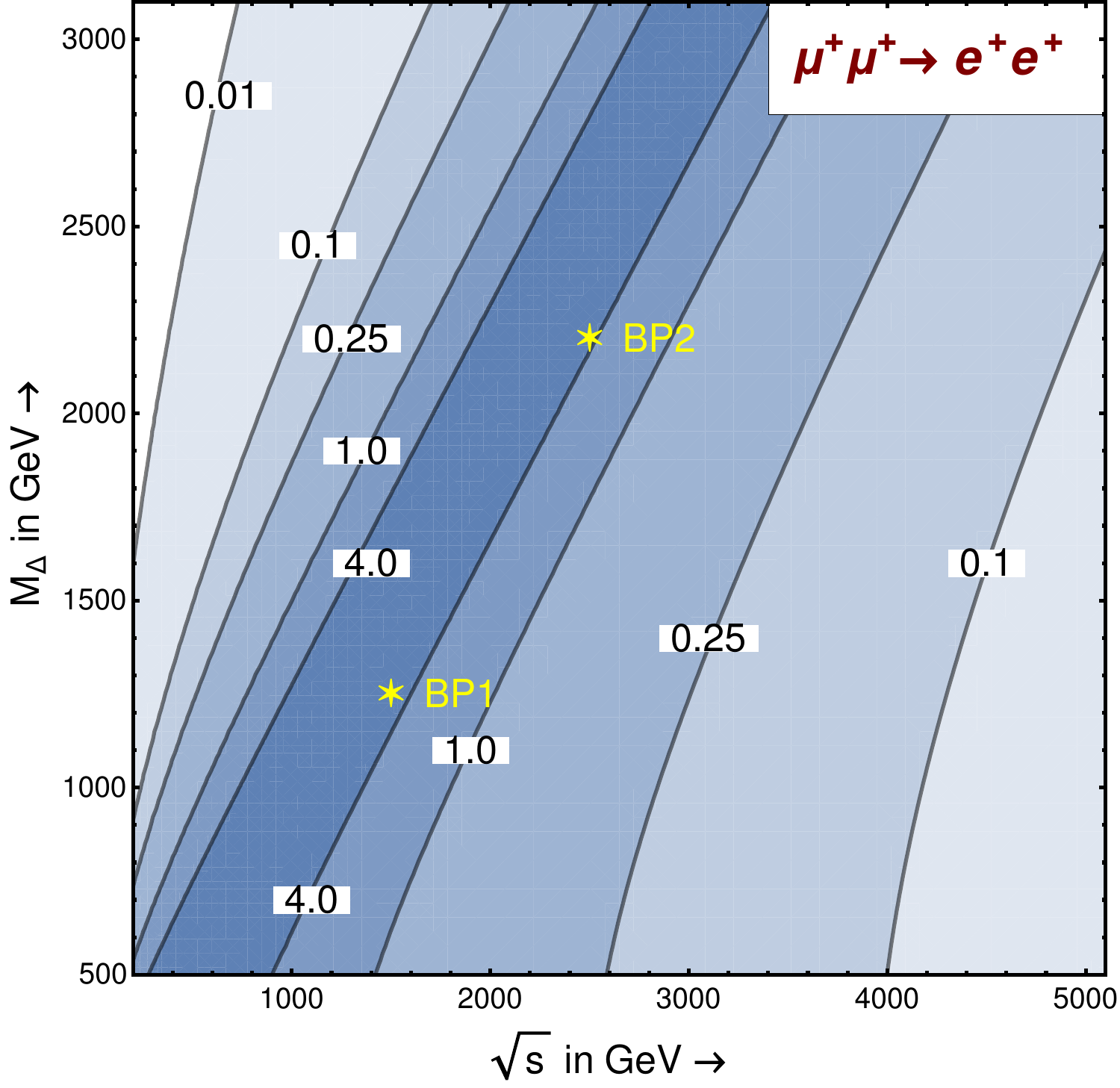}\label{}}
			\hspace*{1.0cm}
			\subfigure[iType-III]{\includegraphics[width=0.35\linewidth,angle=-0]{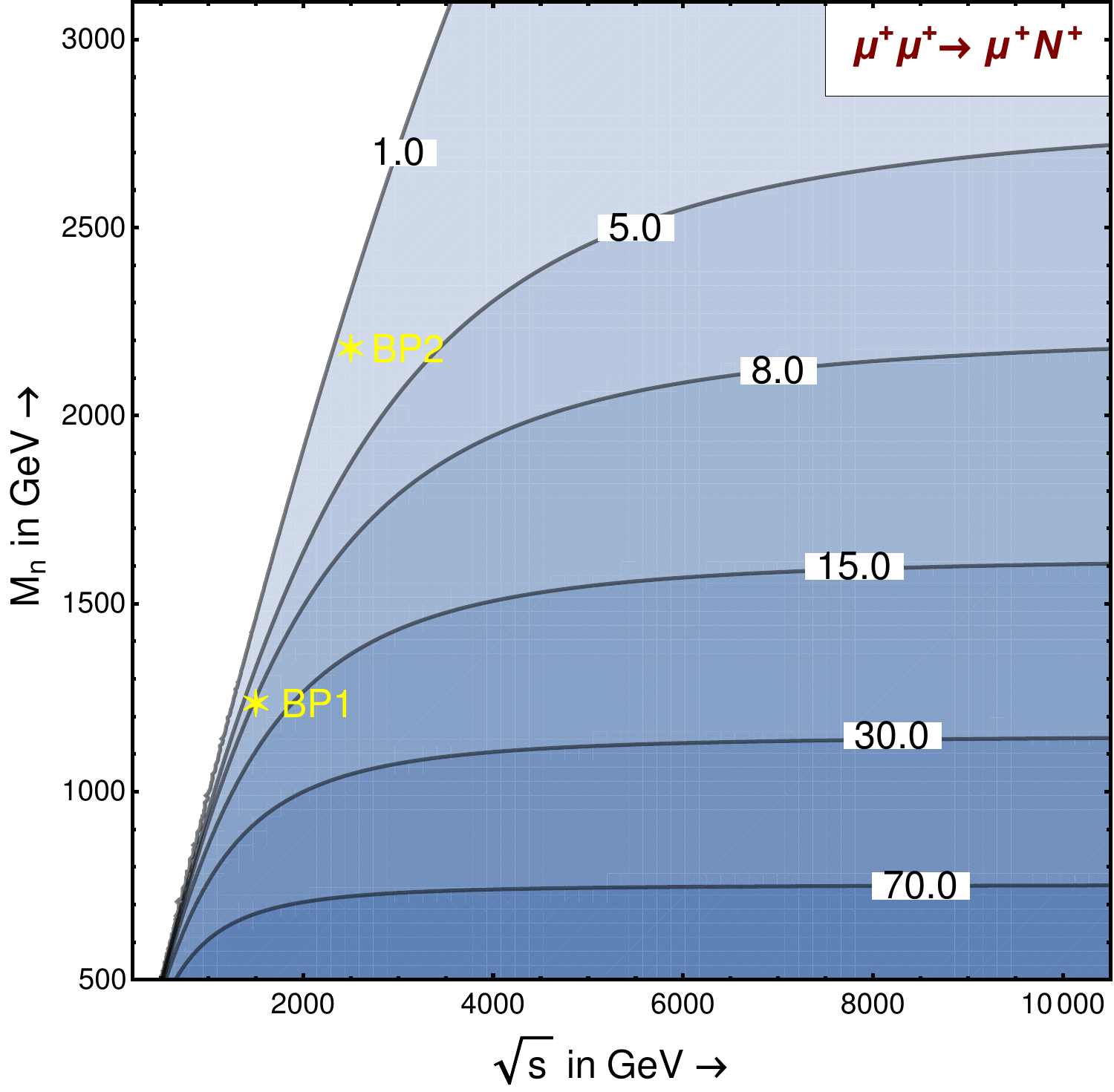}\label{}}}		
		\caption{Contours of the total cross-sections (fb) for $\mu^+ \mu^+$ collider in $M_{\Delta}$ or $M_n$ verses the centre-of-mass energy ($\sqrt{\hat{s}}$) plane for $Y_\Delta$ and $Y_N =0.2$, $\mu_\Delta$ and $\mu=10$ eV eV.}\label{crosec_mu-mu-}
	\end{center}
\end{figure*}

\autoref{crosec_mu-mu-} provides an extensive overview of the cross-section as a function of the BSM particle mass and the centre-of-mass energy, $\sqrt{s}$, illustrated through a contour plot. In \autoref{crosec_mu-mu-}(a), we describe the variation in the total cross-section for the process $\mu^+\mu^+ \to e^+ e^+$. The transition from darker to lighter colours signifies a decrease in the cross-section. Notably, as $M_\Delta$ approaches $\sqrt{s}$, the cross-section significantly increases due to the resonance production of the doubly charged scalar. The two benchmark points selected are near this resonance production, shown by yellow stars. \autoref{crosec_mu-mu-}(b) illustrates a similar analysis for the Type-III scenario, focusing on the process $\mu^+\mu^+ \to \mu^{+}N^{+}$. Here, the cross-section decreases with increasing mass of the triplet fermion, while the centre-of-mass energy remains constant. Conversely, the cross-section initially rises with increasing centre-of-mass energy for a fixed fermion mass, but after a certain point, it remains constant. If $s \leq M_n^2$, the cross-section drops to zero due to insufficient phase space.

\subsection{Final states in reconstructing the resonance particles}

As previously discussed, we are not focusing on the iType-I Seesaw scenario at the $\mu^+ \mu^+$ collider, given that the potential processes involving the heavy neutral lepton have extremely low cross-sections. Instead, as shown in \autoref{crs_mummum}, the processes involving the doubly-charged scalar and singly-charged fermion exhibit cross-sections on the order of 1 fb for TeV-scale BSM particle masses. Therefore, in this section, we aim to study the final state topologies of these processes. Since $\Delta^{++}$ only mediates the process, our focus will be on reconstructing the singly-charged fermion from its decay products. For these analyses, we employ the same simulation cuts mentioned in \autoref{sec:col_sim}.

\subsubsection{Type-II Seesaw at $\mu^+ \mu^+$ Collider} 

Some simple yet intriguing phenomenological aspects emerge from studies of the process $\mu^+ \mu^+ \to e^+ e^+$. The final state involves only two electrons, and given that the muon collider environment is free from initial state QCD radiation, no jets are expected in the transverse direction. This process is notably free from both SM and model-specific backgrounds. The only feasible SM process at a $\mu^+ \mu^+$ collider is M\o ller scattering, which does not involve a flavour change from muons to electrons. Although, in principle, the process $\mu^+ \mu^+ \to \Delta^+ \Delta^+$ could serve as a potential model background, this mode is absent due to the lack of phase space given our choice of benchmark points. Even if the benchmark were adjusted to allow for $\Delta^+$ pair production, the cross-section for this process would still be negligible. {On top of that the reconstruction of $\Delta^{++}$ invariant mass can easily remove all those backgrounds such as coming from muon decaying to electron.  }


\begin{table}[h!]
	\renewcommand{\arraystretch}{1.5}
	\centering
	\begin{tabular}{|c|c|c|}
		\hline
		Final state &  BP1 & BP2  \\ 
		\hline \hline
		$2e^+$ & 7375.1 & 4869.4  \\ 
		\hline	
	\end{tabular}
	\caption{Number of events for $2e^+$ final state for the benchmark points mentioned in \autoref{crs_mummum} \textcolor{black}{with the integrated luminosity of 1000\,fb$^{-1}$.}}  \label{sigType2}
\end{table}

We simulate the channel $\mu^+ \mu^+ \to e^+ e^+$ using PYTHIA8 for the two specified benchmark points with an integrated luminosity of 1000 \fbi, focusing on the same-sign di-electron signature. Since there is no SM or model background for this process, we present only the signal yields for this final state, as shown in \autoref{sigType2}. The results are quite promising, and an early data can probe some BSM signatures via this process.

\subsubsection{iType-III Seesaw at $\mu^+ \mu^+$ Collider} 

To study the angular distribution of an $SU(2)$ triplet singly-charged fermion at the $\mu^+ \mu^+$ collider, we focus on the process $\mu^+ \mu^+ \to \mu^+ N^+$. Although one could also consider the process $\mu^+ \mu^+ \to N^+ N^+$ for detecting the iType-III Seesaw, the cross-section is quite small. This is because the $\mu^+ - N^+$ mixing, which is proportional to the Yukawa coupling $Y$, is required on both the legs of the process. Additionally, the available phase space is more limited in the $N^+ N^+$ mode compared to the $\mu^+ N^+$ mode.

\begin{figure*}[hbt]
	\begin{center}
		\hspace*{-0.5cm}
		\mbox{\subfigure[BP1 (iType-III)]{\includegraphics[width=0.37\linewidth,angle=-0]{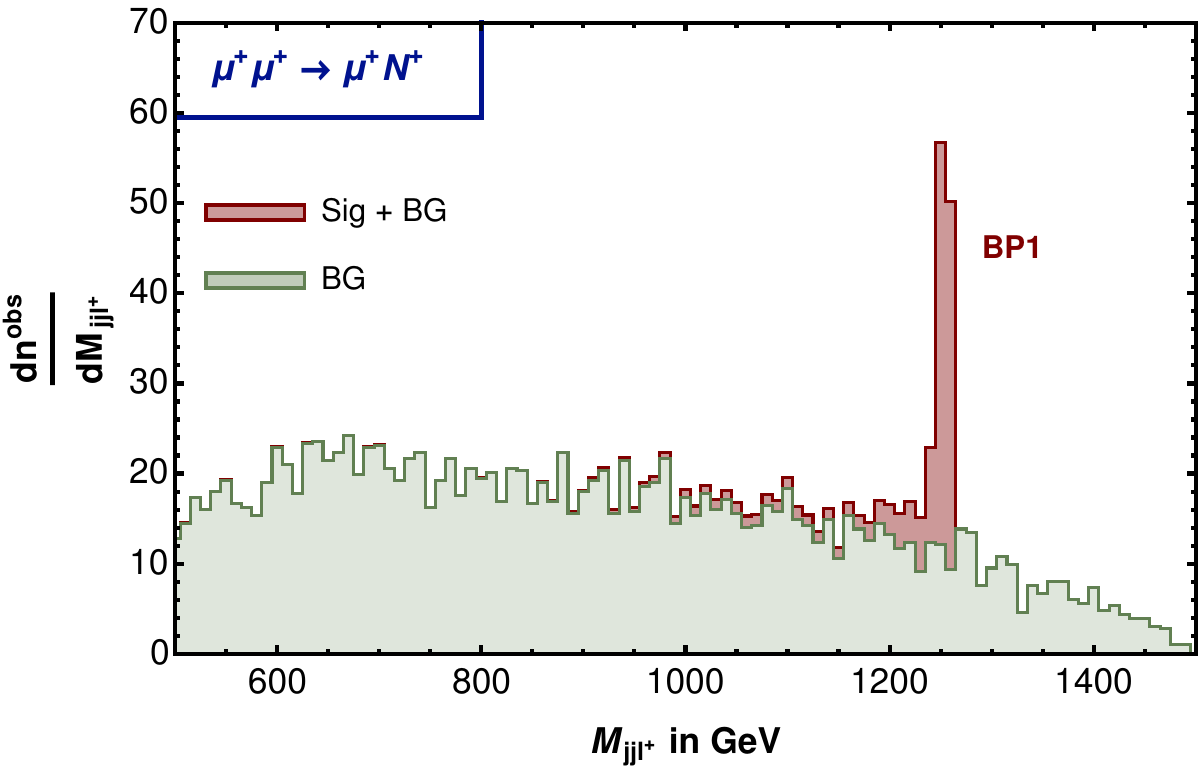}}
			\hspace*{1.0cm}
			\subfigure[BP2 (iType-III)]{\includegraphics[width=0.36\linewidth,angle=-0]{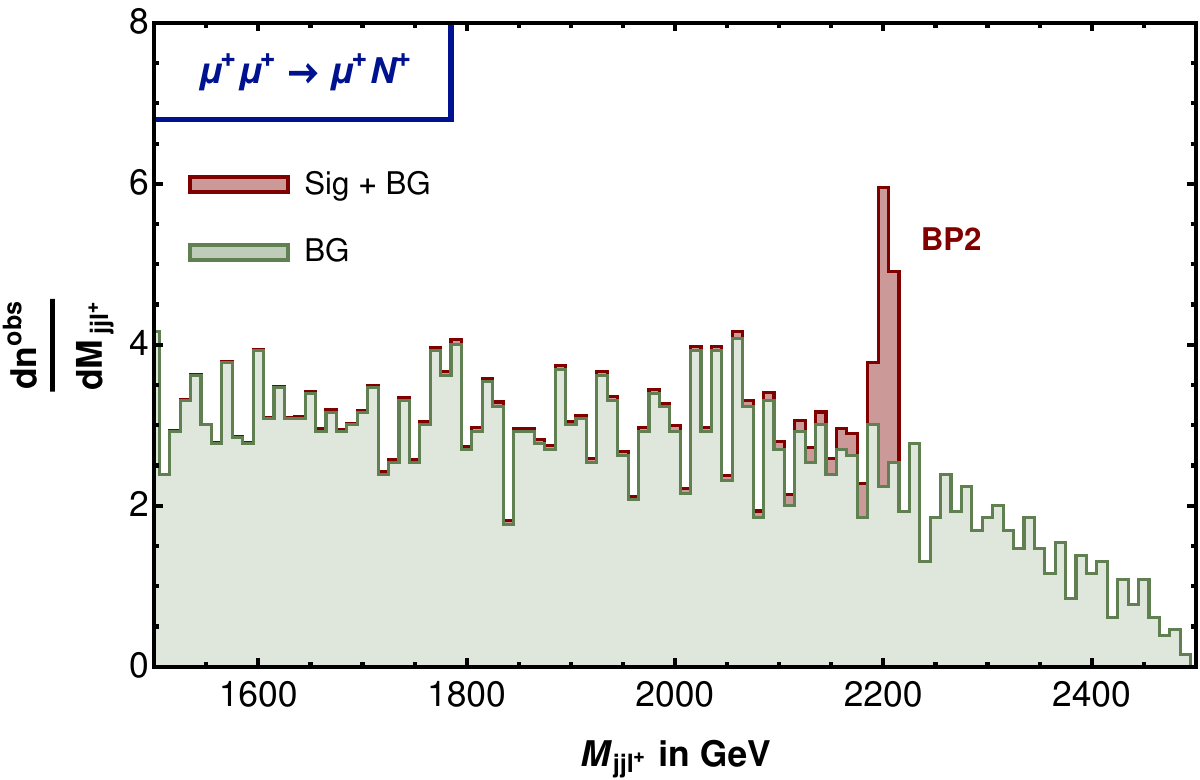}}}		
		\caption{Di-jet-mono-lepton invariant mass distribution ($M_{jjl^+}$) for (a) BP1 and (b) BP2 at the centre-of-mass energies of 1.5, 2.5\, TeV, respectively with the integrated luminosity of 1000\,fb$^{-1}$.  The total (signal $+$ SM background) signature is depicted in brown and the SM background is in olive green.}\label{mu-mu-Ty3invM}
	\end{center}
\end{figure*}

The heavy charged leptons ($N^+$) will eventually decay to either $Z\ell^+$ or $h\ell^+$. Therefore, observing any peak in the invariant mass distribution of $jj \ell^+$ or $bb \ell^+$ around the heavy lepton mass, through the reconstruction of $Z/h$ peaks, would confirm the existence of heavy charged leptons, thus validating the iType-III Seesaw. The dominant SM background in this scenario is expected to come from the process $\mu^+ \mu^+ \to \mu^+ \mu^+ Z(jj)$. However, by constructing the invariant mass of the $jj\ell$ system, the SM background can be significantly reduced. In \autoref{mu-mu-Ty3invM}, we plot the invariant mass of $jj \ell^+$ for the two benchmark points, where the olive green regions represent the SM background contribution, and the red regions indicate signal plus background. It is evident that the red histogram exhibits a sharp peak around $M_n$, while the olive green curves form a continuum. 


\begin{table*}[hbt]
	\renewcommand{\arraystretch}{1.5}
	\centering
	\begin{tabular}{|c|c|c|c|c|}
		\hline 
		Final states&  \multicolumn{2}{c|}{BP1} &\multicolumn{2}{c|}{BP2} \\ 
		\cline{2-5}
		& Sig & BG ($\mu^+ \mu^+ Z$) & Sig & BG ($\mu^+ \mu^+ Z$)\\ 
		\hline \hline
		$2\ell^+ + 2j$ & 2459.21 & $1.48 \times 10^5 $ & 212.80 & $1.13 \times 10^5$ \\ 
		\cline{2-5}
		$+|M_{jj\ell^+} -M_n|\leq 10\,$GeV & 815.02 & 194.78 & 56.19 & 125.78 \\ 
		\hline	\hline	
		$\rm{S}_{\text{sig}} (\mathcal{L}_{\rm int}=1000\, \rm fb^{-1})$& \multicolumn{2}{c|}{25.65 }  & \multicolumn{2}{c|}{4.16} \\
		\hline
		$\int \mathcal{L}_{5\sigma}\,[\rm{fb}^{-1}]$& \multicolumn{2}{c|}{38.00}  & \multicolumn{2}{c|}{1440.86}\\  
		\hline
	\end{tabular}
	\caption{Number of events for signal and background corresponding to $2\ell^+ + 2j$ final state for the benchmark points mentioned in \autoref{crs_mummum}  \textcolor{black}{with the integrated luminosity of 1000\,fb$^{-1}$.}}  \label{sigbg_mummum_Ty3}
\end{table*}

Since we are interested in the process $\mu^+ \mu^+ \to \mu^+ N^+$, it is convenient to search for the same-sign di-lepton plus di-jet signature as a promising fully visible final state to detect the iType-III scenario at a $\mu^+ \mu^+$ collider. The signal-background analysis for this final state, with an integrated luminosity of 1000 \fbi, is presented in \autoref{sigbg_mummum_Ty3}. As observed, a significant SM background arises from the $\mu^+ \mu^+ Z$ mode for the same-sign di-lepton plus di-jet signature. However, by applying an invariant mass cut of $|M_{jj\ell} - M_n| \leq 10$ GeV, the background drops off rapidly. This allows for a signal significance of more than $25\sigma$ for BP1 and $4\sigma$ for BP2 with 1000 \fbi of integrated luminosity. In other words, very early data of around 40 \fbi could provide a $5\sigma$ significance for the first BP, while approximately 1200 \fbi of luminosity would be required to achieve discovery prospects for BP2.

\subsection{Angular distribution at $\mu^+ \mu^+$ collider}

\begin{figure*}[hbt]
	\begin{center}
		\hspace*{-0.7cm}
		\mbox{\subfigure[{Type-II}]{\includegraphics[width=0.4\linewidth,angle=-0]{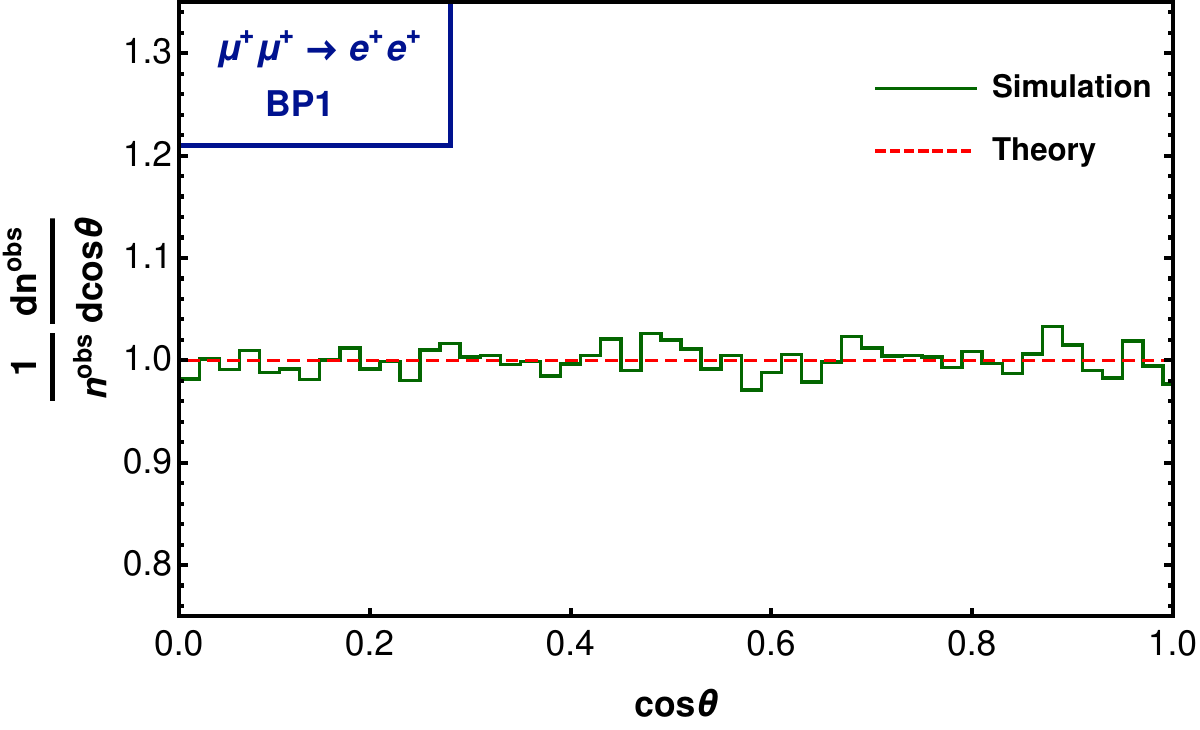}}
			\hspace*{1.0cm}			
			\subfigure[{iType-III}]{\includegraphics[width=0.4\linewidth,angle=-0]{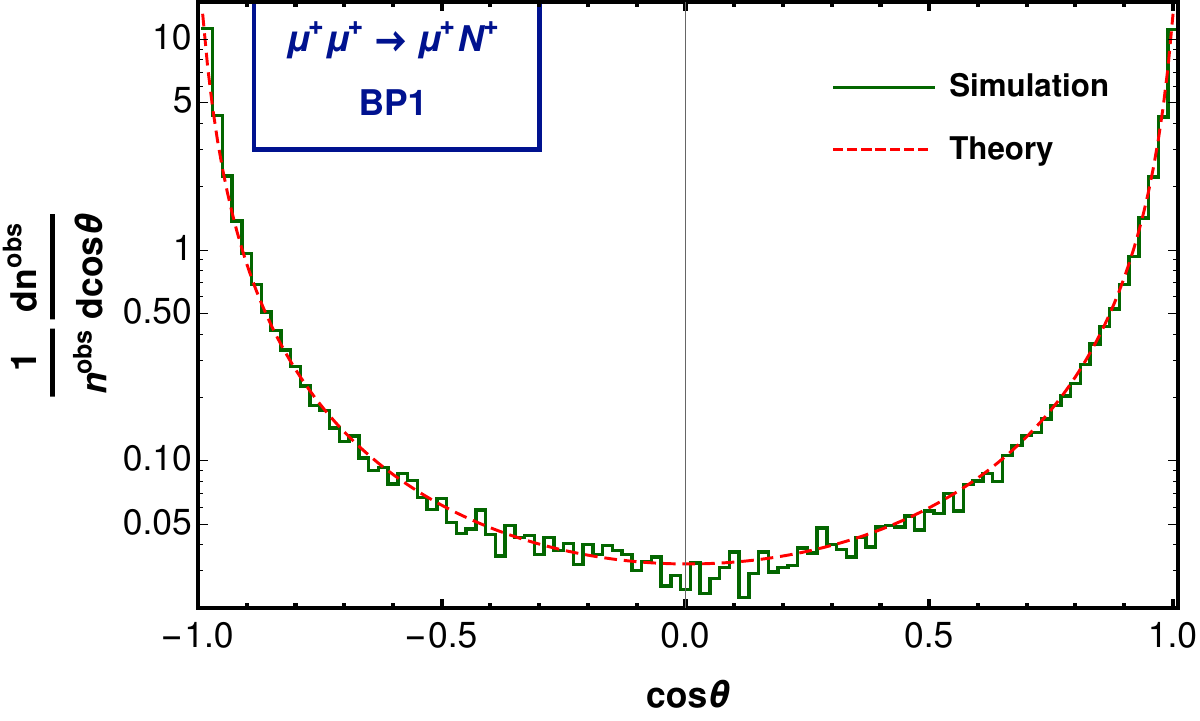}}}				
		\caption{Angular distribution of the theory and simulation  at $\mu^+ \mu^+$ collider for the processes corresponding to Type-II (a) and iType-III (b) Seesaw scenarios for BP1.}\label{angdismummumF}
	\end{center}
\end{figure*}

The normalized angular distribution for the final state $e^+$ is presented in \autoref{angdismummumF}(a) for the process $\mu^+ \mu^+ \to e^+ e^+$, mediated by a doubly-charged scalar ($\Delta^{++}$), a component of the $SU(2)$ triplet scalar field involved in the Type-II Seesaw mechanism. In this plot, the green histogram represents the simulated results, while the red dashed line denotes the theoretical estimation. The simulated results align well with the theoretical prediction, showing a relatively flat angular distribution for the final state $e^+$.

Following this, we present the angular distribution of the reconstructed particle from the $M_{jj \ell^+}$ mass (or equivalently, the muon produced in association with the heavy charged lepton) for BP1 in \autoref{angdismummumF}(b). The color coding follows the same convention, with the theory line and the simulated histogram both in agreement. In contrast to the flat angular distribution of the final state $e^+$ in the Type-II scenario, this distribution resembles a tub shape, which is consistent with the angular distribution expected for a fermionic particle. Since, at a $\mu^+ \mu^+$ collider, only one type of charged heavy fermion field can be produced, the corresponding distribution differs from that of the iType-III scenario at a $\mu^+ \mu^-$ collider, as shown in \autoref{angdismupmumF}(c). {Concerning Type-II seesaw, one can also investigate the angular distribution of $\Delta^{++}$ in the channel $\mu^+\mu^+\to \Delta^{++}\gamma$ showing RAZ at $\cos\theta=0$. However, we do not study this channel since the corresponding angular distribution mimics the same of the $\mu^+\mu^+\to \mu^+ N^+$ in iType-III seesaw.}

\begin{table}[h!]
	\renewcommand{\arraystretch}{1.7}
	\centering
	\begin{tabular}{|c|c|c|c|c|}
		\cline{2-5}
		\multicolumn{1}{c|}{} & \multicolumn{4}{c|}{$\cos \theta$ for iType-III}  \\
		\cline{2-5}
		\multicolumn{1}{c|}{} & ($-0.7 \to -0.35$) & ($-0.35 \to 0.0$) & ($0.0 \to 0.35$)  & ($ 0.35 \to 0.7$) \\
		\hline 
		Signal ($N^+$) & 258.6 & 123.7 & 263.9 & 126.2 \\
		\hline 
		Background ($\mu^+ \mu^+ Z$) & 64.6 & 48.2 & 47.4 & 50.0  \\
		\hline 
	\end{tabular}
	\caption{Number of events in the $\cos\theta$ intervals $(-0.7,0.35)$, $(-0.35,0)$, $(0,0.35)$ and $(0.35,0.7)$ for the BP1 of iType-III seesaw scenario, corresponding to the final state discussed earlier, at an integrated luminosity of 1000~\fbi.} \label{tab:assymtry2}
\end{table}
{In \autoref{tab:assymtry2} we describe the asymmetric number of events that can be observed for two different windows of $\cos{\theta}$ i.e. $\cos{\theta}: (0.00, \,0.35), (0.35,\,0.70$) (for both positive and negative values),  for the final state mentioned in \autoref{sigbg_mummum_Ty3} for BP1 in iType-III. It is evident the symmetry of signal events in both the windows as can be seen from the angular distributions in \autoref{angdismummumF}(b). The asymmetry corresponds the signal events ratio of $2:1$, whereas for backgrounds it is $1.3:1$ for the windows $(-0.70, \, -0.35)$ and $(-0.35, \, 0.00)$, respectively, giving rise to a signal significance of $11 \sigma$. The situation is similar for the positive window with a signal significance for the asymmetry with $12\sigma$. In comparison for Type-II, we have a flat distribution that only exists for positive $\cos{\theta}$ values, as can be seen in \autoref{angdismummumF}(a). The corresponding signal numbers are already summarized in \autoref{sigType2}. If we construct an asymmetry with the number of events in the window for $\cos\theta$ being (0.00, \,0.35), (0.35,\,0.70), it would be vanishing for Type-II.}

\section{At $\mu^+ \gamma$ collider}
\label{sec:mum_gamma}

The $\mu^+ \gamma$ collider (referred as an $e \gamma$ collider in \cite{Velasco:2001fsi}) offers a promising platform for conducting highly precise measurements of SM parameters. Compared to hadron colliders, the interactions in this collider are notably cleaner, allowing for more accurate exploration of particle dynamics. This machine presents unique initial states, enhancing the discovery potential through rare couplings or modes that may be overlooked in hadron colliders. The $\mu^+ \gamma$ collider serves as a complementary tool to the research conducted at hadron colliders like the LHC and lepton-lepton colliders, offering distinct interaction dynamics that open up additional channels for studying processes difficult to probe elsewhere. In this context, we investigate the potential of distinguishing different Seesaw mechanisms by analysing the angular distributions of BSM particles. We first discuss the parton-level processes essential for this analysis, followed by simulations to reconstruct the distribution.

\renewcommand{\arraystretch}{1.5}
\begin{table}[h!]
    \centering
    \begin{tabular}{|c||ccc|}
    \hline\hline
    \multicolumn{4}{|c|}{\textbf{At $\bm{\mu^+\gamma}$ collider}}\\
    \hline\hline
    \multirow{-3.5}{*}{\rotatebox{90}{\textbf{iType-I/III}}} &
\multicolumn{3}{c|}{
\begin{tikzpicture}
			\begin{feynman}
				\vertex (a1);
				\vertex [above left=1cm of a1] (a0){$\mu^+$};
				\vertex [right=1.5cm of a1] (a2);
				\vertex [above right=1 cm of a2] (a3){$W_\rho^+$};
				\vertex [below left=1cm of a1] (b0){$\gamma$};
				\vertex [below right=1cm of a2] (b3){$N^0/\widetilde N^0$};
				\diagram {(a0)--[anti fermion](a1)--[anti fermion,edge label'=$\mu^+$](a2)--[ boson](a3),
					(b0)--[boson](a1),(a2)--[anti fermion](b3)};
			\end{feynman}
		\end{tikzpicture}
        \hfil
		\begin{tikzpicture}
			\begin{feynman}
				\vertex (a1);
				\vertex [left=1cm of a1] (a0){$\mu^+$};
				\vertex [right=1cm of a1] (a2){$N^0/\widetilde N^0$};
				\vertex [below=2 cm of a1] (b1);
				\vertex [left=1cm of b1] (b0){$\gamma$};
				\vertex [right=1cm of b1] (b2){$W^+_\rho$};
				\diagram {(a0)--[anti fermion](a1)--[anti fermion](a2),
					(b0)--[boson](b1)--[boson](b2),
					(a1)--[boson, edge label'=$W$](b1)};
			\end{feynman}
		\end{tikzpicture}}
        \\
        
        \hline 
\multirow{-3.5}{*}{\rotatebox{90}{\textbf{Type-II}}} &

        \begin{tikzpicture}
			\begin{feynman}
				\vertex (a1);
				\vertex [above left=1cm of a1] (a0){$\mu^+$};
				\vertex [right=1.5cm of a1] (a2);
				\vertex [above right=1 cm of a2] (a3){$\Delta^{++}$};
				\vertex [below left=1cm of a1] (b0){$\gamma$};
				\vertex [below right=1cm of a2] (b3){$\mu^-$};
				\diagram {(a0)--[anti fermion](a1)--[anti fermion, edge label'=$\mu^+$](a2)--[anti charged scalar](a3),
					(b0)--[boson](a1),(a2)--[fermion](b3)};
			\end{feynman}
		\end{tikzpicture}

        &
        
		\begin{tikzpicture}
			\begin{feynman}
				\vertex (a1);
				\vertex [left=1cm of a1] (a0){$\mu^+$};
				\vertex [right=1cm of a1] (a2){$\Delta^{++}$};
				\vertex [below=2 cm of a1] (b1);
				\vertex [left=1cm of b1] (b0){$\gamma$};
				\vertex [right=1cm of b1] (b2){$\mu^-$};
				\diagram {(a0)--[anti fermion](a1)--[charged scalar](a2),
					(b0)--[boson](b1)--[fermion](b2),
					(a1)--[fermion, edge label'=$\mu^-\,$](b1)};
			\end{feynman}
		\end{tikzpicture}
        
         &
         
		\begin{tikzpicture}
			\begin{feynman}
				\vertex (a1);
				\vertex [left=1cm of a1] (a0){$\mu^+$};
				\vertex [right=1cm of a1] (a2){$\mu^-$};
				\vertex [below=2 cm of a1] (b1);
				\vertex [left=1cm of b1] (b0){$\gamma$};
				\vertex [right=1cm of b1] (b2){$\Delta^{++}$};
				\diagram {(a0)--[anti fermion](a1)--[fermion](a2),
					(b0)--[boson](b1)--[charged scalar](b2),
					(a1)--[charged scalar, edge label'=$\Delta^{++}\,$](b1)};
			\end{feynman}
		\end{tikzpicture}
        \\
        \hline
        
        \multirow{-3.5}{*}{\rotatebox{90}{\textbf{iType-III}}} &
        \multicolumn{3}{c|}{
        	\begin{tikzpicture}
			\begin{feynman}
				\vertex (a1);
				\vertex [above left=1cm of a1] (a0){$\mu^+$};
				\vertex [right=1.5cm of a1] (a2);
				\vertex [above right=1 cm of a2] (a3){$Z_\rho^0/h$};
				\vertex [below left=1cm of a1] (b0){$\gamma$};
				\vertex [below right=1cm of a2] (b3){$N^+$};
				\diagram {(a0)--[anti fermion](a1)--[anti fermion,edge label'=$\mu^+$](a2)--[ boson](a3),
					(b0)--[boson](a1),(a2)--[anti fermion](b3)};
			\end{feynman}
		\end{tikzpicture}
		\hfil
		\begin{tikzpicture}
			\begin{feynman}
				\vertex (a1);
				\vertex [left=1cm of a1] (a0){$\mu^+$};
				\vertex [right=1cm of a1] (a2){$Z^0_\rho/h$};
				\vertex [below=2 cm of a1] (b1);
				\vertex [left=1cm of b1] (b0){$\gamma$};
				\vertex [right=1cm of b1] (b2){$N^+$};
				\diagram {(a0)--[anti fermion](a1)--[boson](a2),
					(b0)--[boson](b1)--[anti fermion](b2),
					(a1)--[anti fermion, edge label'=$N^+$](b1)};
			\end{feynman}
		\end{tikzpicture}}
        \\
     \hline\hline   
    \end{tabular}
		\caption{Feynman diagrams for dominant  channels distinguishing the seesaw scenarios at $\mu^+\gamma$ collider.}
    \label{tab:feyn_mugam}
\end{table}

\vspace*{3mm}
\noindent
$\bullet$ \underline{\textbf{iType-I seesaw :}}

To explore the iType-I Seesaw mechanism at a $\mu^+ \gamma$ collider, we examine the process $\mu^+\gamma\to W^+_\rho N^0/\widetilde N^0$, where a heavy neutral lepton is produced with a $W^+$ boson. This process primarily occurs through an s-channel and a $W$ boson mediated t-channel diagram, as illustrated in the first row of \autoref{tab:feyn_mugam}.

\vspace*{3mm}
\noindent
$\bullet$ \underline{\textbf{Type-II seesaw :}}

Similarly, for a Type-II Seesaw scenario, the production of $\Delta^{++}$ can be studied through the process $\mu^+\gamma\to\mu^- \Delta^{++}$ at a $\mu^+ \gamma$ collider. The corresponding Feynman diagrams are presented in the second row of \autoref{tab:feyn_mugam}. This process proceeds via an s-channel diagram mediated by $\mu^+$ and t-channel diagrams mediated by $\mu^-$ and $\Delta^{++}$. The interference between these diagrams can influence the angular distribution.

\vspace*{3mm}
\noindent
$\bullet$ \underline{\textbf{iType-III seesaw :}}	

Finally, we consider the iType-III Seesaw scenario, focusing on the production of $N^+\,Z/h$ in a $\mu^+ \gamma$ collider. This production mode involves an s-channel diagram via $\mu^+$ and a t-channel diagram via the charged heavy fermion, as depicted in the third row of \autoref{tab:feyn_mugam}. 


\begin{figure*}[hbt]
	\begin{center}
		\hspace*{-0.7cm}
		\mbox{\subfigure[iType-I/III]{\includegraphics[width=0.35\linewidth,angle=-0]{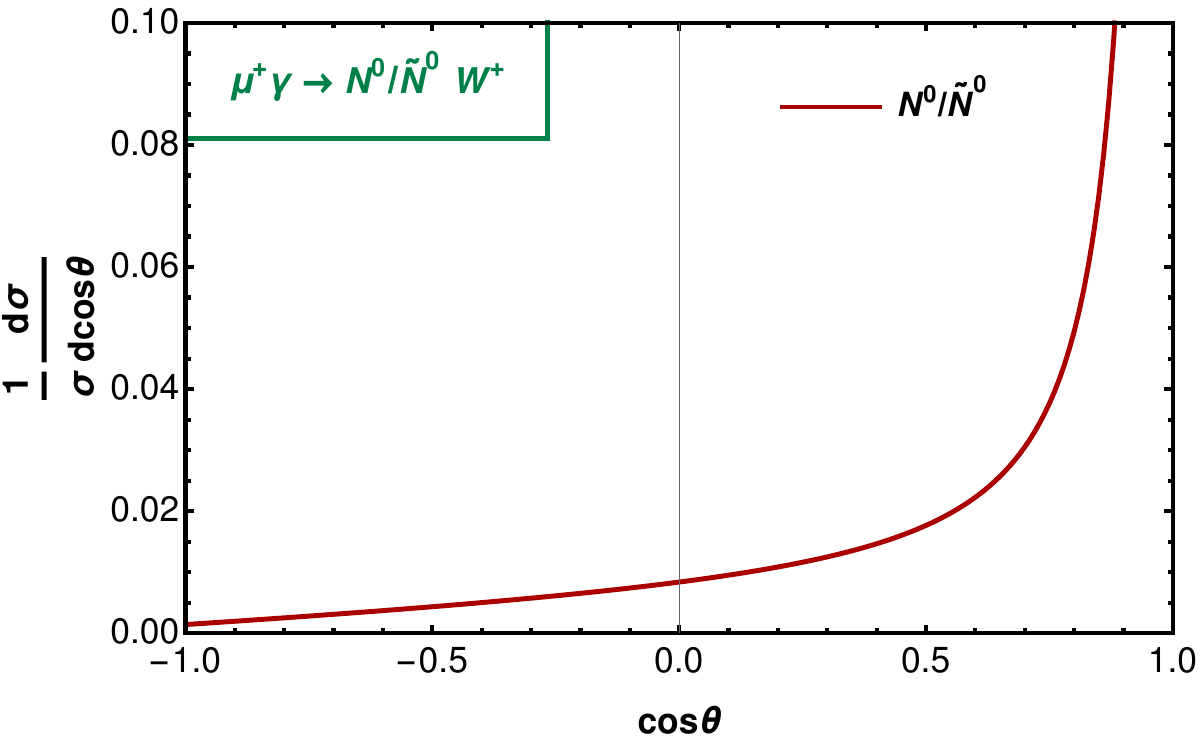}\label{}}
			\subfigure[Type-II]{\includegraphics[width=0.35\linewidth,angle=-0]{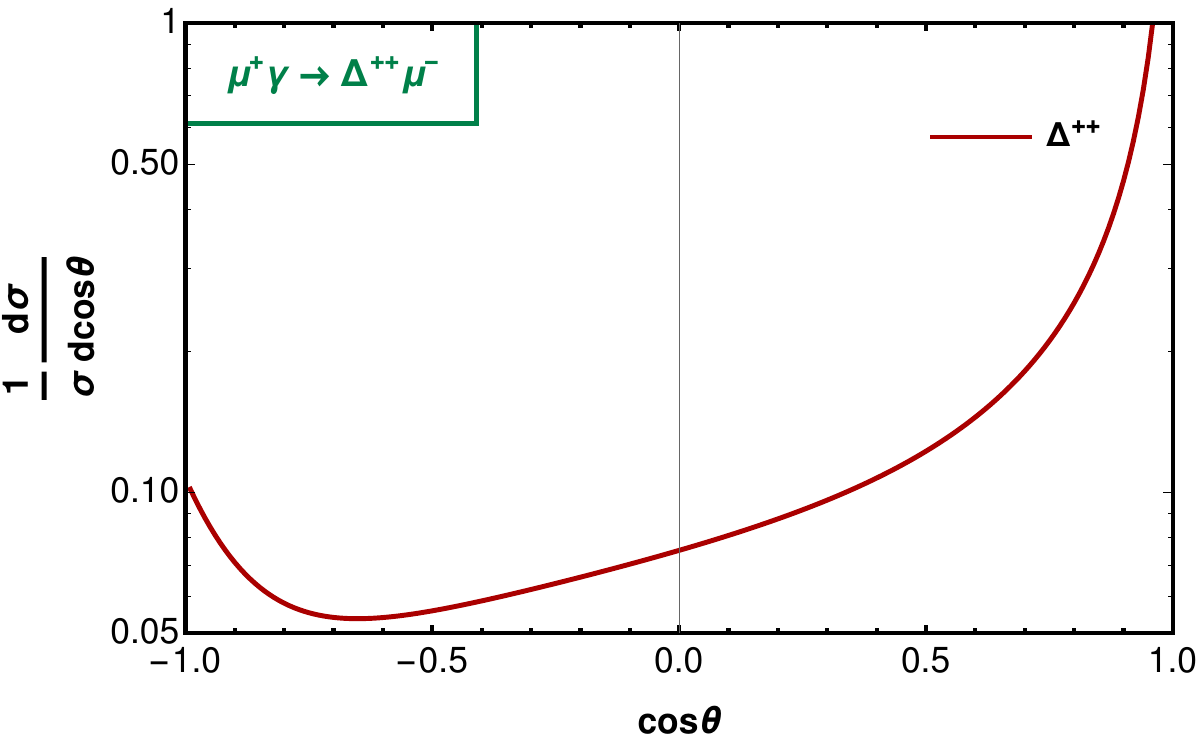}\label{}}
			\subfigure[iType-III]{\includegraphics[width=0.35\linewidth,angle=-0]{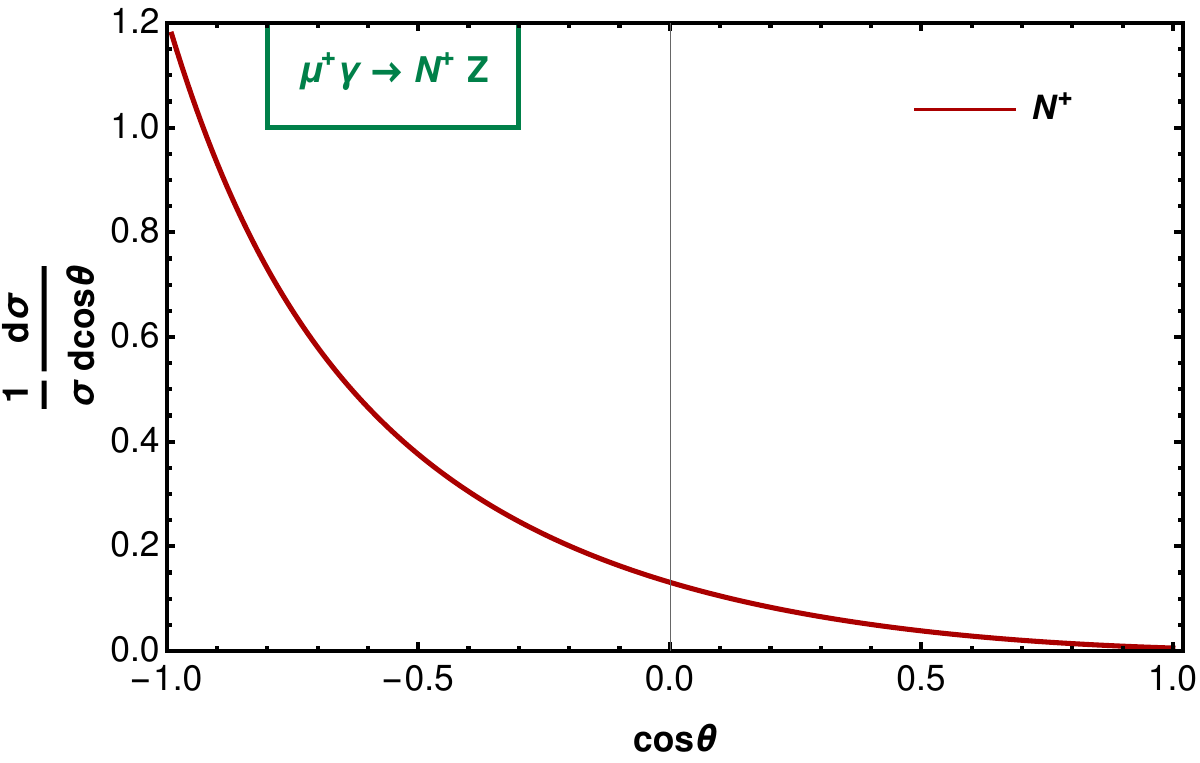}\label{}}}		
		\caption{Angular distributions for the BSM particles in iType-I (a), Type-II (b) and iType-III (c) Seesaw with $Y$ and $Y_\Delta =0.2$, $M_n$ and $M_\Delta=1.25$\,TeV and $\sqrt{s}=3.16$\,TeV in the $\mu^+\gamma$ collider.}\label{mu-photon_ang_anal}
	\end{center}
\end{figure*}

In \autoref{mu-photon_ang_anal}, we analyze the angular distributions for three different Seesaw mechanisms as discussed previously. \autoref{mu-photon_ang_anal} (a) shows the normalized angular distribution of $N^0/\widetilde{N}^0$ with respect to $\cos{\theta}$ for the iType-I case described in \autoref{ty1mg}. It is clear that the distribution drops rapidly until $\cos{\theta} = 0.7$, after which it follows a smooth pattern and approaches zero at $\cos{\theta} = -1$. \autoref{mu-photon_ang_anal} (b) presents the normalized angular distribution corresponding to $\Delta^{++}$ in the Type-II Seesaw scenario. The shallow minimum around $\cos{\theta} = -0.7$ is evident in this plot. In \autoref{mu-photon_ang_anal} (c), we illustrate the distribution for $N^+$ in the iType-III Seesaw. Unlike the iType-I case, here the distribution decreases monotonically as $\cos{\theta}$ moves from -1 to +1. However, the curvature differs between the two cases. The cross-sections and angular distributions for all relevant processes are presented in \autoref{sec:formula}.

\subsection{Collider simulation} \label{sec:collmum_gamma}

In this section, we discuss both the possibilities and challenges associated with reconstructing on-shell SM/BSM particles in the final states, which are essential for examining parton-level angular distributions. Additionally, we address the current limitations in producing monochromatic photon beams with today's technology, necessitating the use of the Laser Backscattering (LBS) technique \cite{LBS1,Telnov:2000ep, Telnov:2000zx, Telnov:1999tb, Telnov:2016lzw, Ginzburg:1981ik, Ginzburg:1981vm, Brinkmann:1997sj} to generate photons. This approach comes with certain trade-offs: first, the resulting photon beam lacks a narrow spectral width and high coherence, and second, the stability of the wavelength is compromised. Despite these challenges, LBS photons are produced through inverse Compton scattering, where low-energy laser photons interact with high-energy electrons, significantly boosting the photon energy, which is an important factor for high-energy collider experiments. Due to the numerous underlying factors in this scenario, it is not possible to accurately reconstruct the parton-level distributions as in other colliders. Nonetheless, the $\mu^+\gamma$ collider remains valuable, as it can probe certain modes that are inaccessible in other colliders.

\begin{table}[h!]
	\renewcommand{\arraystretch}{1.6}
	\centering
	\scalebox{0.9}{\begin{tabular}{|c|c|c|c|c|c|c|c|}
			\cline{6-8}
			\multicolumn{5}{c|}{}&\multicolumn{3}{c|}{Cross-section (in fb) for LBS (monochromatic) photon} \\
			\hline 
			Benchmark& $M_n$ or, $M_{\Delta}$ & $E_\gamma$ & $E_{\mu^+}$ & $E_{CM}$  & Type-I  & Type-II & Type-III \\ 
			Points &in TeV &in TeV &in TeV &in TeV  & $\mu^+ \gamma \to W^+ N^0/\widetilde{N}^0$ & $\mu^+ \gamma \to \Delta^{++} \mu^-$ & $\mu^+ \gamma \to N^+ Z/h$ \\
			\hline \hline
			BP1	& 1.25 &  0.5 & 5.0 & 3.16  & 23.30 (32.89) & 32.56 (23.08) & 1.43 (2.30)  \\ \hline
			BP2	& 1.80 & 0.5 & 10.0 & 4.47  & 11.56 (16.04) & 16.42 (11.77) & 0.67 (1.11) \\ \hline
	\end{tabular}}
	\caption{Masses corresponding to different benchmark points, energy of collision in CM frame and the hard scattering cross-sections (in fb) with the laser backscattering (monochromatic) photon for iType-I, Type-II and iType-III Seesaw models in $\mu^+ \gamma$ collider. ($Y_\Delta$ and $Y_N =0.2$, $\mu_\Delta$ and  $\mu=10$ eV)}  \label{crs_mum_gamma}
\end{table}

In \autoref{crs_mum_gamma}, we select two benchmark points for the collider study, featuring BSM particle masses of 1.25 TeV and 1.80 TeV, which align with all theoretical and experimental constraints \cite{CMS:2017pet, CMS:2019lwf, ATLAS:2017xqs, ATLAS:2020wop}. The table presents the cross-sections corresponding to centre-of-mass energies of 3.16 TeV (BP1) and 4.47 TeV (BP2). The values outside the brackets reflect the cross-sections calculated using an LBS photon beam, while the bracketed values correspond to those obtained with a monochromatic photon beam. Although the associated production cross-section of the $N^+Z/h$ mode in the iType-III Seesaw scenario for TeV-scale $SU(2)$ triplet fermion mass is relatively modest, an integrated luminosity of 1000 \fbi still yields a sufficient number of events for our selected final state, as detailed in \autoref{Tab:FSmuphTy3}.

\begin{figure*}[hbt]
	\begin{center}
		\hspace*{-0.7cm}
		\mbox{\subfigure[iType-I/III]{\includegraphics[width=0.35\linewidth,angle=-0]{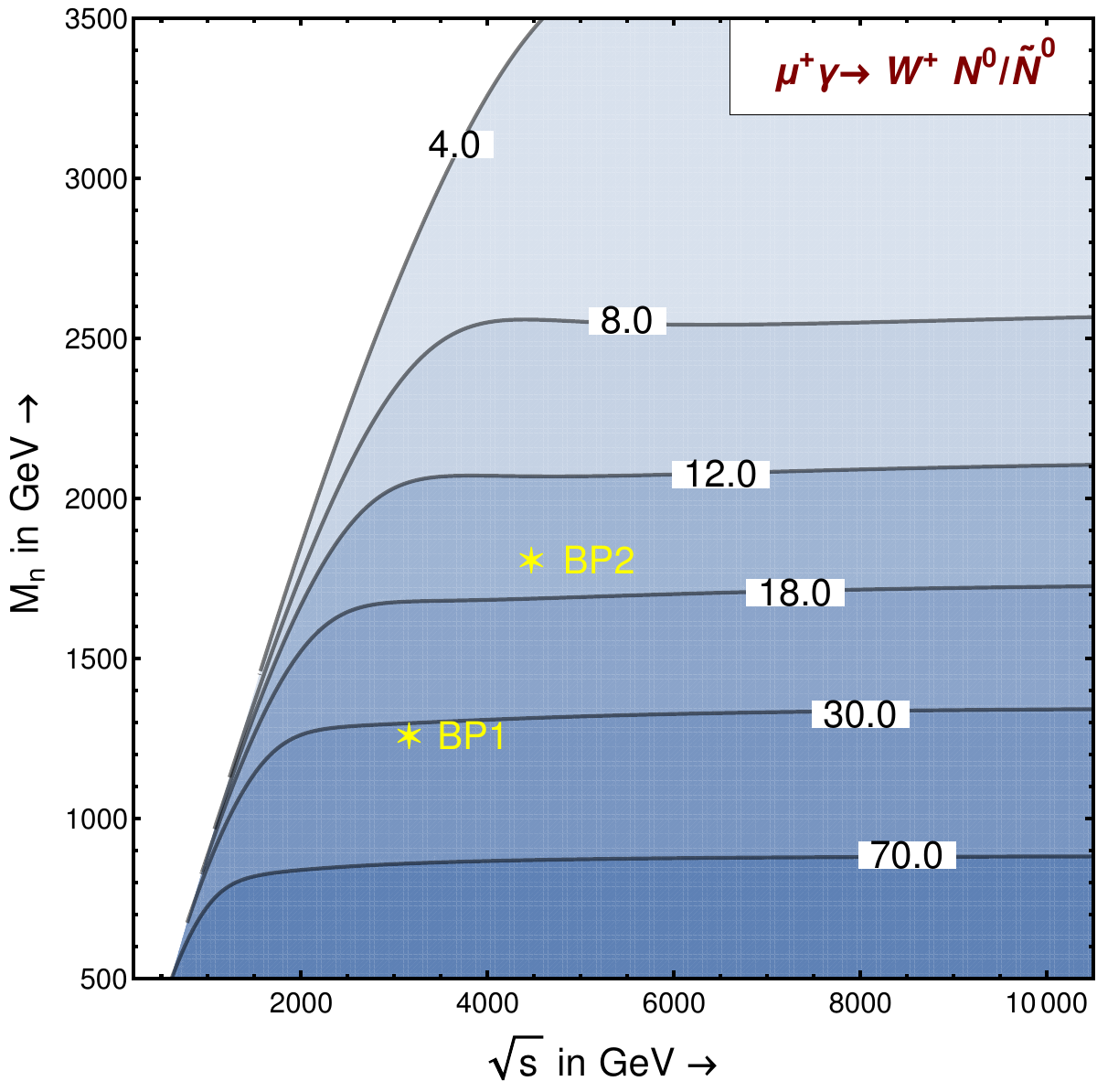}\label{}}
			\subfigure[Type-II]{\includegraphics[width=0.35\linewidth,angle=-0]{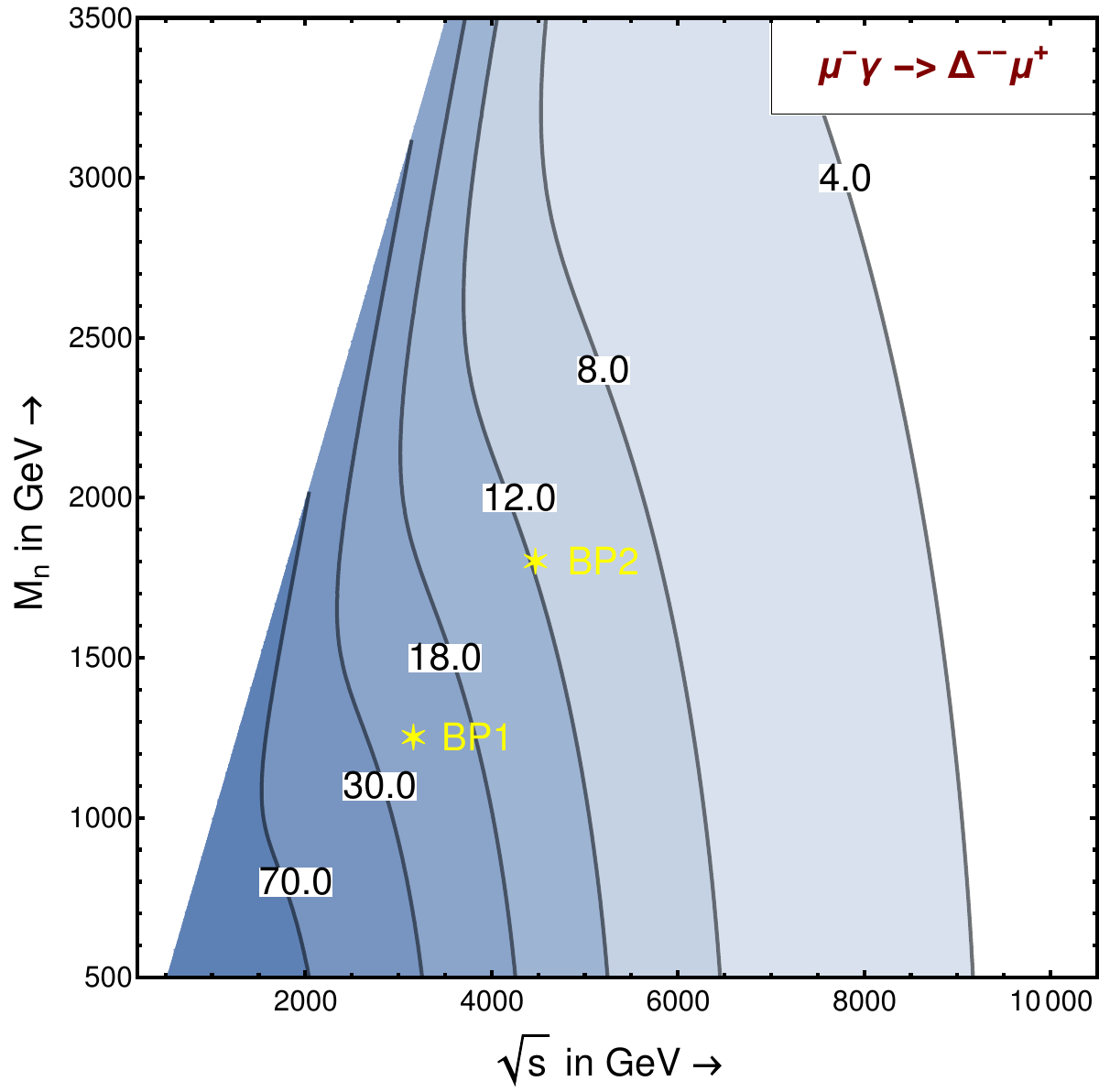}\label{}}
			\subfigure[iType-III]{\includegraphics[width=0.35\linewidth,angle=-0]{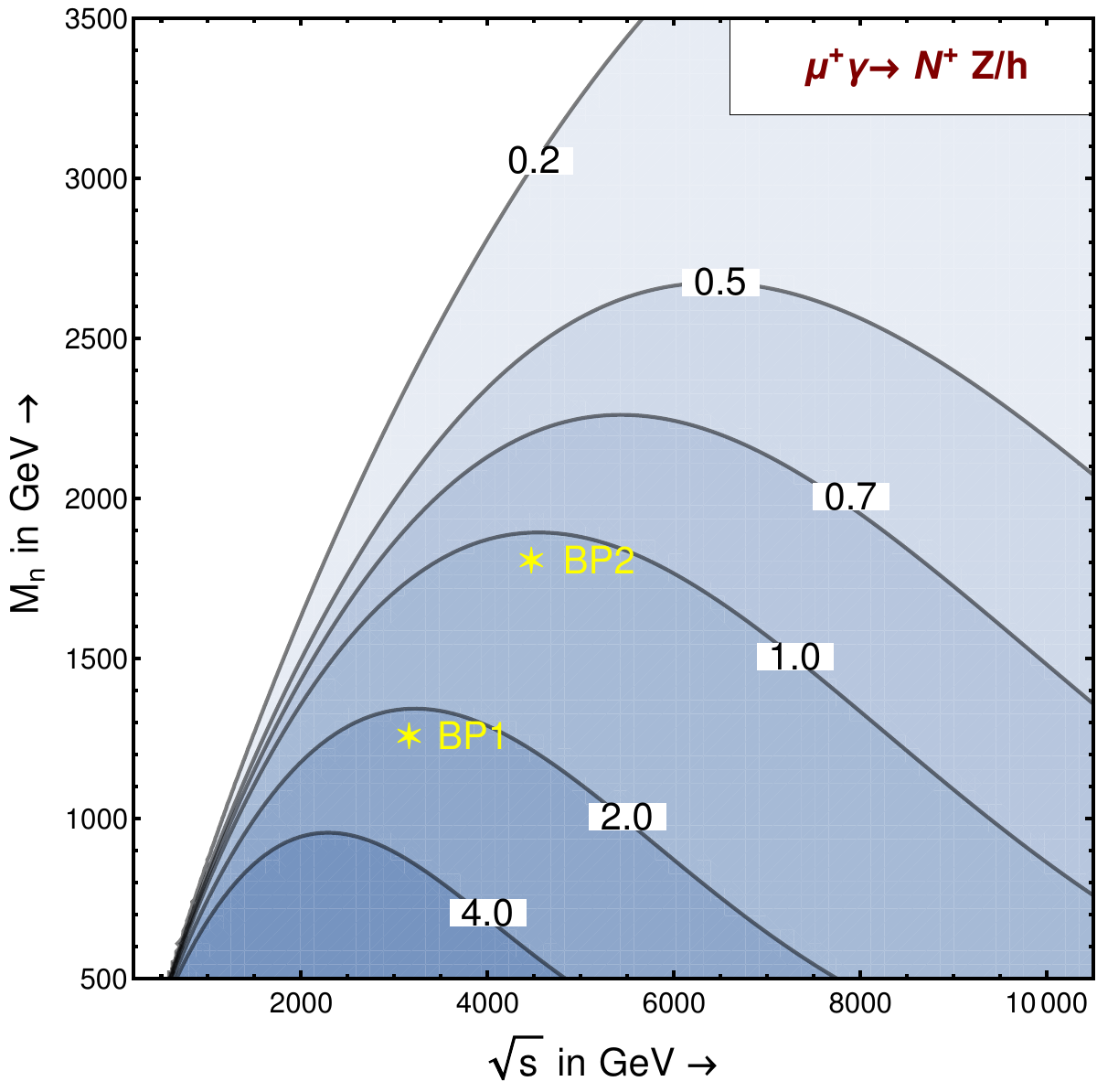}\label{}}}		
		\caption{Contours of the total cross-sections (fb) for $\mu^+ \gamma$ collider in $M_n=M_{\Delta}$ verses the centre-of-mass energy ($\sqrt{\hat{s}}$) plane for $Y_\Delta$ and $Y_N =0.2$, $\mu_\Delta$ and  $\mu=10$ eV.}\label{crosec_mu-gamma}
	\end{center}
\end{figure*}

Figure \autoref{crosec_mu-gamma} illustrates the detailed relationship between cross-sections, centre-of-mass energy, and BSM particle mass for the three Seesaw scenarios in a $\mu^+\gamma$ collider. The contours represent constant cross-section values, with the two benchmark points highlighted as yellow stars. The gradient from deeper to lighter colours reflects a gradual decrease in cross-section values. Notably, the behaviour of cross-section variation with energy differs across the Seesaw scenarios. In the iType-I case, the cross-section stabilizes beyond a certain energy for a fixed heavy neutral fermion mass. For the Type-II case, the cross-section decreases significantly when moving away from resonance. In contrast, the iType-III scenario exhibits very low cross-section values for the selected production mode, with the contours peaking before tapering off.

\subsection{Final states in reconstructing the resonance particles} 

Similar to the approach at the $\mu^+ \mu^-$ collider, we reconstruct the invariant mass of the BSM particles from their most dominant fully visible decay products for all three Seesaw models. Additionally, we discuss the specific final states chosen for our study and the significant SM backgrounds.

\subsubsection{iType-I Seesaw at $\mu^+ \gamma$ Collider} 

\begin{figure*}[hbt]
	\begin{center}
		\hspace*{-0.5cm}
		\mbox{\subfigure[]{\includegraphics[width=0.4\linewidth,angle=-0]{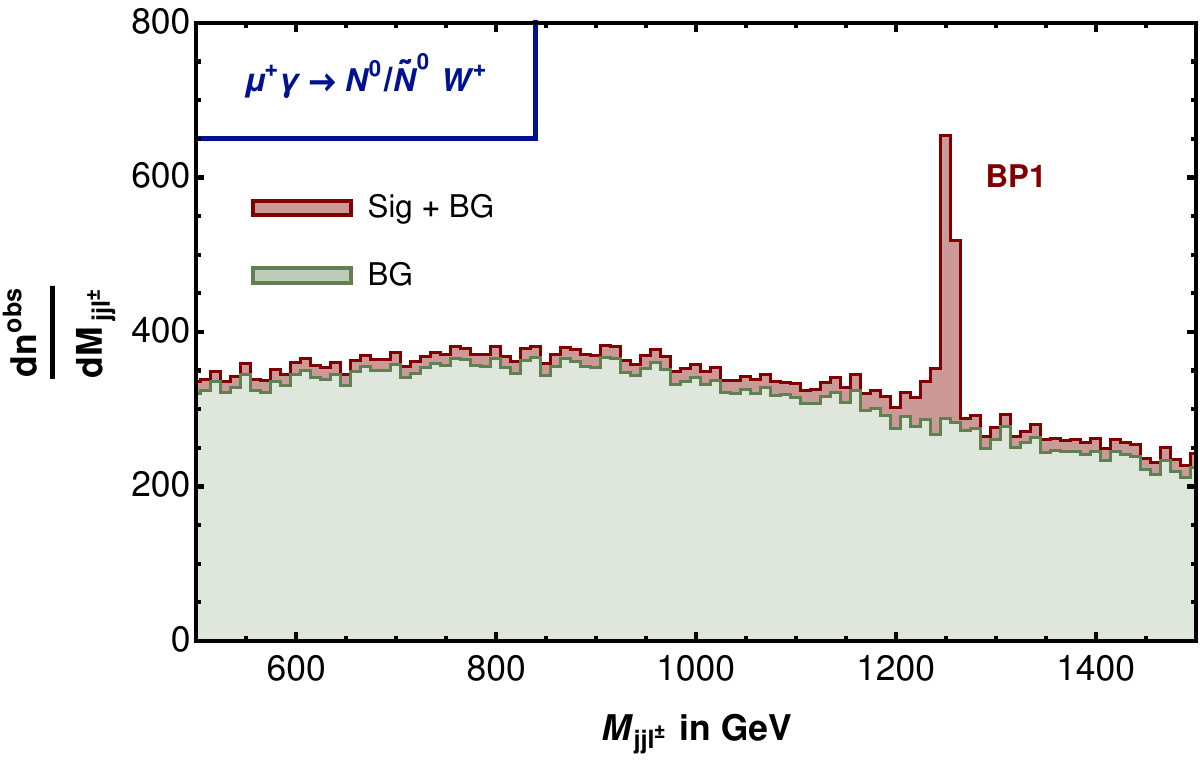}\label{}}
			\hspace*{1.0cm}
			\subfigure[]{\includegraphics[width=0.4\linewidth,angle=-0]{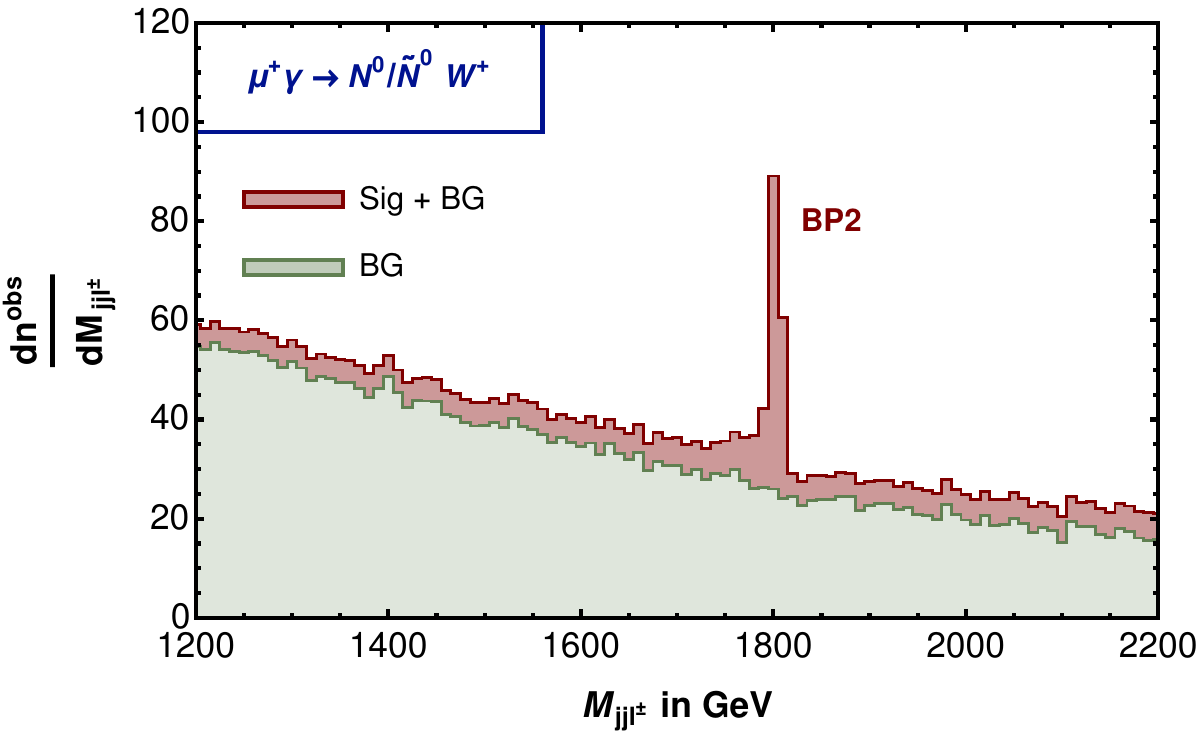}\label{}}}		
		\caption{Di-jet-mono-lepton invariant mass distribution ($M_{jj\ell^{\pm}}$) for (a) BP1 and (b) BP2 at $\mu^+ \gamma$ collider with the integrated luminosity of 1000\,fb$^{-1}$.  The total (signal $+$ SM background) signature is depicted in brown and the SM background is in olive green.}\label{mu-photonTy1invM}
	\end{center}
\end{figure*}

In the iType-I scenario, we focus on the associated production of $N^0/\widetilde{N}^0$ along with a $W^+$ boson. Similar to previous analyses, we reconstruct the invariant mass $M_{jj\ell^\pm}$, as illustrated in \autoref{mu-photonTy1invM} for the benchmark points. To accurately reconstruct the mass of the neutral lepton, we consider its fully visible decay mode  ($N^0 \to W^\pm \ell^\mp \to j j \ell^\pm$). This scenario involves two $W^\pm$ bosons: one produced directly and the other from the $N^0$ decay. Consequently, reconstructing the $N^0$ mass peak from the $M_{jj\ell^\pm}$ invariant mass distribution is somewhat challenging, as it requires correctly identifying the $W^\pm$ boson that originates from the $N^0$ decay. In \autoref{mu-photonTy1invM}, the brown histogram represents the total event contribution, including both signal and SM backgrounds, while the olive green region denotes the background alone. For both benchmark points, we observe a significant peak, clearly indicating the signal over the background.

We focus on the hadronic decay modes of the $W^\pm$ bosons to avoid missing energy in the final state. Consequently, we select the final state $1\ell^{\pm} + 4j $ for our analysis of the iType-I Seesaw mechanism at a $\mu^+ \gamma$ collider. The event numbers for both the signal and the dominant background ($\mu^- VV$, where $V$ represents $Z$ or $W^\pm$) are provided in \autoref{Tab:FSmuphTy1} at an integrated luminosity of 1000 \fbi. Two values are listed for the final state, corresponding to Laser Back Scattering (monochromatic) photon production processes. Additionally, an invariant mass cut of $|M_{jj\ell^{\pm}} -M_n|\leq 10\,$GeV is applied, which significantly reduces background events. A $5\sigma$ signal significance can be achieved with a few hundred to a few thousand \fbi of integrated luminosity for the respective benchmark points.


\begin{table*}[hbt]
	\renewcommand{\arraystretch}{1.4}
	\centering
	\begin{tabular}{|c|c|c|c|c|}
		\hline 
		\multirow{2}{*}{Final states} & \multicolumn{2}{c|}{BP1} &\multicolumn{2}{c|}{BP2} \\ 
		\cline{2-5}
		&   Sig & BG ($\mu^+ V V$) & Sig & BG ($\mu^+ V V$)\\ 
		\hline \hline
		$1\ell^{\pm} + 4j$ & 307.08 (404.68) & 2476.77 (1371.23) & 52.61 (93.73) & 361.00 (257.18) \\ 
		\cline{2-5}
		$+|M_{jj\ell^{\pm}} -M_n|\leq 10\,$GeV & 96.86 (143.84) & 24.81 (13.23) & 17.87 (22.37) & 3.54 (1.16) \\ 
		\hline	\hline	
		$\rm{S}_{\text{sig}} (\mathcal{L}_{\rm int}=1000\, \rm fb^{-1})$&  \multicolumn{2}{c|}{8.78 (11.48) }  & \multicolumn{2}{c|}{3.86 (4.61) } \\
		\hline
		$\int \mathcal{L}_{5\sigma}\,[\rm{fb}^{-1}]$&  \multicolumn{2}{c|}{324.30 (189.69) }  & \multicolumn{2}{c|}{1677.89 (1176.35) }\\  
		\hline
	\end{tabular}
	\caption{Number of events for signal and background corresponding to $1\ell^{\pm} + 4j$ final state with the laser backscattering (monochromatic) photon for the benchmark points mentioned in \autoref{crs_mum_gamma} \textcolor{black}{using the integrated luminosity of 1000\,fb$^{-1}$, at the $\mu^+ \gamma$ collider.}} \label{Tab:FSmuphTy1}
\end{table*}	


\subsubsection{Type-II Seesaw at $\mu^+ \gamma$ Collider} 

\begin{figure*}[h]
	\begin{center}
		\hspace*{-0.5cm}
		\mbox{\subfigure[]{\includegraphics[width=0.4\linewidth,angle=-0]{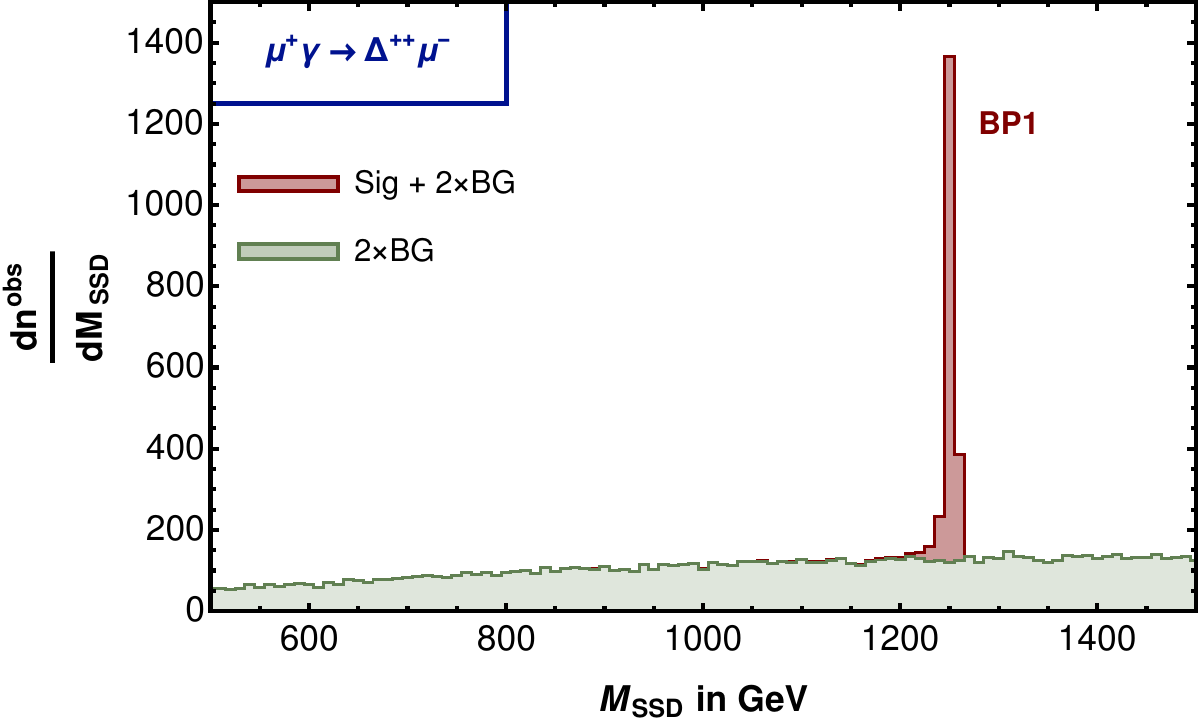}}
			\hspace*{1.0cm}
			\subfigure[]{\includegraphics[width=0.4\linewidth,angle=-0]{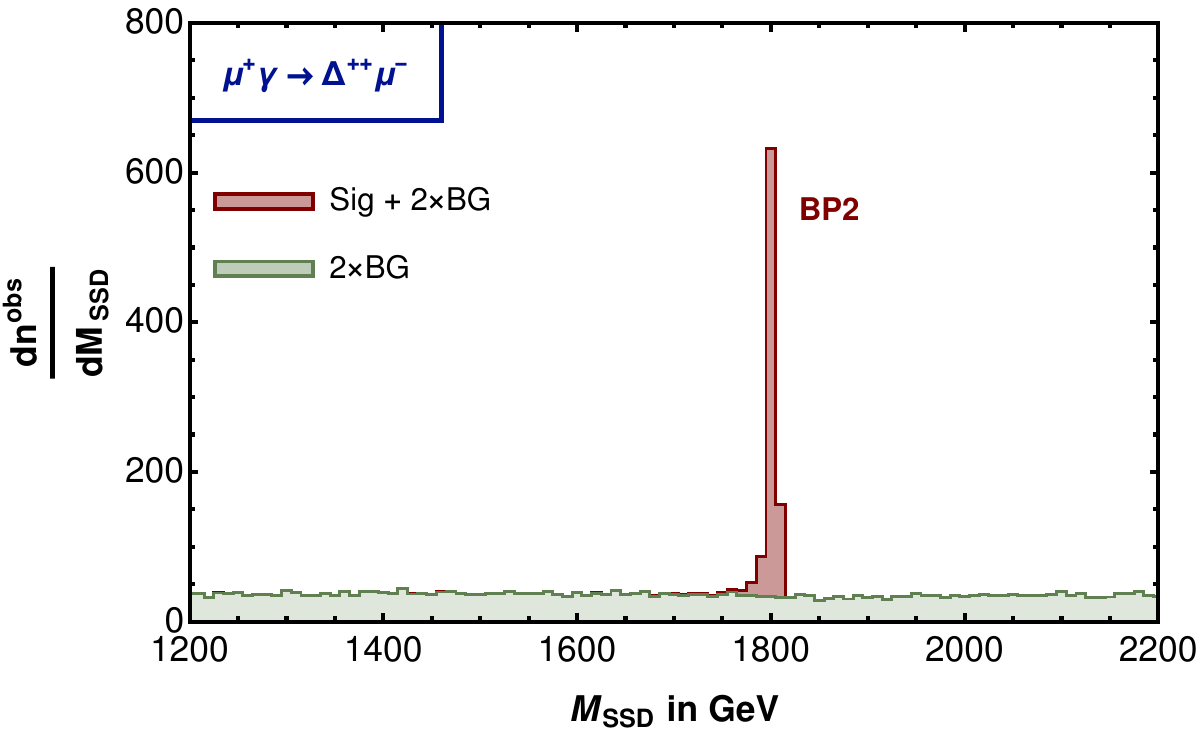}}}		
		\caption{Same sign di-lepton invariant mass distribution ($M_{\text{SSD}}$) for (a) BP1 and (b) BP2 at $\mu^+ \gamma$ collider with the integrated luminosity of 1000\,fb$^{-1}$.  The total (signal $+$ SM background scaled by 2) signature is depicted in brown and the SM background (scaled by 2) is in olive green.}\label{mu-photonTy2invM}
	\end{center}
\end{figure*}

Next, we consider the Type-II Seesaw mechanism, focusing on the production of a doubly-charged scalar in association with a $\mu^-$. As previously discussed, the $\Delta^{++}$ particle decays into two leptons with a 100\% branching ratio, though the flavours of these leptons can vary. The mass of $\Delta^{++}$ can be reconstructed from the $M_{\ell \ell}$ invariant mass peak, requiring correct tagging of two same-sign leptons. Any additional lepton in the event, typically of opposite sign, can be easily excluded when calculating $M_{\ell \ell}$, denoted as $M_{\rm SSD}$ to avoid confusion. \autoref{mu-photonTy2invM} shows this distribution, with the brown histogram representing the combined signal plus $2 \times$ background events, and the olive green curve indicating the background scaled by two. Clear peaks around the masses of the doubly-charged scalar are visible for our chosen benchmark points.

\vspace{-0.1 cm}	

We now focus on the desired final state and the corresponding SM backgrounds that contribute here. Since the $\Delta^{++}$ particle decays into two same-sign leptons and is produced alongside a $\mu^-$, we consider the $1\ell^- +2\ell^+$ final state for our study. The dominant SM backgrounds in this scenario arise from $\ell Z$ and $\ell VV$ processes, particularly when the gauge bosons decay leptonically. However, by applying an invariant mass cut around $|M_{\ell \ell} -M_{\Delta}|\leq 10\,$GeV, the background contribution is significantly reduced, as shown in \autoref{Tab:FSmuphTy2}. Due to the clear environment when focusing on the $1\ell^- +2\ell^+$ final state at a $\mu^+ \gamma$ collider, a $5 \sigma$ significance can be achieved with less than 10 \fbi of integrated luminosity.

\begin{table*}[h!]
	\renewcommand{\arraystretch}{1.3}
	\centering
	\scalebox{0.9}{\begin{tabular}{|c|c|c||c|c|}
			\cline{2-5}
			\multicolumn{1}{c|}{}&\multirow{2}{*}{Final states}& \multirow{2}{*}{Signal} &\multicolumn{2}{c|}{Backgrounds} \\ 
			\cline{4-5}
			\multicolumn{1}{c|}{}&& & $\ell Z$ & $\ell VV$  \\ 
			\hline
			\multirow{5}{*}{BP1} & $1\ell^- +2\ell^+$ & 16802.26 (7700.71) & 5995.97 (2324.02) & 20764.53 (9982.95)   \\ 
			\cline{3-5}
			& $+|M_{\ell \ell} -M_{\Delta}|\leq 10\,$GeV & 14746.53 (5865.59) & 15.43 (2.85) & 153.99 (61.25) \\ 
			\cline{2-5}
			& Total & 14746.53 (5865.59)  &\multicolumn{2}{c|}{169.42 (64.10) } \\
			\cline{2-5}	
			& $\rm{S}_{\text{sig}} \text{ at }\mathcal{L}_{\rm int}=1000\, \rm fb^{-1}$  &  \multicolumn{3}{c|}{120.89 (76.18) } \\  
			\cline{2-5}
			& $\int \mathcal{L}_{5\sigma}\,[\rm{fb}^{-1}]$ &   \multicolumn{3}{c|}{1.71 (4.30) } \\ 
			\hline \hline
			\multirow{5}{*}{BP2} & $1\ell^- +2\ell^+$ & 8089.77 (3969.25) & 2896.60 (954.01) & 7415.90 (3360.53)  \\ 
			\cline{3-5}
			& $+|M_{\ell \ell} -M_\Delta|\leq 10\,$GeV & 7041.58 (2849.73) & 9.56 (1.47) & 30.88 (11.35)   \\ 
			\cline{2-5}
			& Total & 7041.58 (2849.73)  &\multicolumn{2}{c|}{40.44 (12.82) } \\
			\cline{2-5}	
			& $\rm{S}_{\text{sig}} \text{ at } \mathcal{L}_{\rm int}=1000\, \rm fb^{-1}$  &  \multicolumn{3}{c|}{83.67 (53.26) } \\  
			\cline{2-5}
			& $\int \mathcal{L}_{5\sigma}\,[\rm{fb}^{-1}]$ &   \multicolumn{3}{c|}{3.57 (8.81) } \\ 
			\hline 
	\end{tabular}}
	\caption{Number of events for signal and background corresponding to $1\ell^- +2\ell^+$ final state with the laser backscattering (monochromatic) photon for the benchmark points mentioned in \autoref{crs_mum_gamma} \textcolor{black}{using the integrated luminosity of 1000\,fb$^{-1}$.}} \label{Tab:FSmuphTy2}
\end{table*}	

\subsubsection{iType-III Seesaw at $\mu^+ \gamma$ Collider} 

\begin{figure*}[h]
	\begin{center}
		\hspace*{-0.5cm}
		\mbox{\subfigure[]{\includegraphics[width=0.4\linewidth,angle=-0]{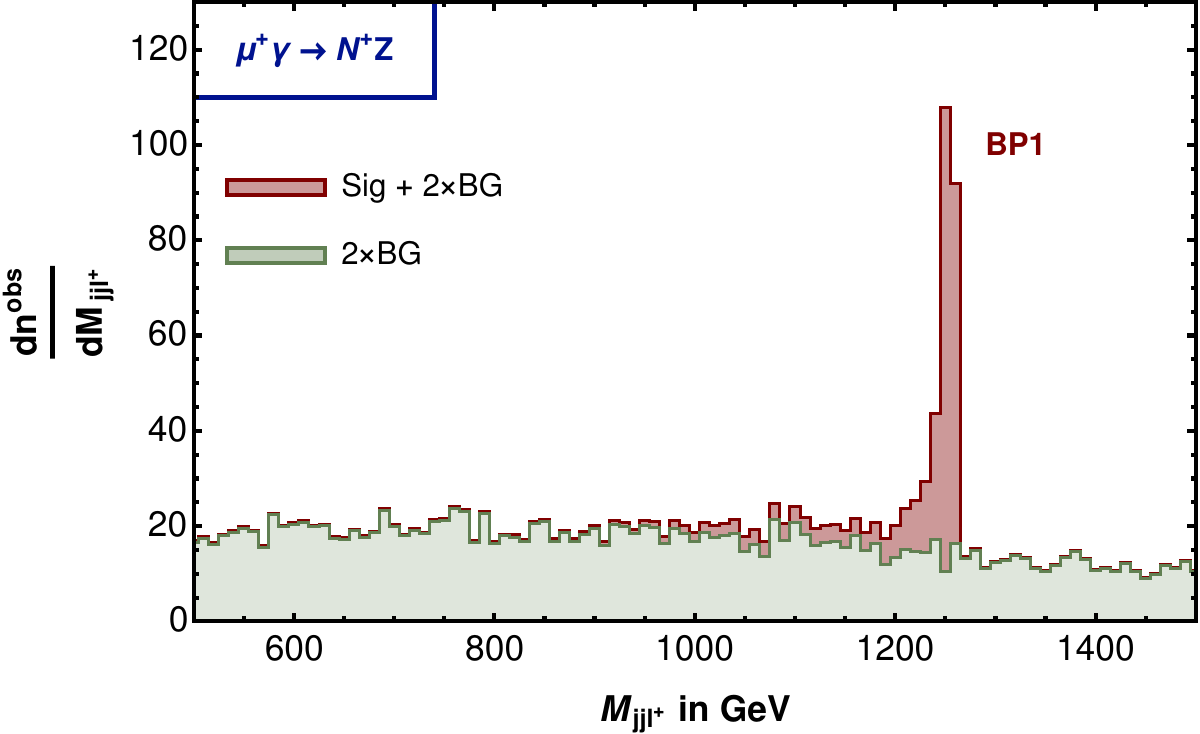}\label{}}
			\hspace*{1.0cm}
			\subfigure[]{\includegraphics[width=0.4\linewidth,angle=-0]{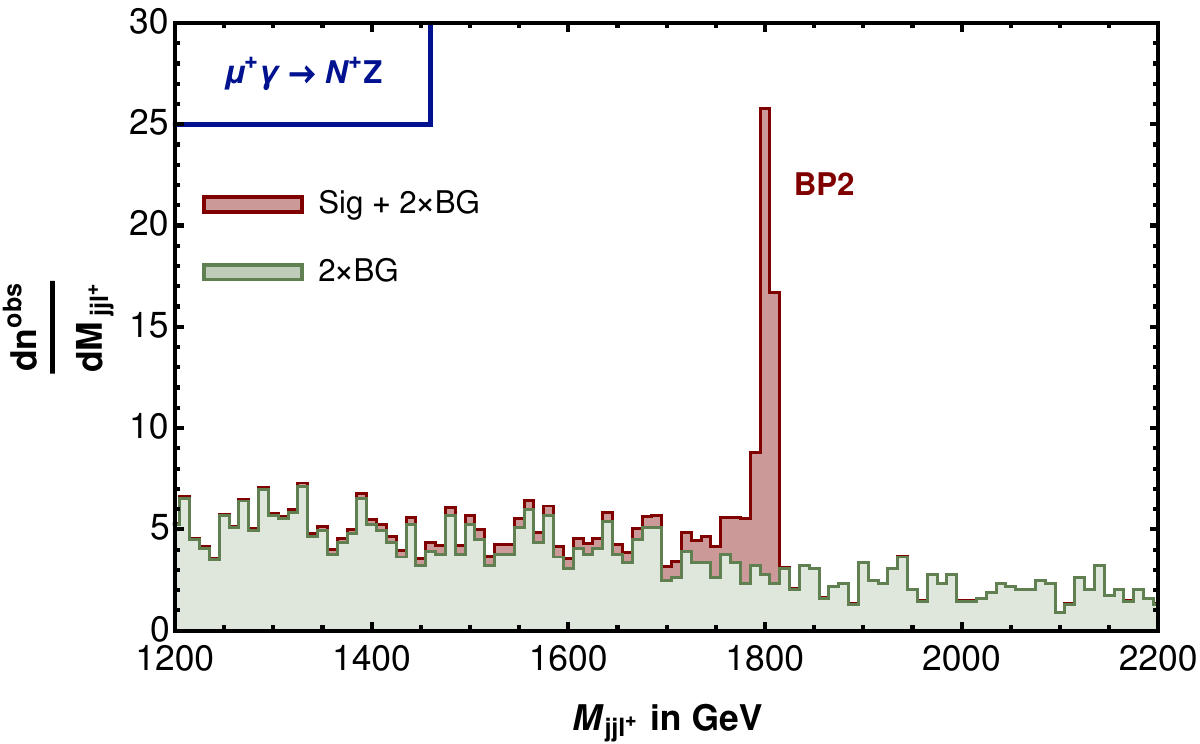}\label{}}}		
		\caption{Di-jet-mono-lepton invariant mass distribution ($M_{jjl^+}$) for (a) BP1 and (b) BP2 at $\mu^+ \gamma$ collider with the integrated luminosity of 1000\,fb$^{-1}$.  The total (signal $+$ SM background scaled by 2) signature is depicted in brown and the SM background (scaled by 2) is in olive green.}\label{mu-photonTy3invM}
	\end{center}
\end{figure*}

After exploring the iType-I and Type-II Seesaw scenarios at a  $\mu^+ \gamma$ collider, we now consider the production of a heavy singly-charged fermion ($N^+$) alongside a $Z/h$ boson. In this case, $N^+$ decays into $Z \ell^+$ and $h \ell^+$ with equal branching ratios of 50\% for each channel. This subsection focuses on the invariant mass distribution $M_{jj \ell^+}$ after reconstructing the $Z/h$ boson mass peak using a window cut of either $|M_{jj} - M_Z|< 10$ GeV or $|M_{bb} - M_h|< 10$ GeV. Since each event typically involves two $Z/h$ bosons, accurately tagging the $Z/h$ bosons that originate from the decay of the $N^+$ particle is crucial for reconstructing the $N^+$ mass. In \autoref{mu-photonTy3invM}, we present these distributions for the signal plus $2\times$ background and for only $2\times$ background, depicted by the brown and olive green histograms, respectively. The clear peak in the brown histogram confirms the $N^+$ mass for the two selected benchmark points.

For the final state analysis, we focus on the process $\mu^+ \gamma \to N^+ Z$, where the $Z$ boson decays hadronically. This leads us to study the $1\ell^{+} + 4j$ final state, with the primary SM background coming from the $\mu^+ V V$ channel. Given the large decay branching ratios of the vector bosons  $V= Z,\,W^\pm$ to hadrons, the background contribution is significant if the invariant mass cut $|M_{jj\ell^{+}} -M_n|\leq 10\,$GeV is not applied. With this cut, a $5 \sigma$ significance can be achieved for the two BPs at integrated luminosities of $\sim 950$ \fbi and $\sim 8450$ \fbi, respectively.


\begin{table*}[hbt]
	\renewcommand{\arraystretch}{1.5}
	\centering
	\begin{tabular}{|c|c|c|c|c|}
		\hline 
		\multirow{2}{*}{Final states} & \multicolumn{2}{c|}{BP1} &\multicolumn{2}{c|}{BP2} \\ 
		\cline{2-5}
		&   Sig & BG ($\mu^+ V V$) & Sig & BG ($\mu^+ V V$)\\ 
		\hline \hline
		$1\ell^{+} + 4j$ & 98.50 (250.72) & 2476.64 (1371.19) & 15.38 (52.11) & 360.97 (257.18) \\ 
		\cline{2-5}
		$+|M_{jj\ell^{+}} -M_n|\leq 10\,$GeV & 31.74 (45.56) & 6.40 (3.63) & 3.83 (7.02) & 1.15 (0.39) \\ 
		\hline	\hline	
		$\rm{S}_{\text{sig}} \text{ at } \mathcal{L}_{\rm int}=1000\, \rm fb^{-1})$&  \multicolumn{2}{c|}{5.14 (6.50) }  & \multicolumn{2}{c|}{1.72 (2.58) } \\
		\hline
		$\int \mathcal{L}_{5\sigma}\,[\rm{fb}^{-1}]$&  \multicolumn{2}{c|}{946.26 (591.71) }  & \multicolumn{2}{c|}{8450.51 (3755.78) }\\  
		\hline
	\end{tabular}
	\caption{Number of events for signal and background corresponding to $1\ell^{+} + 4j$ final state with the laser backscattering (monochromatic) photon for the benchmark points mentioned in \autoref{crs_mum_gamma} \textcolor{black}{using the integrated luminosity of 1000\,fb$^{-1}$.}} \label{Tab:FSmuphTy3}
\end{table*}	


\subsubsection{Reconstructed angular distribution}

 As shown in \autoref{mu-photon_ang_anal} (b), the angular distribution of the Type-II scalar $\Delta^{++}$, produced in association with a $\mu^-$ at a $\mu^+\gamma$ collider, is presented. In this case, the final state $\mu^-$ is a fundamental particle, while $\Delta^{++}$ decays dominantly into $\ell^+\ell^+$. This makes the reconstruction of $\Delta^{++}$ comparatively straightforward, as illustrated in \autoref{mu-photonTy2invM}. Consequently, the reconstruction of the CM frame is significantly simpler than in the iType-I and iType-III scenarios, where the heavy BSM states undergo multi-prong decays. For instance, in the iType-I and iType-III cases, the particles $N^0/\tilde{N}^0$ and $N^{+}$ are produced in association with $W^+$ and $Z/h$, respectively. These associated bosons subsequently decay into jets, and the heavy fermions themselves further decay into gauge or Higgs bosons. Such cascades lead to large combinatorial backgrounds, making it challenging to correctly reconstruct the CM frame on an event-by-event basis. Additional complications arise from asymmetric beams and asymmetric final states. It is worth noting that one-prong decays of BSM particles (such as leptoquarks) in symmetric configurations with symmetric beams can, in contrast, allow for feasible reconstruction even at the LHC \cite{Bandyopadhyay:2020wfv}.

\begin{figure*}[hbt]
	\begin{center}
		\includegraphics[width=0.5\linewidth,angle=-0]{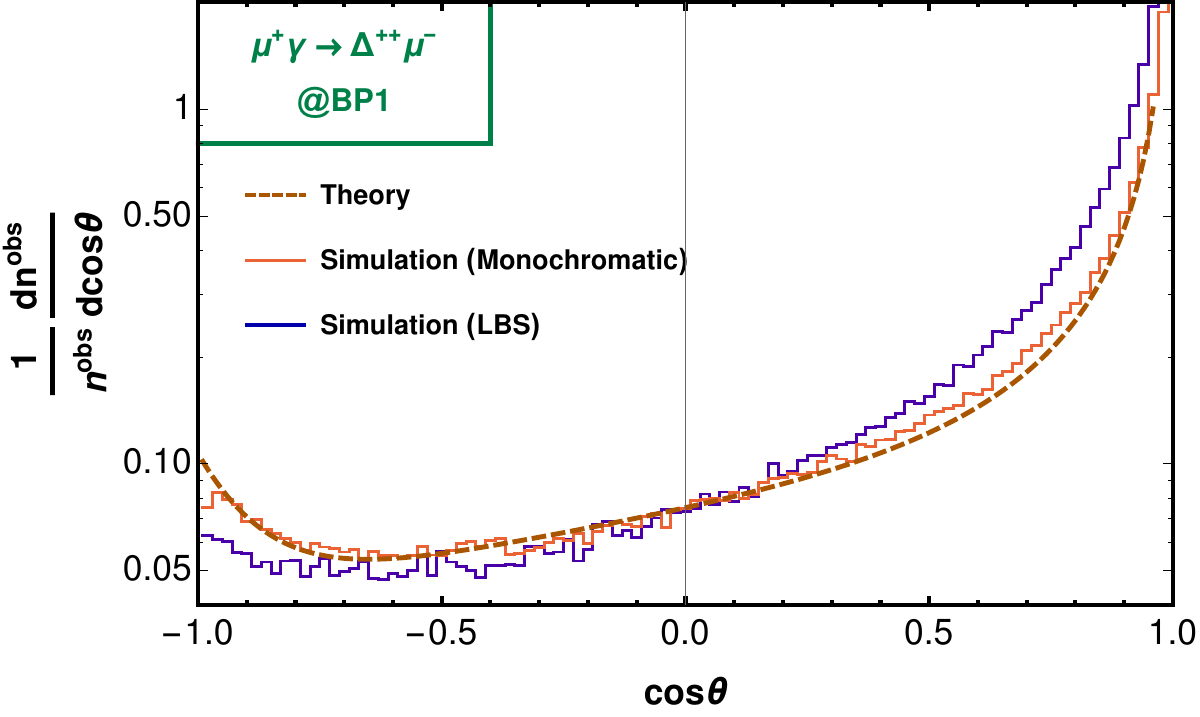}		
		\caption{Angular distribution of  the theory and simulation (with laser backscattering and monochromatic photon)  at $\mu^+ \gamma$ collider for the Type-II Seesaw scenario for BP1.}\label{angdismumPh}
	\end{center}
\end{figure*}

For this reason, we highlight the Type-II Seesaw case as the most promising for reconstructing angular distributions. \autoref{angdismumPh} shows three representative angular distributions, where the $x$-axis denotes $\cos\theta$ (with $\theta$ defined as the angle between $\Delta^{++}$ and the incoming $\mu^+$), and the $y$-axis gives the corresponding differential cross sections. The dark orange dashed line represents the theoretical expectation as described in \autoref{typeIImg}, while the orange solid line corresponds to the monochromatic photon beam and the blue solid line to the laser back-scattered (LBS) photon beam, shown for BP1. It is evident from the figure that the distributions for both beam setups follow the theoretical curve closely, with identifiable minima despite some uncertainties, thereby retaining the distinctive angular features predicted by theory. 

{It is interesting to mention that although the angular distribution of $\Delta^{++}$ shows a dip around $\cos\theta\approx-0.7$, it never vanishes at any particular angle. In order to observe RAZ at $\mu^+\gamma$ collider the absolute value of the electric charge of the BSM particle must be strictly between zero and one \cite{Bandyopadhyay:2020klr}, which cannot be satisfied with $\Delta^{++}$. Due to the same requirement, the processes of $\mu^+\gamma\to W^+N^0/\tilde N^0$ or $\mu^+\gamma\to N^+ Z/h$, arising in iType-I or iType-III Seesaw, also cannot show RAZ at $\mu^+\gamma$ collider. This can be testified from the general condition for the occurrence of RAZ~\cite{Brodsky:1982sh}.}

\begin{table}[h!]
	\renewcommand{\arraystretch}{1.7}
	\centering
	\begin{tabular}{|c|c|c|}
		\cline{2-3}
		\multicolumn{1}{c|}{} & \multicolumn{2}{c|}{$\cos \theta$ for Type-II Seesaw}  \\
		\cline{2-3}
		\multicolumn{1}{c|}{} & ($-0.95 \to 0.0$) & ($0.0 \to 0.95$)  \\
		\hline 
		Signal & 2607.2 (1423.5) & 12749.6 (4811.9) \\
		\hline 
		Background & 115.3 (60.5) & 89.7 (27.2)   \\
		\hline 
	\end{tabular}
	\caption{Number of events in the $\cos\theta$ intervals $(-0.95,0.0)$ and $(0.0,0.95)$ for the BP1 of the Type-II seesaw scenario, at $\mu^+ \gamma$ collider, corresponding to the final states discussed earlier. The non-bracketed (bracketed) values denote results for laser backscattering (monochromatic) photon configurations. All event yields are given for an integrated luminosity of 1000~\fbi.}  \label{tab:assymtry3}
\end{table}
{In \autoref{tab:assymtry3} we provide the number of events at an integrated luminosity of 1000 fb$^{-1}$ for the signal and background for the chosen final states as given in \autoref{Tab:FSmuphTy2} for Type-II seesaw for the chosen bins of BP1 for $\cos{\theta}:(-9.95, \, 0.00), \, (0.00,\, 0.95)$. Here the signal significance for the left bins are of $50\sigma (37 \,\sigma) $ for the laser back scattering(monochromatic) photon configurations. The corresponding numbers for the right bins are $112.5 \sigma\, (69\sigma)$. However, the symmetry of the signal events can be spot on if we see the ratio of signal events for $\cos{\theta}:(-9.95, \, 0.00)$ and $(0.00,\, 0.95)$ bins, which are $1:5$ for the laser back scattering and $1:3.4$ for the monochromatic photon, which are drastically different from the backgrounds numbers as well. Thus it helps to successfully probe the angular distributions coming from Type-II seesaw at the $\mu \gamma$ collider.}

\section{At $\mu^+ e^-$ collider}\label{sec:mum_ep}	

Electron-muon colliders have been proposed as intermediate facilities between $e^+e^-$ and $\mu^+ \mu^-$ colliders, combining the clean experimental environment of electron beams with the higher energy reach of muon beams. Such machines can offer reduced beam-induced backgrounds compared to $\mu^+\mu^-$ colliders and higher luminosities, since the muon beam is not split between two rings. A first-stage realization of this concept is the MUonE experiment \cite{Abbiendi:2022oks}, where high-energy muons scatter elastically on atomic electrons. Extending this idea to a dedicated $\mu^+e^-$ collider opens up a novel environment for probing BSM scenarios, in particular those involving lepton flavour violation. In principle, an asymmetric beam configuration ($E_\mu \gg E_e$) can be advantageous, since the final-state particles are boosted in the muon direction and the beam-induced background is further suppressed. However, for simplicity in this study, we restrict ourselves to symmetric beam energies, which already capture the essential phenomenology and allow a clearer comparison with other lepton collider modes.

We explore the potential of a $\mu^+e^-$ collider \cite{Lu:2020dkx} to distinguish Seesaw scenarios through angular distributions. Specifically, we study $\sqrt{s}=0.3$ and $1.0$ TeV with an integrated luminosity of $1000$ fb$^{-1}$. Given that the mass of inverse Type-III Seesaw fermions is constrained to be above $\sim 1.2$ TeV \cite{CMS:2019lwf}, this scenario cannot be probed at such energies. In contrast, the iType-I Seesaw remains viable, as no strict lower bound exists on the mass of heavy neutrinos. In this case, the $\nu N^0/\widetilde{N}^0$ channel becomes accessible whenever $M_{N^0/\widetilde{N}^0} < \sqrt{s}$. The production cross section is sizable, and the resulting $1\ell+2j+p_T^{\rm miss}$ final state exhibits an angular distribution similar to that at a $\mu^+\mu^-$ collider. To avoid redundancy, we therefore concentrate on the distinct signature of the Type-II Seesaw scenario in this section.


\subsection{Type-II seesaw at $\mu^+e^-$ collider}

\begin{figure}[hbt]
	\begin{center}
		\begin{tikzpicture}
			\begin{feynman}
				\vertex (a1);
				\vertex [left=1.2cm of a1] (a0){$\mu^+$};
				\vertex [right=1.22cm of a1] (a2){$\mu^-$};
				\vertex [below=1.5 cm of a1] (b1);
				\vertex [left=1.2cm of b1] (b0){$e^-$};
				\vertex [right=1.2cm of b1] (b2){$e^+$};
				\diagram {(a0)--[anti fermion](a1)--[fermion](a2),
					(b0)--[fermion](b1)--[anti fermion](b2),
					(a1)--[charged scalar, edge label=$\Delta^{++}$](b1)};
			\end{feynman}
		\end{tikzpicture}	
	\end{center}
	\caption{Feynman diagrams for $\mu^+e^-\to \mu^- e^+$.}
	\label{fig:lplpm}
\end{figure}

\begin{figure}[hbt]
	\begin{center}
		\centering
		\mbox{\subfigure{\includegraphics[width=0.4\linewidth,angle=-0]{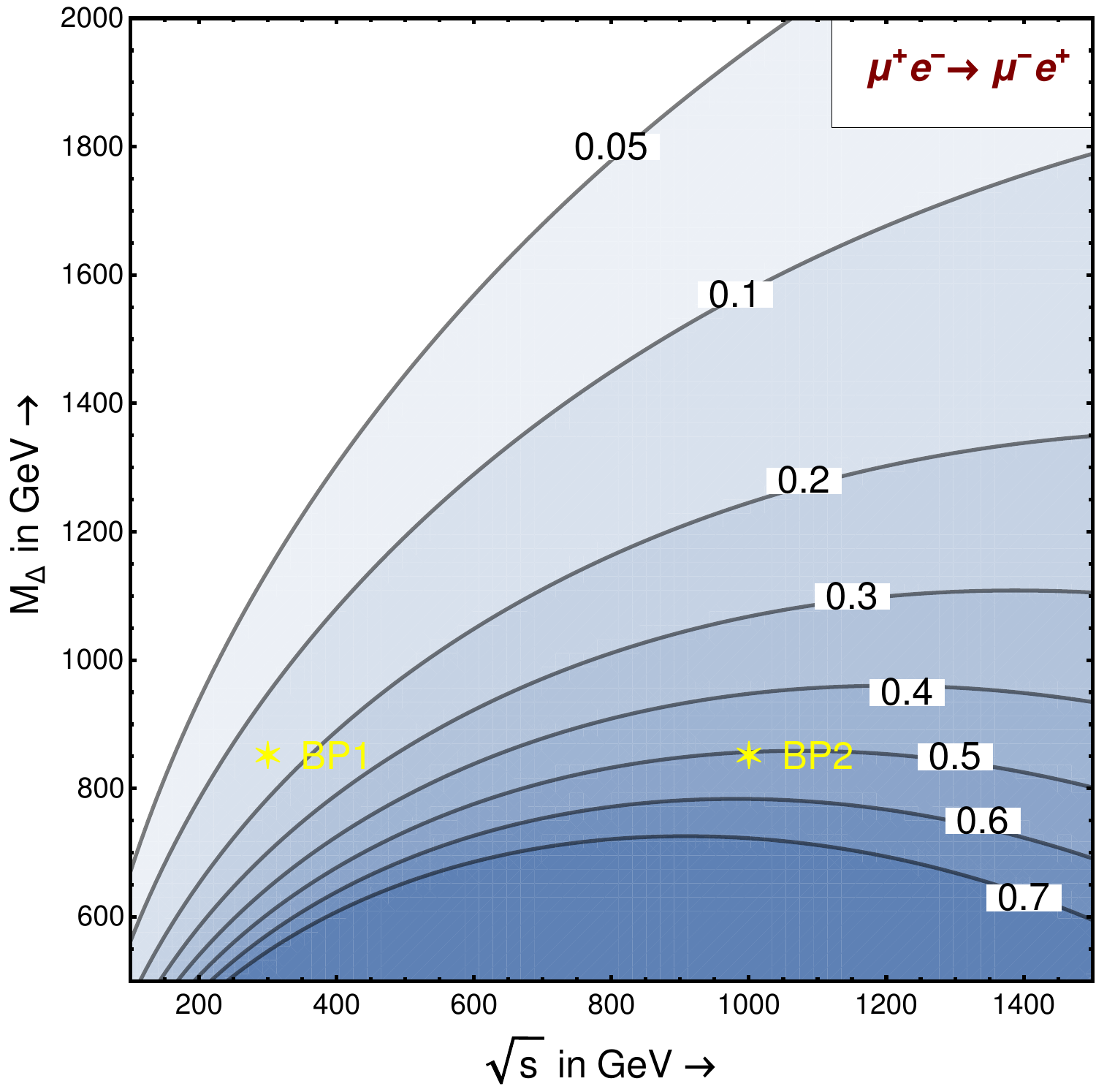}}}		
		\caption{Variation of total cross-section (in fb) for the process $\mu^+e^-\to \mu^- e^+$, with respect to the centre-of-mass energy and mass of the doubly charged scalar in the Type-II Seesaw scenario ($Y_\Delta=0.2$, $\mu_\Delta=10$ eV). The two benchmark points are represented by the yellow stars.}\label{crsecmupemTy2}
	\end{center}
\end{figure}

To detect Type-II case, we search the trace of the process $\mu^+e^-\to \mu^- e^+$, i.e. the charges of $\mu$ and $e$ get swapped, which occurs through the doubly charged scalar mediated t-channel diagram presented in \autoref{fig:lplpm}. The angular distribution and the total cross-section for this process is given by,

 In \autoref{crsecmupemTy2}  we plots  the  contours of the  cross-sections  in $M_{\Delta}-\sqrt{s}$ plane, where the darker to lighter  blue regions depict higher to lower values of cross-sections. The benchmark points  stated in \autoref{muoecrsII}  are shown  by the yellow stars. 

\begin{table*}[h!]
	\renewcommand{\arraystretch}{1.4}
	\centering
	\begin{tabular}{|c|c|c|c|}
		\cline{2-4}
		\multicolumn{1}{c|}{}&\multicolumn{3}{c|}{$\sigma(\mu^+e^- \to \mu^- e^{+} )$} \\
		\hline 
		Benchmark& $M_{\Delta}$ & $E_{CM}$&  Cross-section   \\ 
		Points &in GeV &in GeV & (in fb)\\
		\hline \hline
		BP1	& 850 & 300 & 0.15  \\ \hline
		BP2	& 850 & 1000 & 0.51 \\
		\hline
	\end{tabular}
	\caption{Masses corresponding to different benchmark points, energy of collision in CM frame and the hard scattering cross-sections (in fb) for $\mu^- e^{+}$  final states in Type-II seesaw model and $\mu^{+} e^{-}$ collider. ($Y_\Delta=0.2$, $\mu_\Delta=10$ eV)}  \label{muoecrsII}
\end{table*}
For our simulation, we have used $e^-$ and $\mu^+$ beams with equal energy to collide at a centre-of-mass energy of 0.3 TeV and 1.0 TeV, respectively, for the two benchmark points. Considering $Y_\Delta$ to be 0.2 (with $\mu_\Delta=10$ eV) we get the cross-sections for the above mentioned mode to be 0.15 fb and 0.51 fb, respectively, with $M_\Delta$ being 0.85 TeV, as presented in \autoref{muoecrsII}.

\begin{figure*}[h!]
	\begin{center}
		\mbox{\subfigure[]{\includegraphics[width=0.4\linewidth,angle=-0]{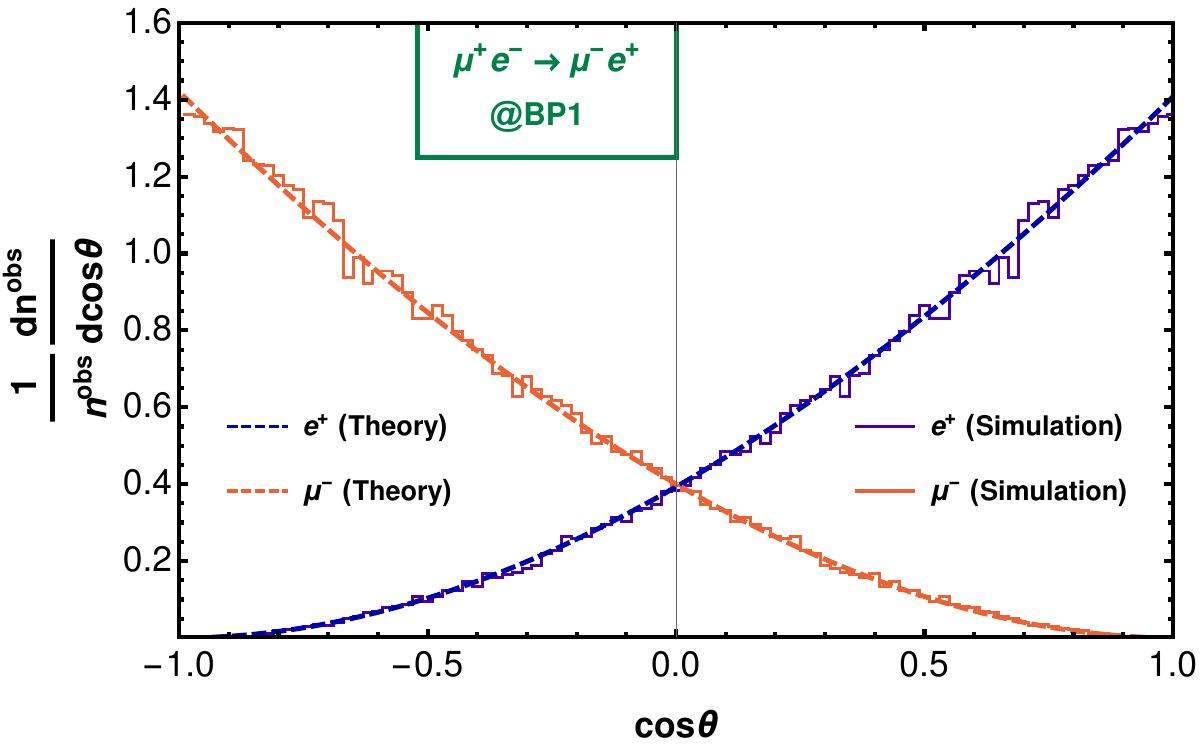}\label{}}
			\hspace*{1.0cm}
			\subfigure[]{\includegraphics[width=0.4\linewidth,angle=-0]{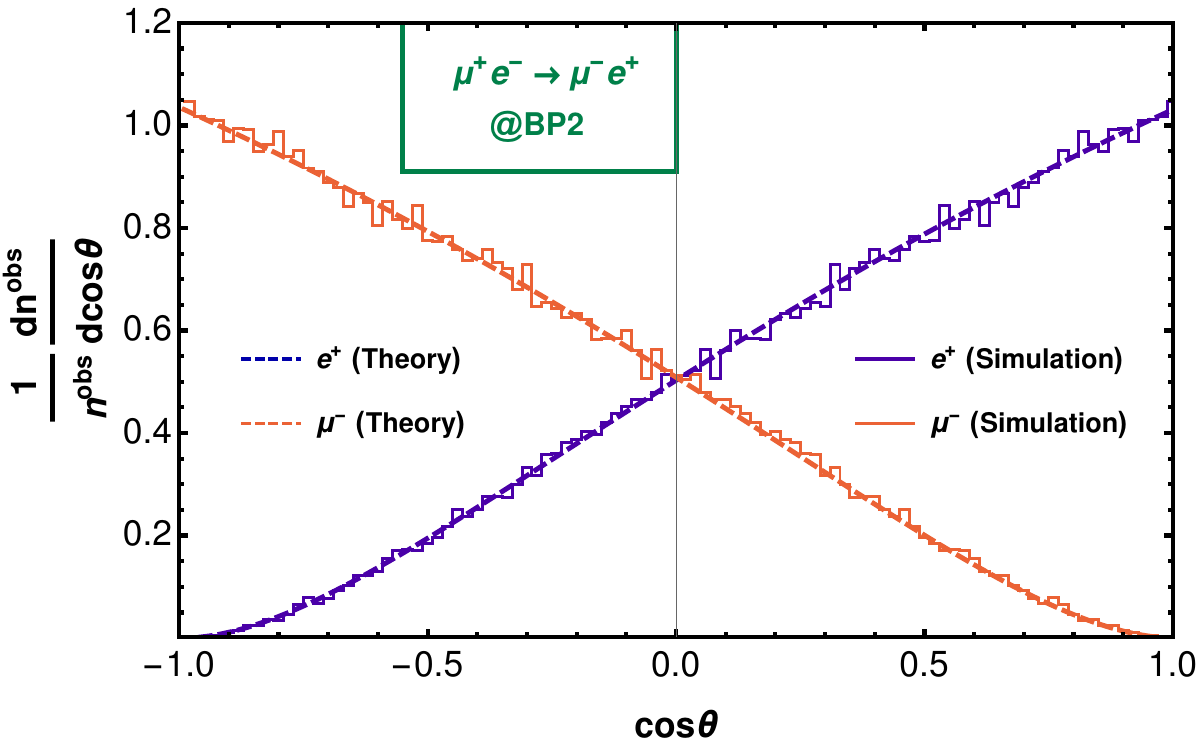}\label{}}}		
		\caption{Angular distribution of the final state charged leptons ($e^+$ in blue and $\mu^-$ in orange) for the benchmark points discussed in \autoref{muoecrsII}. \textcolor{black}{The histogram is for simulated result and the dashed line depicts the theoretical estimation.}} \label{mu+e-Ty2CosAng}
	\end{center}
\end{figure*}

We illustrate the angular distributions for the final state leptons in the CM frame with respect to the angle with the initial state muon ($\mu^+$) beam in \autoref{mu+e-Ty2CosAng}. The two panels show the distributions for the two benchmark scenarios, revealing how the pattern changes with varying centre-of-mass energies. The orange curves represent the angular distributions for the final state $\mu^-$, while the blue curves represent those for the final state $e^+$. In both plots, the solid lines denote the simulated results, and the dashed lines represent the theoretical predictions.

As observed, the blue curves start from zero at $\cos\theta = -1$ and gradually increase with $\cos\theta$, while the orange curves are the mirror image of the blue ones about the $\cos\theta = 0$ line. Notably, the curves remain convex at low centre-of-mass energies (as in BP1); however, as the interaction energy increases, the convexity diminishes, and at even higher energies, the curves become concave (as seen in BP2).

\begin{table}[h!]
	\renewcommand{\arraystretch}{1.5}
	\centering
	\begin{tabular}{|c|c|c|}
		\hline
		Final state &  BP1 & BP2  \\ 
		\hline \hline
		$1\mu ^- 1e^+$ & 149.2 & 508.2   \\ 
		\hline	
	\end{tabular}
	\caption{Number of events for $1\mu ^- 1e^+$ final state with an integrated luminosity of 1000 fb$^{-1}$.}  \label{sig_muOeType2}
\end{table}

Finally, we look for $\mu^-e^+$ final state at the collider. The results for this final state with 1000 fb$^{-1}$ of integrated luminosity are presented in \autoref{sig_muOeType2}, where we find 149 and 508 numbers of events for the two benchmark points. It is interesting to mention that there is no SM background for this final state since this is a lepton flavour violating process. 

\begin{table}[h!]
	\renewcommand{\arraystretch}{1.7}
	\centering
	\begin{tabular}{|c|c|c|c|c|}
		\cline{2-5}
		\multicolumn{1}{c|}{} & \multicolumn{4}{c|}{$\cos \theta$ for Type-II Seesaw}  \\
		\cline{2-5}
		\multicolumn{1}{c|}{} & \multicolumn{2}{c|}{($-0.7 \to 0.0$)} & \multicolumn{2}{c|}{($0.0 \to 0.7$)}  \\
		\cline{2-5}
		\multicolumn{1}{c|}{} & $~~e^+~~$ & $\mu^-$ &  $~~e^+~~$ & $\mu^-$ \\
		\hline 
		Signal & 27.2 & 110.3 & 110.7 & 27.0 \\
		\hline 
	\end{tabular}
	\caption{Number of events in the $\cos\theta$ intervals $(-0.7,0.0)$ and $(0.0,0.7)$ for the BP1 of the Type-II seesaw scenario at $\mu^+ e^-$ collider, corresponding to the final states discussed earlier, at the integrated luminosity of 1000~\fbi.}  \label{tab:assymtry4}
\end{table}
{Finally, in \autoref{tab:assymtry4} we present the number of events in the $\cos\theta$ intervals $(-0.7,0.0)$ and $(0.0,0.7)$ for the BP1 of the Type-II seesaw scenario for the final sates given in \autoref{sig_muOeType2}, at the integrated luminosity of 1000~\fbi. As it is evident from \autoref{mu+e-Ty2CosAng} that there is asymmetry between $\mu^+(e^+$ and $\mu^-(e^-)$, we quantify the same here. For the negative $\cos{theta}$  $\mu^-$ and positive $\cos{theta}$ $\mu^+$ events follow a ratio of $4:1$, where as for $e^+$ in negative $\cos{theta}$ and $e^-$ in positive $\cos{theta}$ follows a reverse ratio of $1:4$. There is no background contamination making it one of the cleanest distributions to be achieved with relatively lower integrated luminosity.  }
\section{Conclusions}

In this article, we aim to differentiate the signatures of various simple tree-level Seesaw scenarios across different leptonic colliders. {We demonstrate that the angular distributions of reconstructed BSM particles (or final state leptons) can provide valuable insights into the nature of the simple tree-level Seesaw models without any information about the other particles in the final-state.} Our analysis primarily focuses on detecting TeV-scale BSM particles with diagonal couplings of $\mathcal{O}(0.1)$. Given that the neutrino mass bounds impose stringent constraints on the couplings for Type-I and Type-III Seesaw models with TeV-scale particles, making them difficult to probe at colliders, we chose to investigate the inverse Seesaw scenarios for both Type-I and Type-III cases. A detailed analysis using {\tt PYTHIA8} is conducted at $\mu^+\mu^-$, $\mu^+\mu^+$, $\mu^+ \gamma$, and $\mu^+ e^-$ colliders, exploring different masses and centre-of-mass energies while considering the diagonal couplings for the BSM particles. 

\begin{figure*}[hbt]
	\begin{center}
	\mbox{{\includegraphics[width=0.9\linewidth,angle=-0]{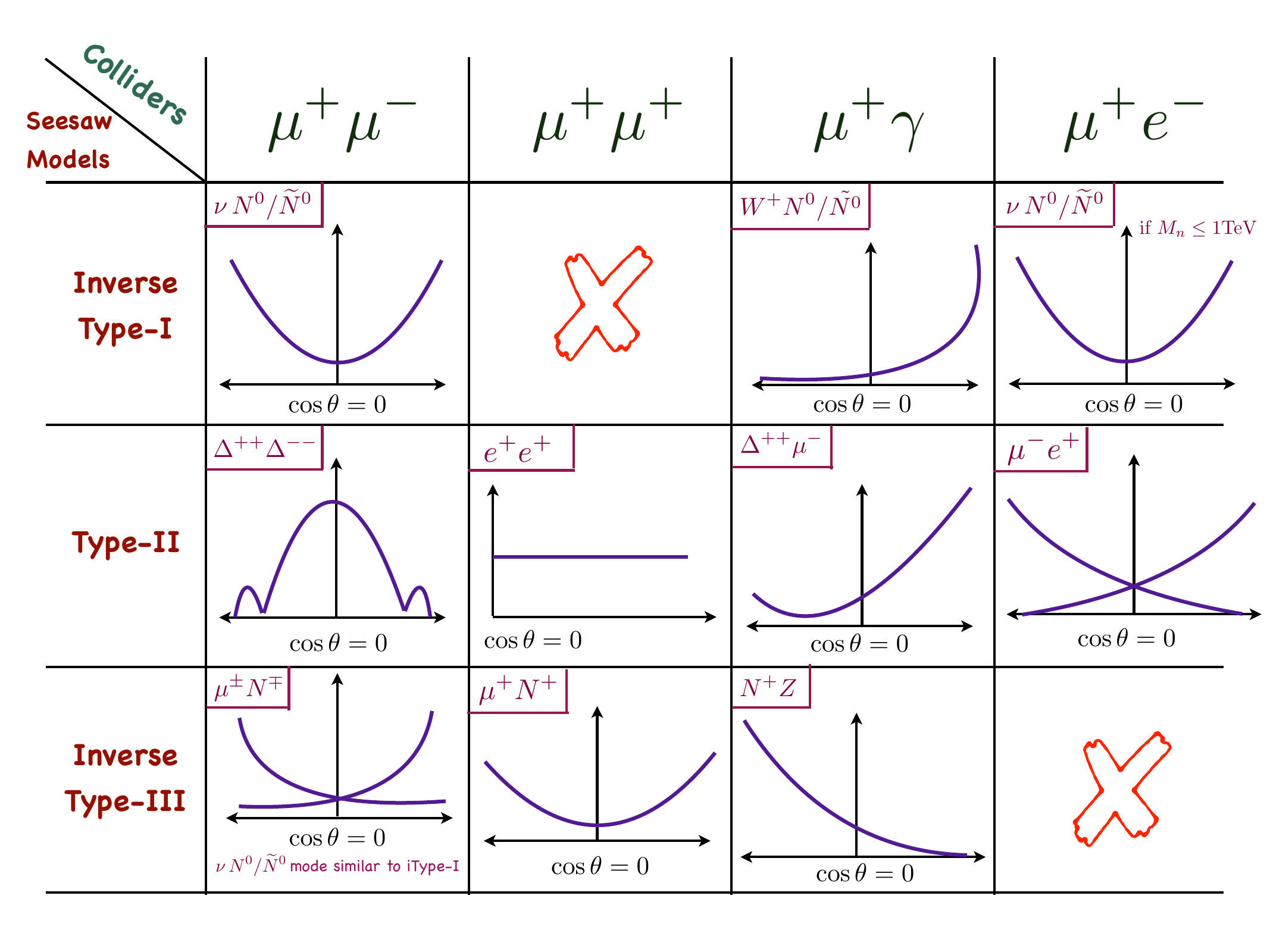}\label{}}}		
		\caption{A table of schematic diagrams on the angular distribution at $\mu^+ \mu^-$, $\mu^+ \mu^+$, $\mu^+ \gamma$ and $\mu^+ e^-$ colliders for iType-I, Type-II and iType-III Seesaw models with the detectable modes. This reflects discerning signatures of Seesaw models at various leptonic colliders.}\label{sch_ang}
	\end{center}
\end{figure*}

The angular distributions, which are instrumental in distinguishing different Seesaw scenarios at various muon(ic) colliders, are summarized in \autoref{sch_ang}. It is evident that {$\mu^+ \mu^-$ and $\mu^+ \gamma$ colliders offer an advantage, as all three types of simple tree-level Seesaw models can be probed, each displaying distinct angular distributions of the BSM particle in the final-state}. In contrast, $\mu^+ \mu^+$ and $\mu^+ e^-$ colliders cannot probe the inverse Type-I and inverse Type-III Seesaw models, respectively. However, they are still capable of distinguishing the remaining Seesaw scenarios. 

{At the $\mu^+\mu^-$ collider, we examined the dominant modes involving BSM particles in the final-states for all the three simple tree-level Seesaw scenarios.}
We first examined the $\nu N^0/\widetilde N^0$ mode, which exhibits {a bowl-like angular distribution for the reconstructed $N^0/\widetilde N^0$ (see Fig. \ref{angdismupmumF}).} {However, both the iType-I and iType-III scenarios share this feature for this particular mode.} Therefore, to confirm the existence of one of these models, one must also investigate the presence or absence of another mode, $\mu^\pm N^\mp$. {In the iType-III Seesaw,} this mode will exist with a {characteristic asymmetric angular distribution for $N^\pm$ (or $\mu^\pm$)}, whereas in the iType-I Seesaw, this channel does not exist. {For the Type-II Seesaw}, we analysed the pair production of the doubly charged scalar, which displays a {uniquely inverted-tub-like angular distribution for the BSM particle}.

{At the $\mu^+\mu^+$ collider, probing the iType-I Seesaw is not feasible}, as the lepton number-violating $W^+W^+$ mode is the only channel available, and its cross-section is very small due to the tiny lepton number-violating parameter $\mu_n$. For {the Type-II scenario}, we focus on the $e^+e^+$ mode, which {exhibits a flat angular distribution of $e^+$ in the CM frame (see Fig. \ref{angdismummumF}).} Conversely, for the {iType-III case}, we investigate the $\mu^+N^+$ mode, which is expected to {show a tub-like angular distribution for the final state muon (or the reconstructed $N^+$) in the CM frame.}

{Next, we examine the angular distribution of reconstructed BSM particles at a $\mu^+ \gamma$ collider, focusing on the three Seesaw scenarios.} In the iType-I Seesaw scenario, the $\nu N^0/\widetilde{N}^0$ particle is produced alongside a $W^+$ boson, while in the iType-III Seesaw scenario, a singly-charged fermion ($N^+$) is produced in association with a $Z/h$ boson. {Reconstructing the angular distribution in these cases is challenging because neither of the produced particles is stable}; instead, both must be reconstructed from their decay products. For instance, in the iType-III scenario, one $Z$ boson is produced directly, while another arises from the decay of the $N^+$ field. Properly tagging the appropriate $Z$ boson to reconstruct the $N^+$ particle is crucial, and we plan to address this in future work. The Type-I scenario faces similar challenges. {However, at the patron level, the angular distribution for these scenarios reveals distinct behaviours: the distribution diverges at $\cos \theta \geq 0.7$ in the iType-I Seesaw model, and the angular distribution of $N^+$ tends toward zero at $\cos \theta \sim 1.0$ (see Fig. \ref{mu-photon_ang_anal}).} For the {Type-II Seesaw scenario}, {the angular distribution of the BSM particle (doubly-charged scalar)} produced in association with a muon {exhibits a minimum at $\cos \theta \sim - 0.7$, which is a distinctive signature}. After reconstructing the doubly-charged scalar, the simulation with a monochromatic photon aligns well with the theoretical predictions, while the simulation with a laser-backscattered photon shows slight deviations, which are noted for practical applications.

Finally, inspired by the {MUonE project}, we explore the potential of a {low-energy muon-electron collider} ($\sqrt{s} \leq 1$ TeV). Given that the mass bound for the inverse Type-III Seesaw with three generations of triplets is approximately 1.2 TeV, higher CM energy would be required to probe this scenario. However, for the {Type-I case}, there is no such collider bound, allowing the search for right-handed neutrinos with masses below 1 TeV through the $\nu N^0/\widetilde{N}^0$ mode, which would exhibit a similar {bowl-like angular distribution} if the final state lepton's flavour is not tagged. The most intriguing feature at this collider is observed in the {Type-II Seesaw scenario} when examining the mode $\mu^+e^- \to \mu^-e^+$. In this case, the angular distribution of the {final state electron or muon displays a very distinctive asymmetric behaviour (see Fig. \ref{mu+e-Ty2CosAng}).}

\section*{Acknowledgements}
The authors thank SERB India, sanction no. CRG/2018/004971  and MATRICS Grant MTR/2020/000668 for the financial support. PB also thank the discussions during Phoenix 2023 (SSY/2023/001078) and Phoenix 2025 (SSY/2025/000420). AK thanks Italian Ministero dell’Universit\`a e Ricerca (MUR) concerning the research grant number 20227S3M3B under the program PRIN 2022. CS gratefully acknowledges the Ministry of Education (MOE), Government of India, for the SRF fellowship received during her PhD, and the Department of Space, Government of India, for supporting her research during her postdoctoral tenure.

\appendix

\section{Decay of $N^\pm/\widetilde{N}^\pm$ in iType-III model with one generation}
\label{appen1}
Here, we briefly outline the typical behaviour of the iType-III Seesaw, focusing on the decays of heavy charged leptons. For clarity, we illustrate the setup with a single generation of leptons and triplet fermions. Upon diagonalizing the mass matrix, the rotation matrices for negatively and positively charged leptons are denoted by $Z^L$ and $Z^R$, respectively, while the corresponding matrix for neutral leptons is denoted by $U$. These rotations determine the mixing between SM and heavy states, and hence directly control the allowed decay channels. Under the hierarchy $M_n \gg v_0 Y \gg v_0 Y_e$, with $Y_e$ the SM lepton Yukawa coupling, the rotation matrices in the iType-III Seesaw can be expressed as:

\begin{equation}
	Z^L\approx\begin{pmatrix}
		1&0&-\frac{v_0 Y}{M_n}\\
		0&1&0\\
		\frac{v_0 Y}{M_n}&0&1
	\end{pmatrix}, \quad  Z^R\approx\begin{pmatrix}
		1&0&0\\
		0&0&1\\
		0&1&0
	\end{pmatrix}, \quad 
	U\approx\begin{pmatrix}
		1&0&-\frac{v_0 Y}{\sqrt2 M_n}\\
		\frac{v_0 Y}{2M_n}&-\frac{1}{\sqrt 2} &\frac{1}{\sqrt 2}\\
		-\frac{v_0 Y}{2M_n}&-\frac{1}{\sqrt 2} &-\frac{1}{\sqrt 2}
	\end{pmatrix}.
\end{equation}
Now, the couplings for the vertices involving the decay of the heavy charged leptons in this scenario can be written as:
\begin{align}
	\bar\nu_j e_i W_\mu^+\equiv-ig_2\bigg[\Big(\frac{1}{\sqrt 2}U_{j1}Z_{i1}^L+U_{j2}Z_{i2}^L+U_{j3}Z_{i3}^L\Big)\gamma_\mu\gamma_L +\Big(U_{j2}Z^R_{i2}+U_{j3}Z^R_{i3}\Big)\gamma_\mu\gamma_R\bigg],
\end{align}
\begin{align}
	\bar e_i e_j Z_\mu\equiv i\Big\{\Big(\frac{g_2}{2}C_w-\frac{g_1}{2}S_w\Big)&Z^L_{i1}Z^L_{j1}+g_2C_w\Big(Z^L_{i2}Z^L_{j2}+Z^L_{i3}Z^L_{j3}\Big)\Big\}\gamma_\mu\gamma_L-i\Big\{g_1S_w Z^R_{i1}Z^R_{j1}\nonumber\\&
	-g_2 C_w\Big(Z^R_{i2}Z^R_{j2} +Z^R_{i3}Z^R_{j3}\Big)\Big\} \gamma_\mu\gamma_R,
\end{align}
\begin{align}
	\bar e_i e_j h\equiv -i Z^L_{j1}\Big(\frac{Y_e}{\sqrt 2}Z^R_{i1}-Y Z^R_{i2}\Big)\gamma_L-i Z^L_{i1}\Big(\frac{Y_e}{\sqrt 2}Z^R_{j1}-Y Z^R_{j2}\Big)\gamma_R,
\end{align}
where $e_i=(e^-, \widetilde N^-, N^-)$ and $\nu_j=(\nu_e,\widetilde N_0, N^0)$ for $(i,j)\in\{1,2,3\}$. One can easily check that the couplings for the vertices $\bar \nu_e N^-W_\mu^+$, $\bar e \widetilde{N}^-Z_\mu$ and $\bar e \widetilde{N}^-h$ regarding the decay of the heavy charged leptons are zero and the rest of the couplings involving transition of heavy charged leptons to SM leptons are non-zero. Therefore, we find the partial decay widths of $N^\pm/\widetilde{N}^\pm$ as:
\begin{align}
	&\Gamma(N^-\to \nu_e W^-)=0,\\
	&\Gamma({N^-}\to {he^-})=\Gamma({N^-}\to {Ze^-})\approx \frac{Y^2 M_n}{32\pi},\\
	&\Gamma(\widetilde N^-\to \nu_e W^-) \approx \frac{Y^2 M_n}{16\pi},\\
	&\Gamma({\widetilde N^-}\to {he^-})=\Gamma({\widetilde N^-}\to {Ze^-})= 0.
\end{align}

\section{Cross-sections and angular distributions}
\label{sec:formula}
Here we describe the mathematical expressions of the total cross-sections and angular distributions for relevant processes. We consider the angle of a final state particle with the initial $\mu^+$ as $\theta$ and define $x=\cos{\theta}$. The quantities $C_w$ and $S_w$ are the cosine and sine of the Weinberg angle or the weak angle, respectively.

\vspace{4mm}
\noindent
$\bullet$ {\large\textbf{At $\bm{\mu^+\mu^-}$ collider:}}

\vspace{2mm}
\noindent
i) \underline{iTpe-I seesaw:}
\begin{align}\label{typ1}
	&\frac{d\sigma}{dx}(\mu^+\mu^-\to \nu N^0/\widetilde{N}^0)
	=\frac{m_W^4 Y^2}{\pi\,s\,v_0^2 M_n^2} \bigg[\frac{\lambda^2(\lambda+2\epsilon_w)^2}{\big\{(\lambda+2\epsilon_w)^2-\lambda^2 x^2\big\}^2}\bigg],\\[5 pt]
	&\sigma(\mu^+\mu^-\to \nu N^0/\widetilde{N}^0)
	=\frac{Y^2 \,\lambda\,m_W^2}{4\pi\,v_0^2\,M_n^2} \bigg[\frac{\lambda}{(\lambda+\epsilon_w)}-\frac{2\epsilon_w\log\big(1+\frac{\lambda}{\epsilon_w}\big)}{(\lambda+2\epsilon_w)}\bigg],
\end{align}
where, $\epsilon_w=m_W^2/s$ and $\lambda=(1-M_n^2/s)$.

\vspace{2mm}
\noindent
ii) \underline{Tpe-II seesaw:}
\begin{align}\label{typII}
	\frac{d\sigma}{dx}(\mu^+\mu^-&\to\Delta^{++}\Delta^{--})=
	\frac{\beta^{3}(1-x^2)}{512\pi sS_w^4 C_w^4(1-\epsilon_z)^2(1+\beta x-2\epsilon_\Delta)^2}
	\;\bigg[e^4\Big\{4\epsilon_\Delta^2-4\epsilon_\Delta(1+\beta x+x^2) \nonumber\\
	&+(1+2\beta x+x^2)\Big\} \; \Big\{32 \epsilon_z^2 S_w^4 C_w^4-8\epsilon_z C_w^2S_w^2(1+2S_w^2)+(1+4S_w^4)\Big\} -4 e^2 Y_\Delta^2 S_w^2 C_w^2 \nonumber\\
	&(1+\beta x-2\epsilon_\Delta) \Big\{1+4\epsilon_z^2 S_w^2 C_w^2- \epsilon_z (1+4S_w^2-4S_w^4)\Big\} +4 Y_\Delta^4 S_w^4 C_w^4 (1-\epsilon_z)^2 \bigg],
\end{align}
\begin{align}
	\sigma(\mu^+\mu^-&\to\Delta^{++}\Delta^{--})=\frac{1}{384\pi s S_w^4 C_w^4(1-\epsilon_z)^2}\; \bigg[-\beta\Big\{12Y_\Delta^4S_w^4 C_w^4(1-\epsilon_z)^2-e^4\beta^2\Big(1-8S_w^2\epsilon_z \nonumber\\
	&+4S_w^4\Big(1+2\epsilon_z \times (-1+2S_w^2+4\epsilon_zC_w^4)\Big)\Big)
	+6e^2Y_\Delta^2S_w^2C_w^2(1-\epsilon_z)(1-2\epsilon_\Delta)(1-4\epsilon_zC_w^2S_w^2)\Big\} \nonumber\\ 
	&+12Y_\Delta^2S_w^2 C_w^2(1-\epsilon_z)\; \Big\{Y_\Delta^2S_w^2C_w^2 (1-\epsilon_z)(1-2\epsilon_\Delta) +2e^2\epsilon_\Delta^2 (1-4\epsilon_zS_w^2C_w^2)\Big\} \coth^{-1}\bigg(\frac{1-2\epsilon_\Delta}{\beta}\bigg)\bigg],
\end{align}
where, $\epsilon_\Delta=M_\Delta^2/s$, $\epsilon_z=m_Z^2/s$,  $\beta=\sqrt{1-4M_\Delta^2/s}$, $x$ is the cosine of the angle of doubly charged particle with the beam axis and $e$ is the electric charge of the positron.

\vspace{2mm}
\noindent
iii) \underline{iTpe-III seesaw:}
\begin{align}\label{lNds}
	&\frac{d\sigma}{dx}(\mu^+\mu^-\to \mu^{\pm} N^{\mp})=\frac{m_Z^4 Y^2 \lambda^2}{16\pi s v_0^2 M_n^2}\bigg[\frac{16S_w^4}{(\lambda-\lambda x+2\epsilon_z)^2}
	+\frac{(1-2S_w^2)^2(1-x^2)(1-\lambda+\frac{1+x}{1-x})}{(\lambda-\lambda x+2\epsilon_z)^2}\bigg],\\[7 pt]
	&\sigma(\mu^+\mu^-\to \mu^{\pm} N^{\mp})=\frac{m_Z^4 Y^2}{8\pi s v_0^2 M_n^2}\bigg[(1-2S_w^2)^2\,\Big(2\lambda+\frac{\lambda}{\epsilon_z}\Big)
	+\frac{4\lambda^2 S_w^4}{\epsilon_z(\lambda+\epsilon_z)}\nonumber\\
    &\hspace{7cm}-(1-2S_w^2)^2\,(1+\lambda+2\epsilon_z)\log\Big(1+\frac{\lambda}{\epsilon_z}\Big)\bigg],
\end{align}
with $\epsilon_z=m_Z^2/s$, $\lambda=(1-M_n^2/s)$, and $x$ is the cosine of the angle of charged heavy fermion with the beam axis.

\vspace{3mm}
\noindent
$\bullet$ {\large\textbf{At ${\mu^+\mu^+}$ collider:}}

\vspace{2mm}
\noindent
i) \underline{Type-II seesaw:}
\begin{align}
	\label{sigmumu}
	\sigma(\mu^+ \mu^+ \to  e^{+} e^{+} )&=\int_{0}^{1}\frac{d\sigma}{dx}(\mu^+ \mu^+ \to e^{+} e^{+} )dx =\frac{s\,Y_\Delta^4}{128\pi[(s-M_\Delta^2)^{2}+M_\Delta^2\Gamma_\Delta^2]},
\end{align}
where, $\sqrt s$ is the centre-of-mass energy, $\Gamma_\Delta\big(\approx\frac{3M_\Delta Y_\Delta^2}{32\pi}\big)$ is the total decay width of $\Delta^{++}$ (branching of di-bosons are negligible, since, $v_\Delta$ is very small for our benchmark scenarios)  and $x$ is cosine of the angle $(\theta)$ between the beam axis and any one of the final state electrons. It is interesting to mention here that the angle $\theta$ in this case will vary from 0 to $\pi/2$ only, since the two particles in final state are indistinguishable.

\vspace{2mm}
\noindent
ii) \underline{iType-III seesaw:}
\small
\begin{align}
	\label{sigeE}
	&\frac{d\sigma}{dx}({\mu^+\mu^+\to \mu^+ N^+})=\frac{m_Z^4 Y^2\lambda^2}{4\pi s v_0^2M_n^2(\zeta^2-\lambda^2 x^2)^2}\Big[2\zeta^2(1-2S_w^2)^2
	+S_w^4\big\{(1-x^2)(2\zeta^2-\lambda\zeta^2\nonumber\\
    &\hspace{10cm}-x^2\lambda^3)+2 x^2(\zeta+\lambda)^2\big\}\Big],\\ 
	&\sigma({\mu^+\mu^+\to \mu^+ N^+})=\frac{m_Z^4 Y^2\lambda}{4\pi s v_0^2 M_n^2}\Big[\frac{\lambda(1-2S_w^2)^2}{\zeta^2-\lambda^2}+2S_w^4\Big(1+\frac{1}{\zeta-\lambda}\Big)
	+\Big\{\frac{1}{\zeta}(1-2S_w^2)^2\nonumber\\
    &\hspace{9cm}-2S_w^4\frac{(1+\zeta)}{\lambda}\Big\}\tanh^{-1}\Big(\frac{\lambda}{\zeta}\Big)\Big]
\end{align}
\normalsize
where, $\lambda=1-\frac{M_n^2}{s}$\,, $\zeta=1+\frac{2}{s}(m_Z^2-M_n^2)$ and $x$ is the cosine of angle between final state muon and beam axis.

\vspace{4mm}
\noindent
$\bullet$ {\large\textbf{At ${\mu^+\gamma}$ collider:}}

\vspace{2mm}
\noindent
i) \underline{iType-I seesaw:}

\begin{align}\label{ty1mg}
	\frac{d\sigma}{d x}\,(\mu^+\gamma &\to W^+ N^0/\widetilde N^0)=\frac{e^4 v_0^2\,Y^2\,|\vec p_w|}{128\pi s^{5/2}S_w^2\epsilon_w \epsilon_n(1+\eta_{x,w}-\epsilon_n)^2}\Big[4\epsilon_w^3(\epsilon_n-\eta_{x,w})
	\nonumber\\
	&+2\epsilon_w^2(2\eta_{x,w}^2+2\eta_{x,w}-\eta_{x,w} \epsilon_n-\epsilon_n^2)+2\epsilon_w\{\epsilon_n-\eta_{x,w}(1+\epsilon_n)-\eta_{x,w}(\eta_{x,w}-\epsilon_n)^2\}\nonumber\\
	&
	-\epsilon_n(1+\eta_{x,w}-\epsilon_n)(\eta_{x,w}^2+\eta_{x,w}+\epsilon_n^2-\epsilon_n)\Big],
\end{align}
\begin{align}\label{ty1mg2}
	\sigma(\mu^+\gamma&\to W^+ N^0/\widetilde N^0)=\frac{e^4 v_0^2\,Y^2\,|\vec p_w|}{128\pi s^{5/2}S_w^2\epsilon_w \epsilon_n}\Big[8+\epsilon_n+10\epsilon_w+7\,(2\epsilon_w^2-\epsilon_n\epsilon_w-\epsilon_w^2)\nonumber\\	&-\frac{4\sqrt s}{|\vec  p_w|}\{\epsilon_n^3-\epsilon_n^2- \epsilon_n\epsilon_w (1+3\epsilon_w)+2\epsilon_w(\epsilon_w^2+\epsilon_w+2)\}\log\bigg|\frac{1+\epsilon_w-\epsilon_n+\frac{2|\vec p_w|}{\sqrt s}}{2\sqrt \epsilon_w}\bigg|\Big],
\end{align}
where,	$\epsilon_n=M^2_n/s$, $\epsilon_w=m_W^2/s$, $\eta_{x,w}=\frac{1}{2}\Big(\epsilon_n+\epsilon_w-1-\frac{2x\,|\vec p_w|}{\sqrt s}\Big)$ and $|\vec p_w|= \frac{\sqrt s}{2}\, \big[(1-\epsilon_n)^2+\epsilon_w^2-2\epsilon_w(1+\epsilon_n)\big]^{1/2}$,
$x$ is the cosine of the angle between muon and $N^0/\widetilde{N}^0$.

\vspace{2mm}
\noindent
ii) \underline{Type-II seesaw:}

\begin{align}\label{typeIImg}
	\frac{d\sigma}{d x}\,&(\mu^+\gamma\to \Delta^{++} \mu^-)=\frac{e^2\, Y_\Delta^2}{128\,\pi\,s}\bigg[\frac{2+\lambda_\Delta(1-x)}{2-\lambda_\Delta(1-x)}\bigg]^2\, \bigg[\frac{\lambda_\Delta^2\,(x^2+2x+5)-8\lambda_\Delta+4}{(1-x)}\bigg],
\end{align}
\begin{align}\label{typeIImg2}
	\sigma\,(\mu^+\gamma\to \Delta^{++} \mu^-)=\frac{e^2\, Y_\Delta^2}{64\,\pi\,s}\Big[(\ln 4+5)+2\,\epsilon_\Delta(7-\ln 4+8\ln \epsilon_\Delta)-\epsilon_\Delta^2(19-2\ln 4)\Big],
\end{align}
where, $\lambda_\Delta=(1-\frac{M_\Delta^2}{s})$, $\epsilon_\Delta=\frac{M_\Delta^2}{s}$ and $x$ being the cosine of the angle between initial state muon and $\Delta^{++}$.

\vspace{2mm}
\noindent
iii) \underline{iType-III seesaw:}

\begin{align}\label{TypeIIIZmg}
	\frac{d\sigma}{d x}\,&(\mu^+\gamma\to Z N^+)=\frac{e^4\,\epsilon_v\, |\vec p_z|}{128\,\pi s^{3/2}\,C_w^2 S_w^2\,\epsilon_n^3\epsilon_z\,(\eta_{x,z}-\epsilon_n)^2}
	\bigg[\epsilon_n^6-\epsilon_n^5(\eta_{x,z}+2-2\epsilon_v)
	-2\epsilon_v^2\epsilon_z\eta_{x,z}\{(1-\epsilon_z)^2\nonumber\\
	&+(\eta_{x,z}-\epsilon_z)^2\}+\epsilon_n^4\{1-4\epsilon_v+\epsilon_v^2 +2\epsilon_v\epsilon_z-2\epsilon_z^2+2\eta_{x,z}
	-4\eta_{x,z}\epsilon_v-2\epsilon_z\eta_{x,z}+\eta_{x,z}^2\}\nonumber\\
	&+\epsilon_n\epsilon_v\{\epsilon_v(4\epsilon_z^3-6\epsilon_z^2-2\epsilon_z\eta_{x,z}+2\epsilon_z\eta_{x,z}^2-\eta_{x,z}^3) -4\epsilon_z\eta_{x,z}(1+2\epsilon_z^2+\eta_{x,z}^2-2\epsilon_z-2\epsilon_z\eta_{x,z})\}\nonumber\\
	&+\epsilon_n^3\{4\epsilon_z^3-\epsilon_v^2(2-2\epsilon_z+3\eta_{x,z})
	-2\eta_{x,z}\epsilon_z^2-\eta_{x,z}(1+\eta_{x,z})^2+\epsilon_z(2-2\eta_{x,z}+4\eta_{x,z}^2)\nonumber\\
	&+2\epsilon_v(1-3\epsilon_z^2+2\epsilon_z-2\epsilon_z\eta_{x,z}+2\eta_{x,z}^2+2\eta_{x,z})\}+\epsilon_n^2\{\epsilon_v^2(1-4\epsilon_z^2+3\epsilon_z-2\epsilon_z\eta_{x,z}+3\eta_{x,z}+3\eta_{x,z}^2)\nonumber\\
	&+2\epsilon_v(4\epsilon_z^3-2\epsilon_z^2-\epsilon_z^2\eta_{x,z}-\eta_{x,z}^2-\eta_{x,z}^3+3\epsilon_z\eta_{x,z}^2-3\epsilon_z\eta_{x,z})-2\epsilon_z\eta_{x,z}(1+2\epsilon_z^2+\eta_{x,z}^2-2\epsilon_z-2\epsilon_z\eta_{x,z})\}\bigg],
\end{align}
\begin{align}\label{TypeIIIZmg2}
	\sigma\,&(\mu^+\gamma\to Z N^+)=\frac{e^4\,\epsilon_v\, |\vec p_z|}{128\,\pi s^{3/2}\,C_w^2 S_w^2\,\epsilon_n^3\epsilon_z\,}\bigg[-7\epsilon_n^4-2\epsilon_z\epsilon_v^2(1-\epsilon_z)
	-\epsilon_n^3(3+6\epsilon_v+7\epsilon_z)\nonumber\\
	&+\epsilon_n\epsilon_v(3\epsilon_v-5\epsilon_v\epsilon_z-4\epsilon_z+20\epsilon_z^2)
	+\epsilon_n^2(3\epsilon_v^2+2\epsilon_v-14\epsilon_v\epsilon_z+2\epsilon_z+14\epsilon_z^2)\nonumber\\
	&-\frac{2\sqrt s}{\, |\vec p_z|}\,\Big\{\epsilon_n^2(2\epsilon_n^3 +2\epsilon_n^2+\epsilon_n+2\epsilon_n^2\epsilon_v-3\epsilon_v^2)+2\epsilon_z\big(\epsilon_v^2+\epsilon_n \epsilon_v(2+\epsilon_v)
+\epsilon_n^3(1+2\epsilon_v)+\epsilon_n^2(1+\epsilon_v)(1+2\epsilon_v)\big)\nonumber\\
	&-2\epsilon_z^2(\epsilon_n+\epsilon_v)(2\epsilon_v+2\epsilon_n+3\epsilon_n^2+4\epsilon_n\epsilon_v)+4\epsilon_z^3(\epsilon_n+\epsilon_v)^2\Big\}
	\ln\bigg|\frac{1+\epsilon_n-\epsilon_z-\frac{2|\vec p_z|}{\sqrt s}}{2\sqrt \epsilon_n}\bigg|\bigg],
\end{align}
\begin{align}\label{TypeIIIhmg}
	\frac{d\sigma}{d x}\,&(\mu^+\gamma\to h N^+)=\frac{e^2\, Y^2\,|\vec p_h|}{32\pi\, s^{3/2}\,\epsilon_n^2\,(\eta_{x,h}-\epsilon_n)^2}\Big[\epsilon_n\{\epsilon_n^2(2\epsilon_h^2-2\epsilon_h\epsilon_n+(1-\epsilon_n)^2)
	+2\epsilon_n\epsilon_v(\epsilon_h^2-\epsilon_h\epsilon_n+(1-\epsilon_n)^2)\nonumber\\
	&+\epsilon_v^2((1-\epsilon_n)^2-\epsilon_h)\}-\epsilon_n\eta_{x,h}\{2\epsilon_h(\epsilon_n+\epsilon_v)(\epsilon_h-1)+(\epsilon_n-1)(\epsilon_n^2+3\epsilon_v^2+4\epsilon_n\epsilon_v-\epsilon_n)\}\nonumber\\	&+\epsilon_n\,\eta_{x,h}^2(\epsilon_n+\epsilon_v)(2\epsilon_h+\epsilon_n+3\epsilon_v-2)
	-\eta_{x,h}^3(\epsilon_n+\epsilon_v)^2\Big],
\end{align}
\begin{align}\label{TypeIIIhmg2}
	\sigma\,(\mu^+\gamma&\to h N^+)=\frac{e^2\, Y^2\,|\vec p_h|}{32\pi\, s^{3/2}\,\epsilon_n^2}\Big[-7\epsilon_n^3-\epsilon_n^2(3+6\epsilon_v-7\epsilon_h)+3(1-\epsilon_h)\epsilon_v^2+\epsilon_n\epsilon_v(2+3\epsilon_v+6\epsilon_h)-\frac{2\,\epsilon_n\sqrt s}{|\vec p_h|}\{2\epsilon_n^3\nonumber\\
	&+2\epsilon_n^2(1+\epsilon_v-2\epsilon_h)-2\epsilon_h\epsilon_n(1+2\epsilon_v-\epsilon_h)+\epsilon_n-\epsilon_v(2\epsilon_h-2\epsilon_h^2+3\epsilon_v)\}\ln\bigg|\frac{1+\epsilon_n-\epsilon_z-\frac{2|\vec p_z|}{\sqrt s}}{2\sqrt \epsilon_n}\bigg|\bigg],
\end{align}
where,	$\epsilon_n=M^2_n/s$, $\epsilon_z=m_Z^2/s$, $\epsilon_h=m_h^2/s$, $\epsilon_v=Y^2 v_0^2/s$, $\eta_{x,z}=\frac{1}{2}\Big(\epsilon_n+\epsilon_z-1-\frac{2x\,|\vec p_z|}{\sqrt s}\Big)$, $\eta_{x,h}=\frac{1}{2}\Big(\epsilon_n+\epsilon_h-1-\frac{2x\,|\vec p_h|}{\sqrt s}\Big)$, $|\vec p_z|= \frac{\sqrt s}{2}\, \big[(1-\epsilon_n)^2+\epsilon_z^2-2\epsilon_z(1+\epsilon_n)\big]^{1/2}$ and  $|\vec p_h|= \frac{\sqrt s}{2}\, \big[(1-\epsilon_n)^2+\epsilon_h^2-2\epsilon_h(1+\epsilon_n)\big]^{1/2}$, $x=$ is cosine of the angle between the initial state muon and $N^+$.

\vspace{4mm}
\noindent
$\bullet$ {\large\textbf{At ${\mu^+e^-}$ collider:}}

\vspace{2mm}
\noindent
i) \underline{Type-II seesaw:}

\begin{align}
	&\frac{d\sigma}{dx}(\mu^+e^-\to \mu^- e^+)=\frac{Y_\Delta^4}{128\pi s} \frac{(1-x)^2}{(1-x+2\epsilon_\Delta)^2}~,\\
	&\sigma (\mu^+e^-\to \mu^- e^+)=\frac{Y_\Delta^4}{64\pi s}\bigg[\frac{1+2\epsilon_\Delta}{1+\epsilon_\Delta}-2\epsilon_\Delta \log\Big(1+\frac{1}{\epsilon_\Delta}\Big)\bigg],
\end{align}
where, $\epsilon_\Delta=M_\Delta^2/s$ and $x$ is the cosine of the angle between $\mu^+$ and $\mu^-$.

\bibliography{Reference_lepcol}

\end{document}